%% file: main.tex
\documentclass[%
nofootinbib,
 amsmath,amssymb,
 aps,
pra,
]{revtex4-2}

\usepackage{graphicx}
\usepackage{dcolumn}
\usepackage{bbold}

\usepackage{mathtools}
\usepackage{xcolor}
\usepackage{blindtext}
\usepackage{subcaption}
\usepackage{caption}
\usepackage{hyperref}
\usepackage{tikz}
\usetikzlibrary{decorations.markings, calc, patterns,patterns.meta}
\usepackage{stmaryrd}

\tikzset{
		propagator/.style={line width=1pt},
		dashed_propagator/.style={dashed,line width=1pt},
		arrowstyle/.style={decoration={markings, mark=at position #1 with {\arrow[scale=1]{stealth}}}, postaction={decorate}},
        reversed_arrow/.style={
        postaction={decorate},
        decoration={markings, mark=at position 1 with {\arrow[scale=1,rotate=180]{stealth}}}
    },
		every path/.style={propagator},
		every node/.style={scale=1.}
	}

\newcommand{\PlainLoopA}{
    \draw[propagator] (0,0) circle[radius=1];
}

\newcommand{\PlainLoopB}{
    \draw[propagator] (0,1) circle[radius=1];
}

\newcommand{\DeltaRBTree}[2]{
	\coordinate (M) at ($(#1)!0.5!(#2)$);
	
	\draw[propagator] (#1) -- (M);
	\draw[dashed_propagator, stealth-] (M) -- (#2);
}
	
\newcommand{\ThreePointVertices}[2]{
	\filldraw[] (#1) circle (2pt);
	\filldraw[] (#2) circle (2pt);
}

\newcommand{\FourPointVertex}[1]{
	\filldraw[] (#1) circle (2pt);
}

\newcommand{\DeltaRRTree}[2]{
	\coordinate (M) at ($(#1)!0.5!(#2)$);
	
	\draw[propagator, decoration={markings, 
		mark=at position 0.2 with {\arrowreversed{stealth}}, 
		mark=at position 0.8 with {\arrow{stealth}}}, 
	postaction={decorate}] 
	(#1) -- (#2);	
	\filldraw[] (M) circle (1.5pt);
}

\newcommand{\DeltaRRTreeDown}{
	\draw[propagator, decoration={markings, 
		mark=at position 0.25 with {\arrowreversed{stealth}}, 
		mark=at position 0.75 with {\arrow{stealth}}}, 
	postaction={decorate}] 
	(-1,0) arc[start angle=180, end angle=360, radius=1];
	\filldraw[] (0,-1) circle (1.5pt);
}	
	
\newcommand{\DeltaRRTreeUp}{
	\draw[propagator,  decoration={markings, 
		mark=at position 0.25 with {\arrowreversed{stealth}}, 
		mark=at position 0.75 with {\arrow{stealth}}}, 
	postaction={decorate}] 
	(1,0) arc[start angle=0, end angle=180, radius=1];
	\filldraw[] (0,1) circle (1.5pt);
}

\newcommand{\DeltaRBTreeUpL}{
	\draw[propagator] 
	(-1,0) arc[start angle=180, end angle=90, radius=1];
	\draw[propagator, dashed, -stealth, 
	postaction={decorate}] 
	(1,0) arc[start angle=0, end angle=90, radius=1];
}

\newcommand{\DeltaRBTreeUpR}{
	\draw[propagator] 
	(1,0) arc[start angle=0, end angle=90, radius=1];
	
	\draw[propagator, dashed, -stealth, 
	postaction={decorate}] 
	(-1,0) arc[start angle=180, end angle=90, radius=1];
	
}

\newcommand{\DeltaRBTreeDownR}{
	\draw[propagator] 
	(1,0) arc[start angle=0, end angle=-90, radius=1];
	
	\draw[propagator, dashed, -stealth, 
	postaction={decorate}] 
	(-1,0) arc[start angle=180, end angle=270, radius=1];
}

\newcommand{\DeltaRBTreeDownL}{
	\draw[propagator] 
	(-1,0) arc[start angle=180, end angle=270, radius=1];
	
	\draw[propagator, dashed, -stealth, 
	postaction={decorate}] 
	(1,0) arc[start angle=360, end angle=270, radius=1];
}

\newcommand{\DeltaRRTreeRound}{
    \draw[propagator, decoration={markings, 
				mark=at position 0.125 with {\arrowreversed{stealth}}, 
				mark=at position 0.375 with {\arrow{stealth}}}, 
			postaction={decorate}] 
			(0, 1) circle (1);
            \filldraw[] (0,2) circle (1.5pt);
}

\newcommand{\DeltaRBTreeLRound}{
    \draw[propagator, dashed, -stealth, postaction={decorate}] 
				(0,0) arc[start angle=270, end angle=450, radius=1];
				
	\draw[propagator] (0,0) arc[start angle=270, end angle=90, radius=1];
				
}

\newcommand{\nvertex}{    
    \draw[propagator, opacity=0.3, reversed_arrow] (0,0) -- ( 40:0.9cm) ;
    \draw[propagator, dashed, opacity=0.3, -stealth] (0,0) -- ( 140:1.2cm) ;
    \draw[loosely dotted, thick] ( 160:1cm) to [out=225,in=135] node [sloped,above] {} ( 200:1cm);
    \draw ( 120:1.3cm) node [left] {$X_1$};
    \draw ( 60:1.3cm) node [right] {$X_{1}^\prime$};
    \draw[propagator, opacity=0.3, dashed, -stealth] (0,0) -- ( 220:1.2cm) ;
    \draw[propagator, opacity=0.3, reversed_arrow] (0,0) -- ( 320:0.9cm) ;
    \draw[loosely dotted, thick] ( 340:1cm) to [out=45,in=-45] node [sloped,above] {} ( 20:1cm);
    \draw ( 240:1.3cm) node [sloped, left] {$X_{r}$};
    \draw ( 300:1.3cm) node [sloped,right] {$X_{s}^\prime$};
    \filldraw[] (0,0) circle (2pt);
}

\newcommand{\ie}{i.\,e.~}

\newcommand{\eg}{e.\,g.~}

\newcommand{\wrt}{w.\,r.\,t.~}

\newcommand{\mi}{\mathrm{i}}
\newcommand{\ms}[1]{\mathsf{#1}}
\newcommand{\md}{\mathrm{d}}

\newcommand{\ini}[1]{#1^{(\mathrm{i})}}
\newcommand{\vini}[1]{\vec{\mathsf{#1}}^{\,(\mathrm{i})}}
\newcommand{\tini}[1]{\boldsymbol{\mathsf{#1}}^{(\mathrm{i})}}

\newcommand{\tens}[1]{\boldsymbol{\mathsf{#1}}}

\newcommand{\RArrow}[1]{\parbox{#1}{\tikz{\draw[->](0,0)--(#1,0);}}}
\newcommand{\ShortRArrow}{\RArrow{0.2cm}}
\newcommand{\IntOp}[4]{\hat{\mathcal{L}}_{#1\ShortRArrow #2}(#3; #4)}

\newcommand{\fin}[1]{#1^{(\mathrm{f})}}

\newcommand{\tfin}[1]{\boldsymbol{\mathsf{#1}}^{(\mathrm{f})}}

\newcommand{\e}{\mathrm{e}}
\newcommand{\mean}[1]{\left\langle #1 \right\rangle}
\newcommand{\dirac}[1]{\delta_\mathrm{D}\left(#1\right)}

\begin{document}


\title{Cosmic Large-Scale Structure Formation from Newtonian Particle Dynamics}

\author{Tristan Daus}
\email{t.daus@thphys.uni-heidelberg.de}
\affiliation{%
 Institute for Theoretical Physics, Heidelberg University, Philosophenweg 12, 69120 Heidelberg 
}%
\author{Elena Kozlikin}%
 \email{elena.kozlikin@uni-heidelberg.de}
\affiliation{%
Institute for Theoretical Physics, Heidelberg University, Philosophenweg 12, 69120 Heidelberg 
}%


\begin{abstract}
We present results for the cosmic non-linear density-fluctuation power spectrum based on the analytical formalism developed in \cite{Daus2024} which allows us to study cosmic structure formation based on Newtonian particle dynamics in phase-space. This framework provides a field-theory approach to a perturbative solution of the BBGKY-hierarchy where the resulting loop-expansion of the theory introduces a natural truncation criterion.
We show that we are able to reproduce structure growth on large scales $k \leq 0.2 \mathrm{h}\,\mathrm{Mpc}^{-1}$ to very high precision while on small and intermediate scales we find deviations of the order of $10\%$ from current numerical simulations. The results strongly suggest that a significant improvement may be achieved by restructuring the perturbation theory. 
\end{abstract}

\maketitle



\section{Introduction}
The large-scale structure of the Universe contains an incredible wealth of information. At high redshift, temperature fluctuations in the cosmic microwave background (CMB) allow us to constrain the energy and matter content of the Universe and provide important evidence for the existence of dark matter and dark energy. Further constraints on their properties can be inferred by studying the evolution of those initial fluctuations. The current generation of large-scale surveys such as Euclid, LSST and SDSS \cite{Euclid, LSST, SDSS} already provide data of an exquisite quality which can only be fully leveraged if our description of cosmic structure formation is sufficiently well understood. So far, numerical $N$-body simulations still provide the most accurate results for cosmic structure formation. The simple idea of propagating particles according to Newtonian dynamics -- supplemented by sophisticated algorithms to reduce computational costs -- appears to reproduce the best picture of the evolution of cosmic structure. However, $N$-body simulations come with a high computational cost and inevitable problems of shot noise, limited resolution, sample variance, and incomplete coverage of the relevant range of scales. 
Without analytical approaches which complement numerical simulations, it will not be possible to reliably analyse large portions of the available data from current and especially from upcoming surveys. In addition, numerical simulations offer only limited insight into the underlying physical processes governing cosmic structure formation. Questions, for instance, about the universality of density profiles of dark matter haloes or the highly non-linear dynamics that leads to the formation of gravitationally bound objects, require a fundamental understanding of cosmic structure formation.
To provide such a framework which allows us to study the fundamental principles governing structure formation for in- and out-of-equilibrium systems is the main goal of the analytical approach presented in \cite{Daus2024}. It describes the evolution of a classical $N$-particles ensemble based on the path-integral formulation of classical mechanics. In contrast to the available analytical approaches based on fluid dynamics, such as SPT, EFT or EFTofLSS \cite{Bernardeau_2001, CrocceScoccimarro_2006, Matarrese_2007, Pietroni_2012, Baumann_2012, Hertzberg_2014, Carrasco_2014, Konstandin_2019}, the path-integral formulation of \cite{Daus2024} is based on particle dynamics. Just as in numerical $N$-body simulations, the ability to describe structure formation even on small scales is based on the fact that we follow the phase-space evolution of the particle ensemble in time. Starting from Liouvilles's equation we have, therefore, rigorously derived a microscopic perturbation theory (PT) which corresponds to an iterative solution of the well-known BBGKY-hierarchy based on the path-integral formulation of classical mechanics. Drawing on the formal resemblance to many-body quantum mechanics, we were able to apply a slightly modified version of the Hubbard-Stratonovich transformation (HST) which leads to an effective field-theory description of the microscopic physics in terms of macroscopic fields. In \cite{Daus2024} we have shown that this field theory presents a novel resummation scheme of microscopic processes. The perturbative loop-expansion of the macroscopic field theory in terms of propagators and vertices contains an infinite number of microscopic interactions already at lowest order. It resums microscopic processes which appear in arbitrary high orders of the microscopic PT. Viewed on a microscopic level, the loop-expansion provides a natural closure relation for the BBGKY-hierarchy of equations.  

The approach presented in \cite{Daus2024} generalises and extends concepts of Kinetic Field Theory (KFT) \cite{Bartelmann_2014, Fabis_2018, Lilow_2019, Pixius_2022}. The key idea of describing the temporal evolution of an out-of-equilibrium $N$-particle ensemble using the path-integral formulation of classical mechanics underlies both formulations. The formulation of \cite{Daus2024} in configuration space -- instead of the $k$-space formulation of KFT -- allows us to easily draw clear connections to established concepts of statistical physics and quantum many-body systems. Derivations presented in \cite{Bartelmann_2014, Fabis_2018, Lilow_2019, Pixius_2022, Viermann_2015} could thus be considerably simplified and put into context of established methods as we show in \cite{Daus2024} and the present work. While KFT is well-adapted for spatially homogeneous systems, the formulation of \cite{Daus2024} presents a framework which is well suited to treat homogeneous as well as inhomogeneous systems in and out of equilibrium. This allows us to not only simplify but also extend the works of \cite{Fabis_2018} and \cite{Lilow_2019} to inhomogeneous systems. 

The connections drawn in \cite{Daus2024} to the BBGKY-hierarchy and the Vlasov equation highlight the advantages of a particle-based approach in comparison to fluid-based approaches such as SPT. The \textit{stosszahlansatz} and the single-stream approximation on which Eulerian SPT is built, cause a permanent loss of information about microscopic degrees of freedom. Such a loss of microscopic information does not occur in the formalism of \cite{Daus2024} or in KFT \cite{Kozlikin_2020} since microscopic processes are included order by order in the perturbative expansion. The particle-based approach \cite{Daus2024} is, therefore, valid on all scales.\\   

In this work we present results for the cosmic density-fluctuation power spectrum for the tree-level and one-loop order in the field-theory framework of \cite{Daus2024}. Before presenting our results, however, we first introduce the notation and conventions for this paper in Sec.\ref{sec:NotationConventions} and familiarise the reader with the theoretical framework in Sec.\ref{sec:Basics}. We then specify the framework to the cosmological application in Sec.\ref{sec:HomSys} and derive the dynamics and initial conditions for our cosmological system in Sec.\ref{sec:cosmoSystem}.
Finally, in Sec.\ref{sec:Results} we present the tree-level and one-loop results for the cosmic density-fluctuation power spectrum and compare them to results from numerical simulations. As expected, we reproduce the tree-level result of \cite{Lilow_2019} and present the first full result\footnote{In \cite{LilowDiss} a first result for the one-loop power spectrum was obtained using a perturbative expansion in the initial power spectrum.} for the one-loop power spectrum.

\section{Notation and Conventions}\label{sec:NotationConventions}
\subsection{$N$-particle Phase Space}
Given that we are working with an ensemble of $N$ indistinguishable particles, it is important to differentiate between the $6N$-dimensional phase space describing all individual particle states and the 6-dimensional phase space representing averaged properties of the system as a whole. We therefore denote the coordinates of individual particles in their respective phase spaces by \textit{sans serif} variables $\ms{x}=(\vec{\ms{q}}, \vec{\ms{p}})$ and use $x=(\vec{q}, \vec{p})$ for coordinates of the physical phase space.

Due to the indistinguishable nature of the particles, most quantities of interest, such as the Hamilton function, depend on the entire particle ensemble. To enhance readability, we therefore introduce a tensorial notation, bundling all objects with a particle index (\eg $\vec{\ms{q}}_j$) into multi-particle tensors,
\begin{equation}
    \tens{A}(t)=\sum_{j=1}^N \ms{A}_j(t)\otimes e_j\,,
\end{equation}
where $e_j$ is the $N$-dimensional canonical unit vector. The set of phase-space coordinates $\{\ms{x}_j\}$ is now, for instance, represented by the tensor $\tens{x}$. Thus, the scalar product between two quantities $\tens{A}$ and $\tens{B}$ decomposes as
\begin{equation}
    \tens{A\cdot B}=\sum_{j,i=1}^N(\ms{A}_i\otimes e_i)\cdot(\ms{B}_j\otimes e_j)=\sum_{j,i=1}^N\ms{A}_i\ms{B}_j\underbrace{e_i\cdot e_j}_{=\delta_{ij}}=\sum_{i=1}^N\ms{A}_i\ms{B}_i\,.
\end{equation}
Functions depending on all particle positions can be compactly written as $F(\tens{x},t)$, and the integration measure of the $N$-particle phase space becomes
\begin{equation}
    \md^{6N}\ms{x}=\md^{3N}\ms{q}\,\md^{3N}\ms{p} \equiv \md\tens{q}\,\md\tens{p}=\md \tens{x}\,.
\end{equation}

\subsection{Fourier Transform}\label{sec:Fourier}
It will often be convenient to work with the Fourier representation of a function $g(\vec{q}, \vec{p},t)$, where $\vec{q}$ and $\vec{p}$ represent the position and momentum variable in phase space. To keep the notation simple, we use the same symbol for both, the original function and its Fourier transform, relying on the context of the arguments to distinguish between them. We thus define
\begin{align}
    \mathcal{F}_{q,p}[g(\vec{q}, \vec{p}, t)](\vec{k}, \vec{l}, t)=&\int\md^3\vec{q}\,\md^3\vec{p}\,\,g(\vec{q}, \vec{p},t)\e^{-\mi\vec{q}\cdot\vec{k}-\mi\vec{p}\cdot\vec{l}}\;,\\
    \mathcal{F}_{k,l}^{-1}[g(\vec{k}, \vec{l}, t)](\vec{q}, \vec{p}, t)=&\int\frac{\md^3\vec{k}}{(2\pi)^3}\,\frac{\md^3\vec{l}}{(2\pi)^3}\,g(\vec{q}, \vec{p},t)\e^{\mi\vec{k}\cdot\vec{q}+\mi\vec{l}\cdot\vec{p}}\;,
\end{align}
where the subscript of $\mathcal{F}_{q,p}$ indicates over which variables the Fourier transform is performed. 

\subsection{Reduced Phase-Space Densities}

The Liouville phase-space density contains the statistical properties of the system and is denoted by $\varrho_N(\tens{x}, t)$. The $s$-particle reduced phase-space density is defined as 
\begin{equation}\label{eq:reduced_f}
   f_s(\vec{{q}}_1, \vec{{p}}_1,\cdots,\vec{{q}}_s,\vec{{p}}_s,t)=\int\md\tens{x}\!\!\!\sum_{i_1\neq\cdots\neq i_s=1}^N\!\!\!\!\!\!\dirac{x_1-\ms{x}_{i_1}}\cdots\dirac{x_s-\ms{x}_{i_s}}\varrho_{N}(\tens{q}, \tens{p}, t)\,,
\end{equation}
and contains the statistical properties of an $s$-particle subset. Furthermore, we make use of the usual decomposition of $f_s$ into a reducible and an irreducible part,
\begin{align}
   f_2({x}_1, {x}_2, t)=&f_1({x}_1, t)f_1({x}_2, t)+g_2({x}_1, {x}_2, t)\,,\label{eq:f2}\\
   \begin{split}
       f_3({x}_1, {x}_2, {x}_3, t)=&f_1({x}_1, t)f_1({x}_2, t)f_1({x}_3, t)+f_1({x}_1, t)g_2({x}_2, {x}_3, t)+f_1({x}_2, t)g_2({x}_1, {x}_3, t)\\&+f_1({x}_3, t)g_2({x}_1, {x}_2, t)+g_3({x}_1, {x}_2, {x}_3, t)\,,
   \end{split}\\
   \vdots &\,\,\,\,\,\,\,\,\nonumber
\end{align}
where $g_s({x}_1, \cdots, {x}_s, t)$ are the \textit{irreducible} reduced phase-space densities which describe by how much the system deviates from an uncorrelated state. 

\subsection{Field Theory Notation}
For convenience, we introduce the short-hand notation for the four-dimensional object $X\coloneqq (\vec{q}, \vec{p}, t)$ and its Fourier conjugate $S\coloneqq (\vec{k},\vec{l},t)$. In addition, we introduce the compact notation,
\begin{equation}
    \int_X \coloneqq \int \md^3 \vec{q} \,\md^3 \vec{p}\,\md t\,.
\end{equation}

\section{Time Evolution of Classical Ensembles via Path Integrals}\label{sec:Basics}

We consider a statistical ensemble of $N$ particles of equal mass $m$ in $6N$-dimensional phase space, with coordinates $\ms{x}_j=(\vec{\ms{q}}_j, \vec{\ms{p}}_j)\,,\,\,\,j=1,\cdots,N$, where $\vec{\ms{q}}_j$ and $\vec{\ms{p}}_j$ refer to the respective particle positions and momenta. At any given time $t$, the microstate of the system is thus described by the set $\{\ms{x}_j(t)\}_{j=1}^N$, \ie the collection of classical trajectories evaluated at time $t$. These are determined by the Hamilton function of the system, 
\begin{equation}\label{eq:Hamiltongeneral}
    H(\tens{q}, \tens{p},t)=\sum_{i=1}^N\frac{\vec{\ms{p}}_i^{\,2}}{2m}+\frac{1}{2}\sum_{\substack{i,j=1 \\ i \neq j}}^Nv(|\vec{\ms{q}}_i-\vec{\ms{q}}_j|,t)\,,
\end{equation}
which describes a system of particles mutually interacting through the homogeneous, possibly time-dependent, pair potential $v(|\vec{\ms{q}}_i-\vec{\ms{q}}_j|,t)$. 
The classical trajectories are subject to the Hamiltonian equations of motion, 
\begin{equation}\label{eq:eomgeneral}
\begin{aligned}
    \begin{pmatrix}
        \dot{\vec{\ms{q}}}_i \\
        \dot{\vec{\ms{p}}}_i
    \end{pmatrix}
    =
    \begin{pmatrix}
        \nabla_{\vec{\ms{p}}_i}H(\tens{q},\tens{p},t) \\
        -\nabla_{\vec{\ms{q}}_i}H(\tens{q},\tens{p},t)
    \end{pmatrix}
    \quad \Leftrightarrow \quad
    \dot{\ms{x}}_i = \omega \cdot \nabla_i H(\tens{q},\tens{p},t)\,,
\end{aligned}
\end{equation}
where on the right hand side we used the symplectic structure $\omega =\begin{pmatrix}
    0 & \mathbb{1}_{3\times3}\\ -\mathbb{1}_{3\times3} & 0
\end{pmatrix}$ and the gradient $\nabla_i=\begin{pmatrix}
    \nabla_{\vec{\ms{q}}_i}\,,\nabla_{\vec{\ms{p}}_i}
\end{pmatrix}^\top$ acting on phase space. Using the tensorial notation introduced earlier, the equations of motion for the $N$-particle system can be expressed in the compact form,
\begin{equation}\label{eq:tensoreom}
    0 = \tens{\mathcal{E}}(\tens{x})=\dot{\tens{x}} - \tens{\omega}\tens{\cdot}\tens{\nabla}_{\tens{x}}H(\tens{x}, t)\,,
\end{equation}
where $\tens{\omega}=\omega\otimes\mathbb{1}_{N\times N}$ is the tensorial generalization of the symplectic structure and $\tens{\nabla}_{\tens{x}}=(\tens{\nabla}_{\tens{q}}, \tens{\nabla}_{\tens{p}})$ generalizes the gradient to the $6N$-dimensional phase space. Within this setting, the current probabilistic state of the system is described by the Liouville phase-space density $\varrho_N(\tens{x}, t)$, whose time evolution is famously governed by Liouville's equation \begin{equation}\label{eq:Liouville}
     \frac{\partial\varrho_N(\tens{x}, t)}{\partial t}=-\sum_{i=1}^N\frac{\vec{\ms{p}}_i}{m}\cdot\nabla_{\vec{\ms{q}}_i}\varrho_N(\tens{x}, t)+\sum_{\substack{i,j=1\\i\neq j}}^N\nabla_{\vec{\ms{q}}_i}v(|\vec{\ms{q}}_i-\vec{\ms{q}}_j|, t)\cdot\nabla_{\vec{\ms{p}}_i}\varrho_N(\tens{x}, t)\equiv-\mi\hat{L}\varrho_N(\tens{x}, t)\,,
\end{equation}
from which all desired quantities can be obtained. The typical procedure is to reduce \eqref{eq:Liouville} to the BBGKY-hierachy and to study the time evolution of those reduced phase-space distribution function. The usual approach in order to solve these evolution equations is to truncate the infinite BBGKY-hierarchy of equations which leads to the Boltzmann equation, or the Vlasov equation if collisions are neglected. These are then used to derive the hydrodynamical equations which form the basis for SPT and similar approaches. 
Throughout this paper, however, we follow a different approach presented in \cite{Daus2024}. It provides an iterative solution for the BBGKY-hierarchy and thereby avoids the need for an explicit truncation. Instead, the order of PT implicitly determines a truncation criterion for the BBGKY-hierarchy. In this way the full microscopic information on the system is retained and can be recovered order by order in PT. In the following sections we will summarise the main ideas and derivations of \cite{Daus2024} and refer the reader to that publication for more details.

\subsection{Path Integral Construction of Classical Time Evolution}\label{sec:theoryFramework}
The form of the Liouville equation \eqref{eq:Liouville} suggests that it may be treated in a similar way as the Schr\"odinger equation in a quantum mechanical context. This idea leads to the Koopman von Neumann formulation of classical mechanics -- a framework that has been extensively studied in the literature \cite{Mauro:2001rm, vonNeumann, Koopman}. Inspired by those striking similarities to many-body quantum systems, we follow \cite{Abrikosov:2004cf, Gozzi:1993tm, kleinert2009path} and set up a path-integral description in order to study the time evolution of the above system. Let us first note, that the Liouville equation \eqref{eq:Liouville} admits a formal solution in terms of a time ordered exponential \cite{Gozzi1989, Abrikosov:2004cf}, given by 
\begin{equation}
    \varrho_N(\tens{x}, t)=\hat{\mathcal{T}}\exp\left[-\mi\int_{\ini{t}}^t\md t^\prime\hat{L}(t^\prime)\right]\varrho_N(\tini{x},\ini{t})=\hat{\mathcal{U}}(t, \ini{t})\varrho_N(\tini{x},\ini{t})\,.
\end{equation}
We can thus follow similar steps as in quantum mechanics to construct a path-integral representation of the above time evolution operator $\hat{\mathcal{U}}(t, \ini{t})$ in order to propagate the initial phase-space density. To keep things short, we outline the main ideas of the derivation and refer the reader to \cite{Daus2024} for further information. \\

 First note, that once an appropriate set of initial coordinates $\tini{x}=\tens{x}(t=t^{(i)})$ for all the particles has been chosen, their respective trajectories are uniquely defined by the classical equations of motion \eqref{eq:eomgeneral}. Thus, by drawing realizations of initial coordinates from $\varrho_N(\tini{x}, \ini{t})$ and evolving them according to \eqref{eq:eomgeneral}, the system evolves into a final phase-space density distribution $\varrho_N(\tfin{x}, \fin{t})$. At any time $t>\ini{t}$, the probability density of the system is given by,
\begin{equation}
    \varrho_N(\tens{x}, t)=\int\md\tini{x}K(\tens{x}, t|\tini{x}, \ini{t})\varrho_N(\tini{x}, \ini{t})\,,\label{eq:evolution}
\end{equation}
where the transition probability $K(\tens{x}, t|\tini{x}, \ini{t})$ evolves the  initial phase-space density of the system to the final one at time $t$. It reads 
\begin{equation}\label{eq:transition}
    K(\tens{x}, t|\tini{x}, \ini{t})=\,\,\delta_D(\tens{x}-\tens{x}_{\mathrm{cl}}(t;\tini{x}))\,, 
\end{equation}
which ensures that at any time the particles are found on their respective classical trajectories. \\

The path integral representation of $K(\tens{x}, t|\tini{x}, \ini{t})$ is now constructed by discretizing the time evolution into $N$ steps. The final result is then 
\begin{equation}\label{eq:PathPropagator}
    K(\tfin{x}, \fin{t}|\tini{x}, \ini{t})=\int\limits_{\tini{x}}^{\tfin{x}} \mathcal{D}\tens{x}(t)\,\delta_D[\tens{x}(t)-\tens{x}_{\mathrm{cl}}(t;\tini{x})]\,,
\end{equation}
where the functional integration measure $\mathcal{D}\tens{x}(t)\delta_D[\tens{x}(t)-\tens{x}_{\mathrm{cl}}(t;\tini{x})]$ gives weight $0$ to all trajectories but the classical one, with fixed end points at $\tini{x}$ and $\tfin{x}$. We can further include the equations of motion in the form \eqref{eq:tensoreom}, with the help of a variable transformation in the functional measure, to obtain
\begin{equation} \label{eq:variabletransform}
    \delta_D[\tens{x}(t)-\tens{x}_{\mathrm{cl}}(t;\tini{x})]=\delta_D[\tens{\mathcal{E}}[\tens{x}(t)]]\cdot \mathcal{J}\,.
\end{equation}
As has been shown in \cite{Daus2024}, we can absorb the constant Jacobian $\mathcal{J}$ into a normalization constant. In a final step, we replace the remaining functional Dirac delta distribution with its corresponding Fourier transformed expression, 
\begin{equation}\label{eq:PropagatorPathIntegral}
     K(\tfin{x}, \fin{t}|\tini{x}, \ini{t})=\int\limits_{\tini{x}}^{\tfin{x}} \mathcal{D}\tens{x}(t)\mathcal{D}\tens{\chi}(t)\,\exp\Bigg[\mi\int\limits_{\ini{t}}^{\fin{t}}\md t\,\tens{\chi}(t)\tens{\cdot}\left(\frac{\md \tens{x}(t)}{\md t}-\tens{\omega\cdot\nabla}H(\tens{x}(t), t)\right)\Bigg]\,,
\end{equation}
where we introduced a doublet $\tens{\chi}=(\tens{\chi_q}, \tens{\chi_p})$ conjugate to $\tens{x}=(\tens{q}, \tens{p})$. The functional integration is carried out over all trajectories $\tens{x}(t)$ with fixed endpoints and all $\tens{\chi}(t)$ trajectories without endpoint restrictions. We may thus write the transition amplitude as 
\begin{equation}
    K(\tfin{x}, \fin{t}|\tini{x}, \ini{t})=\int\limits_{\tini{x}}^{\tfin{x}} \mathcal{D}\tens{x}(t)\mathcal{D}\tens{\chi}(t)\,\exp\Bigg[\mi\mathcal{S}[\tens{x}(t),\tens{\chi}(t)]\Bigg]\,,
\end{equation}
where we defined the action
\begin{equation}\label{eq:effectiveaction}
    \mathcal{S}[\tens{x}(t),\tens{\chi}(t)]=\int\limits_{\ini{t}}^{\fin t}\md t\,\tens{\chi}(t)\tens{\cdot}\left(\frac{\md \tens{x}(t)}{\md t}-\tens{\omega\cdot\nabla}H(\tens{x}(t), t)\right)\,.
\end{equation}
Upon inserting the Hamiltonian \eqref{eq:Hamiltongeneral}, the action can be split into a free and an interacting part 
\begin{align}\label{eq:interactionaction}
    \mathcal{S}[\tens{x}(t),\tens{\chi}(t)]=\mathcal{S}_0[\tens{x}(t),\tens{\chi}(t)] + \mathcal{S}_\mathrm{I}[\tens{x}(t),\tens{\chi}(t)]\,,
\end{align}
respectively given by
\begin{equation}
    \mathcal{S}_0[\tens{x}(t),\tens{\chi}(t)]=\int\limits_{\ini{t}}^{\infty}\md t\,\sum_{i=1}^N\Big\{\vec{\chi}_{\ms{q}_i}(t)\cdot\left(\dot{\vec{\ms{q}}}_i-\frac{\vec{\ms{p}}_i}{m}\right)+\vec{\chi}_{\ms{p}_i}\cdot\dot{\vec{\ms{p}}}_i\Big\}\,,\,\,\,\,\,\,\,\mathcal{S}_\mathrm{I}[\tens{x}(t),\tens{\chi}(t)]=\int\limits_{\ini{t}}^{\infty}\md t\,\sum_{i\neq j=1}^N\vec{\chi}_{\ms{p}_i}\cdot\nabla_{\vec{\ms{q}}_i}v(|\vec{\ms{q}}_i-\vec{\ms{q}}_j|,t)\,.
\end{equation}
As we show in the following, the above representation of the classical transition amplitude enables the application of path-integral methods in order to analytically approach the above $N$-body problem. 

\subsection{Expectation Values and Free Phase-Space Density Cumulants}\label{sec:ExpectationValues}

Our objects of interest are expectation values of observables $A(t)\equiv A(\tens{x}(t))$ that depend on the collective particle phase-space postions at time $t$. We define the time-ordered expectation value as 
\begin{equation}\label{eq:expectationvaluewithPI}
    \mean{\hat{\mathcal{T}}A_1(t_1)\cdots A_k(t_k)}=\int\md\tfin{x}\int\md\tini{x}\varrho_N(\tini{x}, \ini{t})
    \int\limits_{\tini{x}}^{\tfin{x}} \mathcal{D}\tens{x}(t)\mathcal{D}\tens{\chi}(t)\,A_1(\tens{x}(t_1))\cdots A_k(\tens{x}(t_k))\,\exp\Bigg[\mi\mathcal{S}[\tens{x}(t),\tens{\chi}(t)]\Bigg]\,,
\end{equation}
where $\hat{\mathcal{T}}$ is the usual time-ordering operator. It is straightforward to verify \cite{Daus2024} that the above path integral can be split at the respective operator insertions, 
\begin{equation}\label{eq:ExpectationPropagator}
\begin{split}
    \mean{\hat{\mathcal{T}}A_1(t_1)\cdots A_k(t_k)}=
    \int\md\tens{x}_1\cdots\md\tens{x}_k&\md\tini{x}A_1(\tens{x}_1)K(\tens{x}_1, t_1|\tens{x}_2, t_2)\cdots A_k(\tens{x}_k)K(\tens{x}_k, t_k|\tini{x}, \ini{t})\varrho_N(\tini{x}, \ini{t})\,,
    \end{split}
\end{equation}
where $\tens{x}_s=\tens{x}(t_s)$ and we assumed w.\,l.\,o.\,g. the time ordering $\fin{t}>t_1>\cdots>t_k>\ini{t}$. Using \eqref{eq:transition}, the integrals in the above expression can be solved to give 
\begin{equation}\label{eq:expectationvaluewithoutPI}
    \mean{\hat{\mathcal{T}}A_1(t_1)\cdots A_k(t_k)}=\int\md\tini{x}
    \,A_1(\tens{x}_{\mathrm{cl}}(t_1;\tini{x}))\cdots A_k(\tens{x}_{\mathrm{cl}}(t_k;\tini{x}))\,\varrho_N(\tini{x}, \ini{t})\,,
\end{equation}
which corresponds to the usual definition of unequal-time expectation values from statistical physics.\\

Of particular interest to us is the Klimontovich phase-space density, defined by
\begin{equation}\label{eq:Klimontovich}
    f(\vec{q}, \vec{p}, t):=f(\vec{q}, \vec{p};\tens{q}(t), \tens{p}(t))=\sum_{i=1}^N\delta_D(\vec{q}-\vec{\ms{q}}_i(t))\delta_D(\vec{p}-\vec{\ms{p}}_i(t))\,,
\end{equation}
which counts the number of particles at a given position $x=(\vec{q}, \vec{p})$ of the $6$-dimensional physical phase space\footnote{Notice that we denote the coordinates of the 6-dimensional physical phase space by $x=(\vec{q}, \vec{p})$ while we use the \textit{sans serif} version $\ms{x}=(\vec{\ms{q}}, \vec{\ms{p}})$ to denote particle positions in $6N$-dimensional phase-space.}. The Klimontovich phase-space density can be used to obtain further statistical quantities. Especially relevant in the cosmological context are the moments of the momentum distribution, which can generally be written as 
\begin{equation}
    \mathcal{O}(\vec{q}, t)=\int\md^3\vec p\, \,F_{\mathcal{O}}(\vec p)\,f(\vec{q}, \vec{p}t)\,,
\end{equation}
with some appropriate function $F_{\mathcal{O}}(\vec p)$. For instance, we find for the particle number density $\rho(\vec{q}, t)$ or the momentum density $\Pi(\vec{q}, t)$ the expressions
\begin{equation}\label{eq:densitymomentum}
    \rho(\vec{q}, t)=\int\md^3p\,f(\vec{q}, \vec{p}, t)=\sum_{i=1}^N\delta_D(\vec{q}-\vec{\ms{q}}_i(t))\,,\quad\Pi(\vec{q}, t)=\int\md^3p\,\vec{p}\,f(\vec{q}, \vec{p}, t)=\sum_{i=1}^N\vec{\ms{p}}_i(t)\,\delta_D(\vec{q}-\vec{\ms{q}}_i(t))\,.
\end{equation}
We may, therefore, relate an arbitrary $k$-point correlator of the above objects to the respective $k$-point correlation function of the Klimontovich phase-space density,
\begin{equation}\label{eq:generalmomentummoments}
    \mean{\hat{\mathcal{T}}\mathcal{O}_1(\vec{q}_1, t_1)\cdots \mathcal{O}_k(\vec{q}_k, t_k)}=\int\md^3\vec p_1\cdots \md^3\vec p_k\,\,F_{\mathcal{O}_1}(\vec p_1)\cdots F_{\mathcal{O}_k}(\vec p_k)\mean{\hat{\mathcal{T}}f(\vec q_1, \vec p_1, t_1)\cdots f(\vec q_k, \vec p_k, t_k) }\,.
\end{equation}
As a central element of our discussion, we therefore define the $k$-point phase-space density correlation function
\begin{equation}\label{eq:generalCorrelator}
    C_{f\cdots f}(x_1,t_1, \cdots x_k, t_k):=\mean{\hat{\mathcal{T}}f(x_1,t_1)\cdots f(x_k,t_k)}\,,
\end{equation}
and the associated cumulant
\begin{equation}\label{eq:generalCumulant}
    G_{f\cdots f}(x_1,t_1, \cdots x_k, t_k):=\mean{\hat{\mathcal{T}}f(x_1,t_1)\cdots f(x_k,t_k)}_c\,,
\end{equation}
where the subscript $\mean{\cdots}_c$ refers to the connected part of the respective correlator. For instance, we find 
\begin{equation}\label{eq:G_ff}
    G_{ff}(x_1,t_1, x_2, t_2)=C_{ff}(x_1,t_1, x_2, t_2)-C_{f}(x_1,t_1)C_{f}(x_2,t_2)\,,
\end{equation}
as usual.\\

In the absence of interactions, expressions of the form \eqref{eq:expectationvaluewithPI} or \eqref{eq:expectationvaluewithoutPI} are exactly solvable. In that case, the free trajectory for a particle moving from $(\ini{x}, \ini{t})$ to $(x_1, t_1)$ has an analytical solution, which is given by
\begin{equation}\label{eq:freetrajectory}
    {\ms{x}}_{\mathrm{cl}}(t;\ini{x}, \ini{t})=(\vec{{q}}_{\mathrm{cl}}(t;\ini{x}, \ini{t}),\vec{{p}}_{\mathrm{cl}}(t;\ini{x}, \ini{t}))=\left(\ini{\vec{q}\,}+\frac{\ini{\vec{p}\,}}{m}(t-\ini t), \ini{\vec{p}\,}\right)\,.
\end{equation}
The integrals can thus be solved and for the free one- and two-point $f$-cumulants one finds 
\begin{align}
\begin{split}
  G_f^{(0)}(x_1, t_1)=&\, f_1^{(0)}(\vec{q_1}, \vec p_1, t_1) \,,\label{eq:FreeOnePoint}
\end{split}
\\[2ex]
\begin{split}
G_{ff}^{(0)}(x_1, t_1, x_2, t_2)=&\,\dirac{x_1-x_{\text{cl}}(t_1; x_2, t_2)}\,f_1^{(0)}(\vec{q}_1, \vec p_1, t_1)+g_2^{(0)}(\vec{q}_1, \vec p_1, t_1, \vec{q}_2, \vec p_2, t_2)\label{eq:FreeTwoPoint}\,,
\end{split}
\end{align}
with the freely evolved one- and two-particle reduced phase space densities
\begin{align}
    f_1^{(0)}(\vec{q}_1, \vec p_1, t_1)\equiv& \,f_1\left(\vec{q}_1-\frac{\vec p_1}{m}(t_1-\ini{t}), \vec p_1, \ini t\right)\,,\\[2ex]
    g_2^{(0)}(\vec{q_1}, \vec p_1, t_1, \vec{q_2}, \vec p_2, t_2)\equiv&\,g_2\left(\vec{q}_1-\frac{\vec p_1}{m}(t_1-\ini{t}), \vec p_1, \vec{q}_2-\frac{\vec p_2}{m}(t_2-\ini{t}), \vec p_2, \ini t\right)\,.
\end{align}
$f_1^{(0)}$ and $g_2^{(0)}$ generalize the reduced densities defined in \eqref{eq:reduced_f} to the case of unequal times evolved by the free Liouville equation. They are obtained by shifting the respective initial reduced density by the free trajectory. Thus, we can relate all density cumulants, as expected, to the freely evolved initial phase space density. Note, that the difference between the two-point phase-space density cumulant $G_{ff}^{(0)}$ and the true two-particle correlation $g_2$ is given by the first term on the right-hand side of \eqref{eq:FreeTwoPoint} and represents a one-particle contribution to $G_{ff}^{(0)}$ coming from the diagonal part of the double sum. It is merely a consequence of the discreteness of the system and describes the possibility of picking the same particle twice, while measuring the correlation function. The Dirac delta distribution identifies the particle at $\vec{q}_1$ with the particle freely moving from $\vec{q}_2$ at time $t_2$. Such contributions always arise when dealing with discrete particle systems and are referred to as \textit{Poisson shot-noise} effects. The general $r$-point $f$-cumulant therefore has the form 
\begin{equation}\label{eq:general_free_f_cumulants}
    G^{(0)}_{f\cdots f}(x_1, t_1, \dots , x_r, t_r) = \sum_{\text{shot noise}} + \,\,g_r\left(\vec{q}_1-\frac{\vec p_1}{m}(t_1-\ini{t}), \vec p_1, \dots\,,\vec{q}_r-\frac{\vec p_r}{m}(t_r-\ini{t}), \vec p_r, \ini t\right)\,,
\end{equation}
where the sum includes all lower-order particle contributions. Most importantly, \eqref{eq:general_free_f_cumulants} shows that the free density-cumulants can be expressed in terms of the initial reduced phase-space densities.\\

The exact computation of expectation values as in \eqref{eq:expectationvaluewithPI} is generally not possible due to the presence of an interaction potential. It is, therefore, inevitable to employ some sort of approximation procedure. The most straightforward approach is canonical perturbation theory, which consists of expanding the exponential of $\mathcal{S}_\mathrm{I}$ in \eqref{eq:expectationvaluewithPI} in a Taylor series and sorting the resulting terms by powers of the interaction potential. However, since it is not the subject of this work, we refer the interested reader to \cite{Daus2024} for technical details and to \cite{Pixius:2022hqs, Heisenberg:2022uhb} for similar cosmological applications. Instead, we will construct a perturbation theory based on a re-formulation of the theory in terms of macroscopic fields.

\subsection{Construction of the Field Theory}\label{sec:macroscopicFieldTheory}
Despite being the most straightforward approach, canonical perturbation theory is not very suitable for our purposes, due to the vast number of diagrams that have to be evaluated in order to obtain the desired accuracy (see \cite{Daus2024} and \cite{Pixius:2022hqs, Heisenberg:2022uhb} for a discussion). Therefore, in \cite{Daus2024} we proposed a different approach based on the Hubbard-Stratonovich transformation (HST), well-known from equilibrium statistical mechanics. It allows us to convert the particle-based description to a field theory based on macroscopic fields. We give a brief summary of the derivation leading to said field theory here and refer the interested reader to \cite{Daus2024} for details. 

We begin by introducing the so-called response field $\mathcal{B}(X)$, in addition to the Klimontovich phase-space density $f(X)$, defined as
\begin{align}\label{eq:ResponseField}
   \mathcal{B}(X)\equiv \mathcal{B}(\vec{  {q}}_1, \vec{  {p}}_1, t_1)&=\mi\sum_{i=1}^{N}\vec{\chi}_{p_i}(t_1)\cdot\nabla_{\vec{q}_i}v(|\vec{q}_i-\vec{  {q}}_1|, t_1)\,,
\end{align}
where $X=(\vec{q}, \vec{p}, t)$ denotes the combined phase-space and time coordinates. The interaction part of the action \eqref{eq:interactionaction} can then be written as
\begin{align}
    \mi\mathcal{S}_\mathrm{I}[\tens{x}(t),\tens{\chi}(t)]=\int_{X_1, X_2} \,\mathcal{B}(X_1)\,\delta_D(X_1-X_2)\,f(X_2)\,,\label{eq:ActionBeforeHST}
\end{align}
with $\delta_D(X_1-X_2)=\delta_D(\vec{  {q}}_1-\vec{  {q}}_2)\delta_D(\vec{  {p}}_1-\vec{  {p}}_2)\delta_D(t_1-t_2)\equiv\mathbb{1}(X_1, X_2)$. In order to apply the HST, we bring the action \eqref{eq:ActionBeforeHST} into the quadratic form,
\begin{equation}
    \mi\mathcal{S}_\mathrm{I}[\tens{x}(t),\tens{\chi}(t);J]=\frac{1}{2}\int_{X_1, X_2}\Big(\Phi(X_1)+{J}(X_1)\Big)^\top\cdot\sigma(X_1,X_2)\cdot\Big(\Phi(X_2)+{J}(X_2)\Big)\,,
\end{equation}
where we have collected the fields into a doublet $\Phi(X)=(\mathcal{B}(X),\,\,  f(X))^\top$ and introduced the sources ${J}(X)=(J_f(X),\,\, J_{\mathcal{B}}(X))^\top$ for the $f$- and $B$-fields, as well as the involutory matrix
\begin{equation}
    \sigma(X_1, X_2)=\begin{pmatrix}
    0 & \mathbb{1}(X_1, X_2) \\ 
    \mathbb{1}(X_1, X_2) & 0
    \end{pmatrix}\,.
\end{equation}
The HST now consists of writing the exponential of the quadratic form as a Gaussian functional integral. Following \cite{Daus2024}, we get  
\begin{equation}\label{eq:macroscopicGenFuncAfterTrafo}
\begin{split}
\mathcal{Z}[J]=\mathcal{N}\int\mathcal{D}\Psi\,\exp\Bigg[-\frac{1}{2}\int_{X_1, X_2}\Psi^\top(X_1)\cdot\sigma^{-1}(X_1,X_2)\cdot\Psi(X_2)+\int_{X_1}{J}^\top(X_1)\cdot\Psi(X_1)\Bigg] \cdot\mathcal{Z}^{(0)}_{f\mathcal{B}}\big[\Psi\big]\,,
    \end{split}
\end{equation}
with 
\begin{equation}
    \mathcal{Z}^{(0)}_{f\mathcal{B}}[\Psi]=\int\md \tfin{x}\int\md\tini{x}\varrho_N(\tini{x}, \ini{t})
    \int\limits_{\tini{x}}^{\tfin{x}} \mathcal{D}\tens{x}(t)\mathcal{D}\tens{\chi}(t)\exp\Bigg[\mi\mathcal{S}_0[\tens{x}(t),\tens{\chi}(t)] + \int_{X_1}\Psi^\top(X_1)\cdot\Phi(X_1)\Bigg]\,.
\end{equation}
In the above, we have introduced the macroscopic-field doublet $\Psi(X)=(\Psi_f(X),\,\,\Psi_\mathcal{B}(X))$ conjugate to $\Phi(X)$ and used that $\sigma(X_1,X_2)=\sigma^{-1}(X_1,X_2)$. Notice that the macroscopic fields $\Psi_f(X)$ and $\Psi_\mathcal{B}(X)$ are independent of microscopic degrees of freedom, but couple linearly to $\Phi(X)$ in $\mathcal{Z}^{(0)}_{f\mathcal{B}}$. This makes $\mathcal{Z}^{(0)}_{f\mathcal{B}}$ the generating functional of \textit{free} $f$-$\mathcal{B}$ correlation functions, as one can easily verify by taking functional derivatives \wrt $\Psi_f$ and $\Psi_\mathcal{B}$. The generating functional $\mathcal{Z}^{(0)}_{f\mathcal{B}}$, thus, contains the full microscopic statistics. Introducing the Schwinger functional $\mathcal{W}^{(0)}_{f\mathcal{B}}[\Psi]\equiv \ln\Big[\mathcal{Z}_{f\mathcal{B}}^{(0)}[\Psi]\Big]$, we can bring \eqref{eq:macroscopicGenFuncAfterTrafo} into the compact form 
\begin{align}\label{eq:genFuncWithW}
\mathcal{Z}[J]=\mathcal{N}\int\mathcal{D}\Psi\,\exp\Bigg[-\frac{1}{2}\int_{X_1, X_2}\Psi^\top(X_1)\cdot\sigma^{-1}(X_1,X_2)\cdot\Psi(X_2)+\int_{X_1}{J}^\top(X_1)\cdot\Psi(X_1)+ \mathcal{W}^{(0)}_{f\mathcal{B}}\big[\Psi\big]\Bigg] \,.
\end{align}
The Schwinger functional has the following series expansion,
\begin{align}\label{eq:Schwinger}
    \mathcal{W}_{f\mathcal{B}}^{(0)}[\Psi] = \sum_{r,s=0}^{\infty}\frac{1}{r!s!}\Bigg[\prod_{m=1}^r\int_{ X_m}\Psi_\mathcal{B}(X_m)\Bigg]\Bigg[\prod_{n=1}^{s}&\int_{ X_n^\prime}\Psi_f(X_n^\prime)\Bigg]\,  G_{f\cdots f \mathcal{B}\cdots \mathcal{B}}^{(0)}(X_1,\cdots X_r,X_1^\prime,\cdots X_s^\prime)\,,
\end{align}
which introduces the free $f\mathcal{B}$-cumulants $G_{f\cdots f\mathcal{B}\cdots\mathcal{B}}^{(0)}$, containing $r$ phase-space densities and $s$ response fields,
\begin{equation}\label{eq:timeorderedfB}
    G_{f\cdots f \mathcal{B}\cdots \mathcal{B}}^{(0)}(X_1,\cdots X_r,X_1^\prime,\cdots X_s^\prime)=\mean{\hat{\mathcal{T}}f(X_1)\cdots f(X_r)\mathcal{B}(X_1^\prime)\cdots \mathcal{B}(X_s^\prime)}^{(0)}_c\,.
\end{equation}
Note, that the mean has to be taken \wrt the \textit{free} microscopic action as the potential is included in the $\mathcal{B}$ field. The $G_{f\cdots f\mathcal{B}\cdots\mathcal{B}}^{(0)}$'s thus generalize the pure $f$-cumulants discussed in the end of Sec.\ref{sec:ExpectationValues} to include the response field. General expressions for the $f\mathcal{B}$-cumulants have been derived in \cite{Daus2024} and we discuss their form and physical meaning in the next section. Furthermore, we provide expressions for the cumulants required in the course of this paper in Appendix \ref{app:freeCumulants}.\\

Together with \eqref{eq:genFuncWithW}, equation \eqref{eq:Schwinger} defines the action of a statistical field theory where the generating functional has the form 
\begin{equation}\label{eq:Generatingfunctional}
    \mathcal{Z}[J]=\mathcal{N}\int\mathcal{D}\Psi\exp\Big[-\mathcal{S}[\Psi]+\int_{X_1}{J}^\top(X_1)\cdot\Psi(X_1)\Big]=\mathcal{N}\int\mathcal{D}\Psi\exp\Big[-\mathcal{S}_0[\Psi]-\mathcal{S}_\mathrm{I}[\Psi]+\int_{X_1}{J}^\top(X_1)\cdot\Psi(X_1)\Big]\,,
\end{equation}
with the free macroscopic action containing all terms which are at most quadratic in the fields\footnote{Note that there exists no pure $\mathcal{B}$-field cumulant. See \cite{Daus2024} or the following discussion for more details.}, 
\begin{equation}\label{eq:treelevelaction1}
\begin{split}
    \mathcal{S}_0[\Psi_{f},\Psi_\mathcal{B}]=&\int_{X_1,X_2}\Psi_\mathcal{B}(X_1)\left[\mathbb{1}(X_1, X_2) - G^{(0)}_{f\mathcal{B}}(X_1,X_2)\right]\Psi_f(X_2)-\int_{X_1}\Psi_\mathcal{B}(X_1)G^{(0)}_f(X_1)\\&-\frac{1}{2!}\int_{X_1,X_2}\Psi_\mathcal{B}(X_1)\Psi_\mathcal{B}(X_2)G^{(0)}_{ff}(X_1,X_2)\;,
    \end{split}
\end{equation}
and the remaining terms defining the self-interactions of the fields,
\begin{align}\label{eq:Vertexpart}
    \mathcal{S}_\mathrm{I}[\Psi]=&-\sum_{r+s>2}^{\infty}\frac{1}{r!s!}\Bigg[\prod_{m=1}^r\int_{ X_m}\Psi_\mathcal{B}(X_m)\Bigg]\Bigg[\prod_{n=1}^{s}\int_{ X_n^\prime}\Psi_f(X_n^\prime)\Bigg]\, G_{f\cdots f \mathcal{B}\cdots \mathcal{B}}^{(0)}(X_1,\cdots X_r,X_1^\prime,\cdots X_s^\prime)\,.
\end{align}
Introducing the inverse tree-level propagator
\begin{equation}
    \Delta^{-1}(X_1, X_2)=\begin{pmatrix}
        0 & \mathbb{1}-G^{(0)}_{f\mathcal{B}} \\ \mathbb{1}-G^{(0)}_{\mathcal{B}f} & G^{(0)}_{ff}
    \end{pmatrix}(X_1, X_2)\,,\quad\text{with}\quad G^{(0)}_{\mathcal{B}f}(X_1, X_2)=G^{(0)}_{f\mathcal{B}}(X_2, X_1)
\end{equation}
we can write the quadratic part of the action \eqref{eq:treelevelaction1} as
\begin{equation}\label{eq:treelevelaction}
\begin{split}
    \mathcal{S}_0[\Psi]=&\frac{1}{2}\int_{X_1, X_2}\Psi(X_1)\cdot\Delta^{-1}(X_1, X_2)\cdot\Psi(X_2)-\int_{X_1}\Psi_\mathcal{B}(X_1)G^{(0)}_f(X_1)\,.
    \end{split}
\end{equation}
Note, that by construction we have
\begin{equation}
    \mean{f(X_1)\cdots f(X_k)}=\frac{\delta^k}{\delta {J}_f(X_1)\cdots\delta{J}_f(X_k)}\mathcal{Z}[J]\Bigg\rvert_{{J}=0}\,,
\end{equation}
where the above expectation value is exact as no approximations have been made so far. We may now apply the usual field theoretic methods in order to compute correlation functions. 

First, we replace the fields $\Psi_f$ and $\Psi_\mathcal{B}$ in \eqref{eq:Vertexpart} by the functional derivative \wrt their respective source field and find the usual expression
\begin{equation}
     \mathcal{Z}[J]=\mathcal{N}^\prime\exp\bigg[\mathcal{S}_\mathrm{I}\bigg[\frac{\delta}{\delta J_f},\frac{\delta}{\delta J_\mathcal{B}}\bigg]\bigg]\cdot\mathcal{Z}_0[J]\;,
     \label{eq:generating_functional}
\end{equation}
where $\mathcal{Z}_0[J]$ is given by the quadratic part of the action,\footnote{The normalization $\mathcal{N}=\mathcal{N}^\prime\mathcal{N}^{\prime\prime}$ is chosen such that $\mathcal{Z}[0]=1=\mathcal{Z}_0[0]$.}
\begin{align}
    \label{eq:FreeMacrosGeneratingFunctionalPathInt}
    \mathcal{Z}_0[J]=\mathcal{N}^{\prime\prime}\int\mathcal{D}\Psi_f\mathcal{D}\Psi_\mathcal{B}\exp\bigg[-\mathcal{S}_0[\Psi_\mathcal{B},\Psi_{f}]+\int_{X_1} J_f(X_1)\Psi_f(X_1)+\int_{X_1} J_\mathcal{B}(X_1)\Psi_\mathcal{B}(X_1)\bigg]\,.
\end{align}
The above path integration for the tree-level generating functional can be performed analytically (see \cite{Daus2024} for details) and yields
\begin{equation}\label{eq:FreeMacroGeneratingFunctional}
    \mathcal{Z}_0[J]=\exp\Bigg[\frac{1}{2}J\cdot\Delta\cdot J+J_f\cdot\Delta_{f\mathcal{B}}\cdot G_{f}^{(0)}\Bigg]=\exp\Bigg[\frac{1}{2}\begin{pmatrix}
        J_f \\ J_\mathcal{B}
    \end{pmatrix}^\top\cdot\begin{pmatrix}
        \Delta_{ff} & \Delta_{f\mathcal{B}}\\ \Delta_{\mathcal{B}f} & 0
    \end{pmatrix}\cdot\begin{pmatrix}
        J_f \\ J_\mathcal{B}
    \end{pmatrix}+J_f\cdot\Delta_{f\mathcal{B}}\cdot G_{f}^{(0)}\Bigg]\,,
\end{equation}
which allows us to express the field theory in terms of the propagators
\begin{align}
    \Delta_{f\mathcal{B}}(X_1,X_2)&=\Big[\mathbb{1} - G^{(0)}_{f\mathcal{B}}\Big]^{-1}(X_1,X_2)\,\label{eq:causal}\\[2ex]
     \Delta_{\mathcal{B}f}(X_1,X_2)&\equiv\Delta_{f\mathcal{B}}(X_2,X_1)\,\\[2ex]
    \Delta_{ff}(X_1,X_2)&=\int_{\Bar{X}_1,\Bar{X}_2}\,\Delta_{f\mathcal{B}}(X_1,\bar{X}_1)\,G^{(0)}_{ff}(\bar{X}_1,\bar{X}_2)\,\Delta_{\mathcal{B}f}(\bar{X}_2,X_2)
    \label{eq:statistical}\;,
\end{align}
and its vertices $G^{(0)}_{f\dots f\mathcal{B}\dots \mathcal{B}}(X_1,\dots,X_r,X_{1}^\prime,\dots,X_{s}^\prime)$.
We have, thus, developed a field theoretic formulation for the $N$-particle system in which the phase-space density plays the role of the fundamental degree of freedom, as opposed to the path integral in \eqref{eq:PropagatorPathIntegral} which is constructed entirely on microscopic degrees of freedom. We will, therefore, refer to the HS transformed formulation as the \textit{macroscopic} (field) theory. 

\subsection{Free Cumulants}

As we have seen above, the free $f$-$\mathcal{B}$-cumulants play the role of fundamental building blocks of the field theory, as they appear in both, the propagators and the vertices. They contain information on the \textit{microscopic} degrees of freedom. The exact derivation of free $f$-$\mathcal{B}$-cumulants is rather involved and has been presented in detail in \cite{Daus2024}. Let us therefore only summarize the main results here focussing on their structure and physical properties. \\

The general $f$-$\mathcal{B}$-cumulant is defined as the connected correlation function 
\begin{equation}
    G_{f\cdots f \mathcal{B}\cdots \mathcal{B}}^{(0)}(X_1,\cdots X_r,X_1^\prime,\cdots X_s^\prime)=\mean{\hat{\mathcal{T}}f(X_1)\cdots f(X_r)\mathcal{B}(X_1^\prime)\cdots \mathcal{B}(X_s^\prime)}^{(0)}_c\,.
\end{equation}
The prescription of how this object is generally computed is given by equation \eqref{eq:ExpectationPropagator} where the free propagator has to be inserted. The pure $f$-cumulants \eqref{eq:general_free_f_cumulants} have already been discussed at the end of section\ref{sec:ExpectationValues} and describe the free evolution of a connected $k$-point density cluster, propagating the initial correlations, if present. If one or more $\mathcal{B}$-fields are involved, however, the expressions become more complex. Note, that in order to derive \eqref{eq:ExpectationPropagator}, the $\tens{\chi}$-path integrals have been carried out. Since the response field \eqref{eq:ResponseField} contains a $\vec{\chi}_{ p_i}$, it is promoted to an operator,
\begin{equation}
    \mathcal{B}\rightarrow\hat{\mathcal{B}}(X)=\sum_{i=1}^N\nabla_{\vec{q}_i}v(|\vec{q}_i-\vec{  {q}}_1|, t_1)\cdot\nabla_{\vec p_i}\,,
\end{equation}
acting on all Liouville propagators and response fields on its right in \eqref{eq:ExpectationPropagator}. The time ordering in which the operators in \eqref{eq:ExpectationPropagator} appear becomes relevant since the operators do not commute. Furthermore, if the operator associated with the latest time is a $\mathcal{B}$-field, the corresponding expression vanishes identically. This can easily be verified by performing a partial integration and requiring that the Liouville phase-space density vanishes at infinite momenta.\footnote{This is required for the Liouville phase-space density to be normalizable.} In particular, we find
\begin{equation}\label{eq:noGBBB}
    G_{\mathcal{B}\cdots\mathcal{B}}^{(0)}(X_1, \dots\,,X_s)=0\,,
\end{equation}
and the rule 
\begin{equation}\label{eq:causalityrule}
    G_{f\dots f \mathcal{B}\dots \mathcal{B}}^{(0)}(X_1,\dots X_r,X_1^\prime,\dots X_s^\prime)=0\,, \quad \text{if}\,\,\,\,\exists\,\,\,\bar{s}\in[1, \dots\,,s]:t_{\bar{s}}\geq t_{\bar{r}} \,\,\forall \,\,\bar{r}\in[1, \dots\,,r]\,.
\end{equation}
The above rule is simply a causality restriction. A given density cluster can only respond to interactions that happen at earlier times. Consider for instance the simplest mixed cumulant $G_{f\mathcal{B}}^{(0)}$ that appears in the propagator $\Delta_{f\mathcal B}$. It is defined as 
\begin{equation}\label{eq:HowToG_fB}
    G_{f\mathcal{B}}^{(0)}(X_1, X_2)=\mean{\hat{\mathcal{T}}f(X_1)\mathcal{B}(X_2)}_c=\mean{f(X_1)\mathcal{B}(X_2)}_c\theta(t_1-t_2)+\mean{\mathcal{B}(X_2)f(X_1)}_c\theta(t_2-t_1)\,.
\end{equation}
Following the discussion above, the second term on the right-hand side of \eqref{eq:HowToG_fB} vanishes identically. The integration in the first term yields 
\begin{equation}\label{eq:G_fB}
    G_{f\mathcal{B}}^{(0)}(X_1, X_2)=\IntOp{2}{1}{t_2}{\ini{t}}\,G_f^{(0)}(X_1)\,\theta(t_1-t_2)\,,
\end{equation}
with the interaction operator defined as
\begin{equation}\label{eq:InteractioOperator}
    \IntOp{b}{a}{t_b}{\ini{t}}:=\nabla_{\vec{q}_a}v\left(\Big\rvert\vec{q}_a-\frac{\vec{p}_a}{m}(t_a-t_b)-\vec{q}_b\Big\rvert, t_b\right)\cdot\left[-\frac{t_b-\ini{t}}{m}\nabla_{\vec{q}_a}+\bar{\nabla}_{\vec{p}_a}\right]\,,
\end{equation}
where $\bar\nabla_{\vec{p}}$ only acts on the explicit momentum argument of $G_f^{(0)}$. The form of $G_{f\mathcal{B}}^{(0)}$ is closely related to the $G_f$-cumulant at first-order in canonical perturbation theory and corresponds to the time evolved Liouville operator. Thus, it describes how the free one-point cumulant evolving to position $\vec{q}_1$ at time $t_1$ \textit{responds} to a force exhibited at an \textit{earlier} time $t_2$ by a particle at position $\vec{q}_2$. This can be generalized to higher-order mixed cumulants as follows: $G_{f\cdots f\mathcal{B}\cdots\mathcal{B}}^{(0)}(X_1, \cdots, X_{s}, X_{1^\prime}, \cdots, X_{r^\prime})$ describes the response of the freely evolved $s$-point cumulant to $r$ interactions at times $t_{1^\prime}, \cdots,t_{r^\prime}$ with external particles located at $\vec{q}_{1^\prime}, \cdots , \vec{q}_{r^\prime}$. The cumulant thus consists of a differential operator acting on the freely evolved \textit{pure} $f$-cumulant,
\begin{equation}
    G_{f\cdots f\mathcal{B}\cdots\mathcal{B}}^{(0)}(X_1, \cdots, X_{s}, X_{1^\prime}, \cdots, X_{r^\prime})\simeq\hat{\mathcal{D}}\cdot G_{f\cdots f}^{(0)}(X_1, \cdots, X_{s})\,.
\end{equation}
Importantly, due to \eqref{eq:causalityrule} the cumulant introduces a time flow from the external particles sourcing the interaction to the densities being deflected. This fact is incorporated via the Heaviside functions introduced by the time-ordering in the definition of \eqref{eq:generalCumulant}. By inserting the general expression of pure $f$-cumulants \eqref{eq:general_free_f_cumulants}, it can be immediately seen that the mixed $f\mathcal{B}$-cumulants can also be expressed in terms of initial reduced phase-space densities. This is an important property of the free cumulants which shows again that once the initial conditions and the particle trajectories are fixed, the evolution of the system is fully determined.

In \cite{Daus2024} we provide a detailed discussion on the free cumulants, including the case of multiple response fields, which is beyond the scope of this paper. In Appendix \ref{app:freeCumulants} we present a list of the Fourier representation of all cumulants relevant for the results of the present work.

\subsection{Propagators and Feynman Diagrams}\label{sec:propagators}
 
\begin{figure}
     \centering
     \begin{subfigure}[b]{0.45\textwidth}
         \centering
         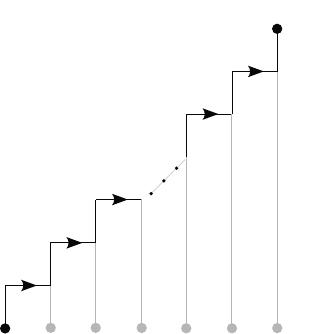
     \end{subfigure}
     \hfill
     \begin{subfigure}[b]{0.5\textwidth}
         \centering
         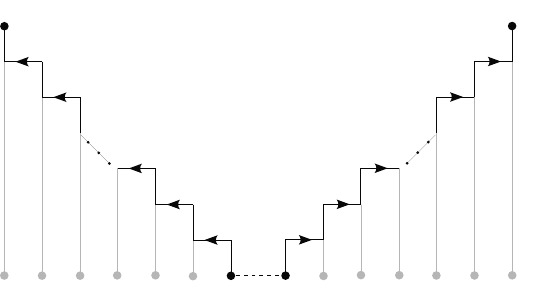
     \end{subfigure}
     \caption{Graphical illustration of the processes contributing to the causal propagator $\Delta_{f\mathcal{B}}$ (\textit{left panel}) and to the statistical propagator $\Delta_{ff}$ (\textit{right panel}). Phase-space densities sourcing the interactions are presented in gray. Dashed lines indicate the connected part of the correlation between these points.}
     \label{fig:StatisticalProp}
\end{figure}

We now turn to the interpretation of the propagator equations \eqref{eq:causal} and \eqref{eq:statistical}. We first note, that the inverse in \eqref{eq:causal} is to be understood in the functional sense, \ie as the solution of the integral equation 
\begin{equation}
    \int_{\bar{X}_1}\Big[\mathbb{1}(X_1,\bar{X}_1) - G^{(0)}_{f\mathcal{B}}(X_1,\bar{X}_1)\Big]\cdot\Delta_{f\mathcal{B}}(\Bar{X}_1,X_2)=\mathbb{1}(X_1,X_2)\,.
\end{equation}
By making the ansatz 
\begin{equation}\label{eq:Delta_rB}
    \Delta_{f\mathcal{B}}(X_1,X_2) = \mathbb{1}(X_1,X_2)+\widetilde{\Delta}_{f\mathcal{B}}(X_1,X_2)\,,
\end{equation}
we find that the corresponding integral equation for $\widetilde{\Delta}_{f\mathcal{B}}$ is given by 
\begin{equation}\label{eq:tildeDelta}
    \widetilde{\Delta}_{f\mathcal{B}}(X_1, X_2)=G^{(0)}_{f\mathcal{B}}(X_1,X_2)+\int_{\bar{X}}G^{(0)}_{f\mathcal{B}}(X_1,\bar{X})\widetilde{\Delta}_{f\mathcal{B}}(\bar{X}, X_2)\,.
\end{equation}
Due to the Heaviside function in \eqref{eq:G_fB} we also find $\widetilde{\Delta}_{f\mathcal{B}}(X_1, X_2)\propto\theta(t_1-t_2)$, which preserves the causal propagation. We therefore call ${\Delta}_{f\mathcal{B}}(X_1, X_2)$ and ${\Delta}_{\mathcal{B}f}(X_1, X_2)$ the \textit{retarded} and \textit{advanced causal propagators}, respectively, and define the Feynman rules,
\begin{align}
    \Delta_{f\mathcal{B}}(X_1,X_2)&=
    \begin{tikzpicture}[baseline={(0,-0.6ex)}, scale=0.55]
         \DeltaRBTree{-3,0}{0,0}
     \end{tikzpicture}\,, \\[3ex]
     \Delta_{\mathcal{B}f}(X_1,X_2)&=
     \begin{tikzpicture}[baseline={(0,-0.6ex)}, scale=0.55]
         \DeltaRBTree{3,0}{0,0}
     \end{tikzpicture}  \,,
\end{align}
where the arrow indicates the time flow from earlier to later times. Equation \eqref{eq:tildeDelta} is of Volterra-Fredholm type and has to be solved numerically in general. However, we will see shortly, how symmetry assumptions of the system under consideration can simplify the above integral equation. By iterating \eqref{eq:tildeDelta} we find the Neumann series representation of the causal propagator, 
\begin{align}\label{eq:Neumannseries}
    \Delta_{f\mathcal{B}}(X_1,X_2)=&\mathbb{1}(X_1,X_2)+G^{(0)}_{f\mathcal{B}}(X_1,X_2)+\int_{\Bar{X}_1}G^{(0)}_{f\mathcal{B}}(X_1,\Bar{X}_1)\,G^{(0)}_{f\mathcal{B}}(\Bar{X}_1,X_2)+\cdots
    =\sum_{n=0}^\infty\Big[G^{(0)}_{f\mathcal{B}}\Big] ^n(X_1,X_2)\,,
\end{align}
where the power of $n$ on the right hand side implies integration. The interpretation of the causal propagators, therefore, becomes clear: each $G_{f\mathcal{B}}^{(0)}$ describes how a one-point phase-space density responds to an incoming interaction, which is sourced by a phase-space density that has itself already been influenced by an earlier interaction. This process can repeat arbitrarily many times within the time interval $[t_1, t_2]$. Figure \ref{fig:StatisticalProp} provides a graphical representation of this iteration.
Thus, objects involving $\Delta_{f\mathcal{B}}$ propagators are, by construction, non-perturbative in terms of the microscopic interaction potential as they involve an analytical or a numerical resummation of the interaction series \eqref{eq:Neumannseries}. Note, that the potential does not depend on the particle momenta. $G_{f\mathcal{B}}^{(0)}(X_1, X_2)$ is, therefore, independent of $\vec{p}_2$ as well, which is inherited by $\Delta_{f\mathcal{B}}$. We can, thus, perform the integration over the momenta which simplifies the integral equation in \eqref{eq:tildeDelta},
\begin{align}\label{eq:TrueIteration}
    \widetilde{\Delta}_{f\mathcal{B}}(X_1, X_2)=&\,G^{(0)}_{f\mathcal{B}}(X_1,X_2)+\int_{\vec{q}^{\,\prime}, t^\prime}G^{(0)}_{f\mathcal{B}}(X_1,\bar{X})\widetilde{\Delta}_{\rho\mathcal{B}}(\vec{q}^{\,\prime}, t^\prime, \vec q_2, t_2)\,,\\[2ex]
    \widetilde{\Delta}_{\rho\mathcal{B}}(\vec{q}_1, t_1, \vec{q}_2, t_2)=&\,G^{(0)}_{\rho\mathcal{B}}(\vec{q}_1, t_1, \vec{q}_2, t_2)+\int_{\vec{q}^{\,\prime}, t^\prime}G^{(0)}_{\rho\mathcal{B}}(\vec{q}_1, t_1, \vec{q}^{\,\prime}, t^\prime)\widetilde{\Delta}_{\rho\mathcal{B}}(\vec{q}^{\,\prime}, t^\prime,\vec{q}_2, t_2)\,.
\end{align}
This reduces the dimensionality of the integral equation from seven to four, since interactions are only sourced by the particle density\footnote{In principle, we could have performed the whole derivation only using the particle density instead of the full phase space density. However, this would loose the momentum information which is now restored in the final iteration in the above equation.} $\rho(\vec{q}, t)$. Therefore, all momentum integrals are trivially performed during the iteration procedure, apart from the final one which restores the information on the momentum distribution. Nevertheless, the solution of the system \eqref{eq:TrueIteration} is rather involved and requires numerical discretization in time and space. However, in the next sections we will see how symmetry considerations will simplify the above system.\\

Similarly, we find an interpretation of the propagator \eqref{eq:statistical}: Two initially correlated phase-space densities, described by $G^{(0)}_{ff}$, source the interaction chain included in the $\Delta_{f\mathcal{B}}$ propagators which we described above. The two external densities at $X_1$ and $X_2$ thus remain fully connected, see Figure \ref{fig:StatisticalProp}. Since $\Delta_{ff}$ thus carries statistical information, we refer to it as the \textit{statistical propagator} and equip it with the Feynman rule
\begin{equation}
     \Delta_{ff}(X_1,X_2)=\int_{\Bar{X}_1,\Bar{X}_2}\,\Delta_{f\mathcal{B}}(X_1,\bar{X}_1)\,G^{(0)}_{ff}(\bar{X}_1,\bar{X}_2)\,\Delta_{\mathcal{B}f}(\bar{X}_2,X_2)
     =\; \begin{tikzpicture}[baseline={(0,-0.6ex)}, scale=0.55]
         \DeltaRRTree{3,0}{0,0}
     \end{tikzpicture} \,.
\end{equation}
The two outgoing arrows indicate that $\Delta_{ff}$ contains interaction chains propagating outwards in both directions which are sourced by the $G^{(0)}_{ff}$-cumulant, represented by the solid dot. By means of \eqref{eq:Delta_rB}, equation \eqref{eq:statistical} can be expanded such that
\begin{equation}\label{eq:Delta_ffRealSpace}
\begin{split}
    \Delta_{ff}(X_1,X_2)=G^{(0)}_{ff}(X_1,X_2)+&\int_{\bar{X}_1}\widetilde{\Delta}_{f\mathcal{B}}(X_1, \bar{X}_1)G^{(0)}_{ff}(\bar{X}_1,X_2)+\int_{\bar{X}_2}G^{(0)}_{ff}(X_1,\bar X_2)\widetilde{\Delta}_{\mathcal{B}f}(\bar X_2, X_2)\\&+\int_{\bar{X}_1, \bar{X}_2}\widetilde{\Delta}_{f\mathcal{B}}(X_1, \bar{X}_1)G^{(0)}_{ff}(\bar{X}_1,\bar X_2)\widetilde{\Delta}_{\mathcal{B}f}(\bar X_2, X_2)\,.
    \end{split}
\end{equation}
Since there is no pure $G_{\mathcal{B}\mathcal{B}}^{(0)}$-cumulant, we also have $\Delta_{\mathcal{B}\mathcal{B}}=0$. 

Last but not least, the vertices $G_{f\cdots f\mathcal{B}\cdots\mathcal{B}}^{(0)}$ describe how $r$ interactions affect the future evolution of a connected $s$-particle cluster. We therefore represent vertices by the following Feynman rule, 
\begin{equation}\label{eq:vertex}
    G^{(0)}_{f\dots f\mathcal{B}\dots \mathcal{B}}(X_1,\dots,X_r,X_{1}^\prime,\dots,X_{s}^\prime) = \;\begin{tikzpicture}[baseline={(0,-0.6ex)}, scale=0.55]
        \nvertex
    \end{tikzpicture} 
    \;,
\end{equation}
where we represent the time flow by arrows attached to the amputated external legs. The solid line will be attached to incoming interaction chains, while the dashed lines will source outgoing interactions, sourced by the cumulant itself. Importantly, due to the rule in \eqref{eq:causalityrule} there has to be at least one outgoing line, representing the density that is being deflected, meaning that there is no sink in the time flow. However, a pure phase-space density cumulant $G_{f\cdots f}^{(0)}$ will only source outgoing interaction chains. We can now understand the general construction of a Feynman diagram. All components of the diagram, \ie propagators and vertices, are attached to each other such that the densities of one component act as sources for the response functions of another. This ensures a continuous time flow, as indicated by the arrows, which can only terminate at the external phase-space densities. Within vertices, the time flow can split or merge, indicating how a single density influences the future evolution of multiple densities or is affected through interactions with multiple densities from the past.

\subsection{Macroscopic Self-Energy}\label{sec:selfenergy}

With the generating functional \eqref{eq:Generatingfunctional} at hand, we now wish to construct a systematic way in order to compute full correlation functions. As we mainly aim at the computation of cumulants, \ie connected correlation functions, we define the Schwinger functional $\mathcal{W}[J]$ as 
\begin{equation}
    \mathcal{W}[J]:=\ln\Big[\mathcal{Z}[J]\Big]
\end{equation}
which generates the connected correlation functions upon taking functional derivatives \wrt the source field $J$,
\begin{equation}
    \frac{\delta^r}{\delta J(X_1)\cdots\delta J(X_r)}\mathcal{W}[J]=G(X_1, \cdots X_r)
\end{equation}
where $G$ denotes the full cumulant.\footnote{In the above equation we have suppressed the tensor indices standing for $f$ and $\mathcal{B}$. $G$ is therefore a tensorial object. } In particular, the full propagator of the theory is given by 
\begin{equation}
    \frac{\delta^2}{\delta J(X_1)\delta J(X_2)}\mathcal{W}[J]=G(X_1, X_2)\,.
\end{equation}
As ususal, we introduce the statistical effective action $\Gamma[\Psi]$ as the functional Legendre transform of the Schwinger functional, 
\begin{equation}\label{eq:effectiveaction}
    \Gamma[\Psi]=\int_{X_1}J(X_1)\cdot\Psi(X_{1})-\mathcal{W}[J]\,,\quad \text{with}\quad \Psi(X)= \frac{\delta \mathcal{W}[J]}{\delta J(X)}\,,\quad \text{and}\quad  \frac{\delta \Gamma[\Psi]}{\delta \Psi(X)}=J(X)\,.
\end{equation}
$\Gamma[\Psi]$ will help us relate the full, unknown propagator $G$ to the bare, known propagator $\Delta$. As in the QFT analogue, the effective action $\Gamma[\Psi]$ generalizes the bare action $\mathcal{S}[\Psi]$ in \eqref{eq:Generatingfunctional} to the full statistical theory, in which all vertices already contain the full statistical information. Using the functional relations in \eqref{eq:effectiveaction}, one finds
\begin{equation}
    \int_{X_2}\frac{\delta^2\mathcal{W}[J]}{\delta J(X_1)\delta J(X_2)}\cdot\frac{\delta^2\Gamma[\Psi]}{\delta \Psi(X_2)\delta \Psi(X_3)}=\dirac{X_1-X_3}\,,
\end{equation}
which implies 
\begin{equation}
    \frac{\delta^2\Gamma[\Psi]}{\delta \Psi(X_1)\delta \Psi(X_2)}=G^{-1}(X_1, X_2)\,,
\end{equation}
\ie giving the full inverse propagator in contrast to 
\begin{equation}
    \frac{\delta^2\mathcal{S}_0[\Psi]}{\delta \Psi(X_1)\delta \Psi(X_2)}=\Delta^{-1}(X_1, X_2)\,,
\end{equation}
which follows form \eqref{eq:treelevelaction} and only contains tree-level information. We therefore define the macroscopic \textit{self-energy} $\Sigma(X_1, X_2)$ as the difference 
\begin{equation}\label{eq:selfenergy}
    \Sigma(X_1, X_2)=\frac{\delta^2\mathcal{S}_0[\Psi]}{\delta \Psi(X_1)\delta \Psi(X_2)}-\frac{\delta^2\Gamma[\Psi]}{\delta \Psi(X_1)\delta \Psi(X_2)}=\Delta^{-1}(X_1, X_2)-G^{-1}(X_1, X_2)\,.
\end{equation}
One can show that by its definition \eqref{eq:effectiveaction}, $\Gamma[\Psi]$ only contains 1PI diagrams\footnote{This can, for example, be shown by considering the equations of motion for the full correlation functions, \ie the respective \textit{Dyson-Schwinger} equations, relating lower-order full correlators to higher-order full correlators.}, \ie diagrams that cannot be separated by cutting one internal line. Consequently, the self-energy $\Sigma$ only consists of 1PI diagrams with two amputated external legs. Equation \eqref{eq:selfenergy} can formally be solved by multiplication with $G$ and $\Delta$ respectively, and leads to the \textit{Dyson equation},
\begin{equation}\label{eq:Dyson}
    G(X_1, X_2)=\Delta(X_1, X_2)+\int_{X_3, X_4}\Delta(X_1, X_3)\cdot\Sigma(X_3, X_4)\cdot G(X_4, X_2)\,,
\end{equation}
which iterates all 1PI diagrams by connecting them through $\Delta$. Equation \eqref{eq:Dyson} is a matrix integral equation. Its respective component equations read
\begin{equation}\label{eq:FullCumulants}
    \begin{split}
        G_{ff}(X_1, X_2)=&\,\,\Delta_{ff}(X_1, X_2)+\int_{X_3, X_4}\Delta_{ff}(X_1, X_3)\Sigma_{\mathcal{B}f}(X_3, X_4) G_{\mathcal{B}f}(X_4, X_2)\\&+\int_{X_3, X_4}\Delta_{f\mathcal{B}}(X_1, X_3)\bigg[\Sigma_{f\mathcal{B}}(X_3, X_4) G_{ff}(X_4, X_2)+\Sigma_{ff}(X_3, X_4) G_{\mathcal{B}f}(X_4, X_2)\bigg]\,,\\[2ex]
        G_{f\mathcal{B}}(X_1, X_2)=&\,\,\Delta_{f\mathcal{B}}(X_1, X_2)+\int_{X_3, X_4}\Delta_{f\mathcal{B}}(X_1, X_3)\Sigma_{f\mathcal{B}}(X_3, X_4) G_{f\mathcal{B}}(X_4, X_2)\,,\\[2ex]
        G_{\mathcal{B}f}(X_1, X_2)=&\,\,G_{f\mathcal{B}}(X_2, X_1)\,,
    \end{split}
\end{equation}
and represent coupled integral equations that have to be solved numerically. In the above equations we have used that $\Delta_{\mathcal{B\mathcal{B}}}=0$ as well as $\Sigma_{\mathcal{B}\mathcal{B}}=0$, which follows from the fact that all pure $\mathcal{B}$-cumulants vanish identically. This implies that also $G_{\mathcal{B}\mathcal{B}}=0$. Furthermore, $G_{f\mathcal{B}}$ inherits the causal structure of $\Delta_{f\mathcal{B}}.$\\

We can now turn to the respective diagrams that contribute to the self-energy. Therefore note, that the interaction part of the action \eqref{eq:Vertexpart} consists of an infinite tower of $k$-point self-interactions. Thus, to one-loop order there exist two different topologies of diagrams that contribute to the self energy, given by 
\begin{align}\label{eq:1loop_topologies}
    \begin{tikzpicture}[baseline={-0.6ex}, scale=0.55]
        \draw[propagator, opacity=0.3] (-2,0)--(-1,0); 
        \ThreePointVertices{-1,0}{1,0}
        \PlainLoopA
        \draw[propagator, opacity=0.3] (1,0)--(2,0);
    \end{tikzpicture} \quad\text{and}\quad \begin{tikzpicture}[baseline={-0.6ex}, scale=0.55]
        \draw[propagator, opacity=0.3] (-1,0)--(0,0);
        \FourPointVertex{0,0}
        \PlainLoopB
        \draw[propagator, opacity=0.3] (0,0)--(1,0);
    \end{tikzpicture}\;,
\end{align}
coming from the three- and four-point self-interactions. Indeed, the highest vertex contributing to the self-energy at $s$-loop order is an $2s+2$-point self-interaction, as it connects all propagator lines to the same vertex yielding generalizations of the diagram on the right in \eqref{eq:1loop_topologies}. Having determined the possible topologies at a given loop-level, we can now construct all contributing diagrams by inserting the respective vertices appearing in the action \eqref{eq:Vertexpart} and connecting them with the respective propagator. For our one-loop example, the three- and four point vertices are $G^{(0)}_{fff}\,,G^{(0)}_{ff\mathcal{B}}\,,G^{(0)}_{f\mathcal{B}\mathcal{B}}\,,G^{(0)}_{fff\mathcal{B}}\,,G^{(0)}_{ff\mathcal{B}\mathcal{B}}\,,G^{(0)}_{f\mathcal{B}\mathcal{B}\mathcal{B}} $. Keeping the above Feynman rules in mind, the one-loop macroscopic self-energy is given by the following set of diagrams\footnote{Note, that we have suppressed the tadpole diagrams, that contribute to the self energy. Technically, those diagrams are present due to the possibly inhomogeneous background represented by $G^{(0)}_\rho$. In our case of interest however, we focus on a homogeneous setting, where those contributions identically vanish, as discussed in the next section.},
\begin{align}
\begin{split}
    \Sigma_{ff}^{(\text{1-loop})}(X_1, X_2)=&\,\,\,\,\frac{1}{2}\,\,\begin{tikzpicture}[baseline={-0.6ex}, scale=0.55]
            \draw[dashed_propagator, stealth-, opacity=0.3] (-2,0)--(-1,0);
            \ThreePointVertices{-1,0}{1,0};
            \DeltaRRTreeDown;
            \DeltaRRTreeUp;
            \draw[dashed_propagator, -stealth, opacity=0.3] (1,0)--(2,0);
        \end{tikzpicture} \,\,\,\,+ \,\,\,\,
        \begin{tikzpicture}[baseline={-0.6ex}, scale=0.55]
            \draw[dashed_propagator, stealth-, opacity=0.3] (-2,0)--(-1,0);
            \ThreePointVertices{-1,0}{1,0};
            \DeltaRBTreeDownL;
            \DeltaRRTreeUp;
            \draw[dashed_propagator, -stealth, opacity=0.3] (1,0)--(2,0);
        \end{tikzpicture}\,\,\,\,+ \,\,\,\,
        \begin{tikzpicture}[baseline={-0.6ex}, scale=0.55]
            \draw[dashed_propagator, stealth-, opacity=0.3] (-2,0)--(-1,0);
            \ThreePointVertices{-1,0}{1,0};
            \DeltaRBTreeDownR;
            \DeltaRRTreeUp;
            \draw[dashed_propagator, -stealth, opacity=0.3] (1,0)--(2,0);
        \end{tikzpicture}\,\,\,\,+ \,\,\,\,
        \begin{tikzpicture}[baseline={-0.6ex}, scale=0.55]
            \draw[dashed_propagator, stealth-, opacity=0.3] (-2,0)--(-1,0);
            \ThreePointVertices{-1,0}{1,0};
            \DeltaRBTreeDownL;
            \DeltaRBTreeUpR;
            \draw[dashed_propagator, -stealth, opacity=0.3] (1,0)--(2,0);
        \end{tikzpicture}\\[2ex] &+\,\,\,\,
        \frac{1}{2}\,\,\begin{tikzpicture}[baseline={-0.6ex}, scale=0.55]
            \draw[dashed_propagator, stealth-, opacity=0.3] (-2,0)--(-1,0);
            \ThreePointVertices{-1,0}{1,0};
            \DeltaRBTreeDownL;
            \DeltaRBTreeUpL;
            \draw[dashed_propagator, -stealth, opacity=0.3] (1,0)--(2,0);
        \end{tikzpicture}\,\,\,\,+ \,\,\,\,
        \frac{1}{2}\,\,\begin{tikzpicture}[baseline={-0.6ex}, scale=0.55]
            \draw[dashed_propagator, stealth-, opacity=0.3] (-2,0)--(-1,0);
            \ThreePointVertices{-1,0}{1,0};
            \DeltaRBTreeDownR;
            \DeltaRBTreeUpR;
            \draw[dashed_propagator, -stealth, opacity=0.3] (1,0)--(2,0);
        \end{tikzpicture}\,\,\,\,+ \,\,\,\,
        \frac{1}{2}\,\,\begin{tikzpicture}[baseline={-0.6ex}, scale=0.55]
            \draw[dashed_propagator, stealth-, opacity=0.3] (-1,0)--(0,0);
            \FourPointVertex{0,0};
            \DeltaRRTreeRound;
            \draw[dashed_propagator, -stealth, opacity=0.3] (0,0)--(1,0);
        \end{tikzpicture} \,\,\,\,\,\,\,\,+\,\,\,\,\,\,\,\,
        \begin{tikzpicture}[baseline={-0.6ex}, scale=0.55]
            \draw[dashed_propagator, stealth-, opacity=0.3] (-1,0)--(0,0);
            \FourPointVertex{0,0};
            \DeltaRBTreeLRound;
            \draw[dashed_propagator, -stealth, opacity=0.3] (0,0)--(1,0);
        \end{tikzpicture}\,,
    \end{split}\label{eq:1LoopSelfenergyff}\\[6ex]
\begin{split}
    \Sigma_{f\mathcal{B}}^{(\text{1-loop})}(X_1, X_2)=&\,\,\,\,\begin{tikzpicture}[baseline={-0.6ex}, scale=0.55]
            \draw[dashed_propagator, stealth-, opacity=0.3] (-2,0)--(-1,0);
            \ThreePointVertices{-1,0}{1,0};
            \DeltaRBTreeDownL;
            \DeltaRRTreeUp;
            \draw[propagator, reversed_arrow, opacity=0.3] (1,0)--(1.8,0);
        \end{tikzpicture} \,\,\,\,+ \,\,\,\,
        \frac{1}{2}\,\,\begin{tikzpicture}[baseline={-0.6ex}, scale=0.55]
            \draw[dashed_propagator, stealth-, opacity=0.3] (-2,0)--(-1,0);
            \ThreePointVertices{-1,0}{1,0};
            \DeltaRBTreeUpL;
            \DeltaRBTreeDownL;
            \draw[propagator, reversed_arrow, opacity=0.3] (1,0)--(1.8,0);
        \end{tikzpicture}\,\,\,\,+ \,\,\,\,
       \begin{tikzpicture}[baseline={-0.6ex}, scale=0.55]
            \draw[dashed_propagator, stealth-, opacity=0.3] (-2,0)--(-1,0);
            \ThreePointVertices{-1,0}{1,0};
            \DeltaRBTreeUpR;
            \DeltaRBTreeDownL;
            \draw[propagator, reversed_arrow, opacity=0.3] (1,0)--(1.8,0);
        \end{tikzpicture}\\[2ex] &+\,\,\,\,
        \frac{1}{2}\,\,\begin{tikzpicture}[baseline={-0.6ex}, scale=0.55]
            \draw[dashed_propagator, stealth-, opacity=0.3] (-1,0)--(0,0);
            \FourPointVertex{0,0};
            \DeltaRRTreeRound;
            \draw[propagator, reversed_arrow, opacity=0.3] (0,0)--(0.8,0);
        \end{tikzpicture} \,\,\,\,\,\,\,\,+\,\,\,\,\,\,\,\,
        \begin{tikzpicture}[baseline={-0.6ex}, scale=0.55]
            \draw[dashed_propagator, stealth-, opacity=0.3] (-1,0)--(0,0);
            \FourPointVertex{0,0};
            \DeltaRBTreeLRound;
            \draw[propagator, reversed_arrow, opacity=0.3] (0,0)--(0.8,0);
        \end{tikzpicture}\,.
    \end{split}\label{eq:1LoopSelfenergyfB}
\end{align}
To illustrate how the symmetry factors appearing in \eqref{eq:1LoopSelfenergyff} and \eqref{eq:1LoopSelfenergyfB} are determined, let us consider the first diagram in \eqref{eq:1LoopSelfenergyff}: 
First, there are $2\times2=4$ ways to attach the $\Delta_{ff}$ propagators to the vertices. This number must be divided by the symmetry factors associated with the $G_{f\mathcal{B}\mathcal{B}}^{(0)}$ cumulants at each vertex and the second-order Taylor expansion of the exponential. Specifically, each vertex contributes a factor of $\frac{1}{2!}$ from the cumulants, and the second-order expansion contributes an additional factor of $\frac{1}{2!}$. Thus, the overall prefactor is $\frac{4}{(2!)^3}=\frac{1}{2}$. 
Finally, the analytical expression for the first diagram reads 
\begin{equation}\label{eq:diagram1}
    \frac{1}{2}\,\,\begin{tikzpicture}[baseline={-0.6ex}, scale=0.55]
            \draw[dashed_propagator, stealth-, opacity=0.3] (-2,0)--(-1,0);
            \ThreePointVertices{-1,0}{1,0};
            \DeltaRRTreeDown;
            \DeltaRRTreeUp;
            \draw[dashed_propagator, -stealth, opacity=0.3] (1,0)--(2,0);
        \end{tikzpicture} = \frac{1}{2}\,\,\int_{X_3,X_4, X_5, X_6}G_{f\mathcal{B}\mathcal{B}}^{(0)}(X_1, X_3, X_4)\Delta_{ff}(X_3, X_5)\Delta_{ff}(X_4, X_6)G_{f\mathcal{B}\mathcal{B}}^{(0)}(X_2, X_5, X_6)\,.
\end{equation}\\

With a given loop approximation of the self-energies, \eqref{eq:Dyson} represents a non-perturbative equation for the propagator, as it formally resums an infinite tower of 1PI loop diagrams that are connected by tree-level propagators. The numerical solution of this coupled integral equation, therefore, contains information up to infinite loop-order. Beyond this approach, other non-perturbative methods such as the 2PI (two-particle-irreducible) formalism and Dyson-Schwinger master equations provide more sophisticated resummation prescriptions that capture higher-order effects. The 2PI formalism, derived from the 2PI effective action (or \textit{Luttinger-Ward functional}), resums two-particle-irreducible diagrams and provides self-consistent equations for the two-point cumulant by involving only full propagators in the self-energy. This approach is particularly useful for studying non-equilibrium phenomena. Similarly, Dyson-Schwinger equations (DSE) provide equations of motion for the full statistical $k$-point functions. As they couple lower-order cumulants to higher-order cumulants, they represent an infinite tower of coupled integral equations. This hierarchical framework therefore requires a truncation procedure for the full statistical $k$-point functions that is different from the one in the 2PI approach. Last but not least, the functional renormalization group approach (FRG) represents yet another, conceptually different, non-perturbative approach based on flow equations for the resepective full statistical $k$-point function. As in the Dyson-Schwinger case, it provides an infinite tower of coupled integro-differential equations which, however, only includes full statistical propagators and vertices. \\

At the other end of the spectrum, the simplest solution to the Dyson equation emerges from truncating the series to the lowest loop-order. This corresponds to the one-loop perturbative expansion, where the self-energy is approximated by only including the one-loop contributions given by \eqref{eq:1LoopSelfenergyff} and \eqref{eq:1LoopSelfenergyfB}. The iteration in \eqref{eq:Dyson} is then performed once, leading to 
\begin{equation}\label{eq:1LoopPropagator}
    G^{(\text{1-loop})}(X_1, X_2)=\Delta(X_1, X_2)+\int_{X_3, X_4}\Delta(X_1, X_3)\cdot\Sigma^{(\text{1-loop})}(X_3, X_4)\cdot \Delta(X_4, X_2)\,.
\end{equation} 

\section{Homogeneous and Isotropic Systems}\label{sec:HomSys}

In the following, we will apply the above framework to homogeneous and isotropic systems, as both conditions are required by the cosmological principle. The symmetries implied by these conditions will introduce considerable simplifications. Furthermore, the notions of homogeneity and isotropy plays a significant role in various physical applications beyond cosmology, making it an interesting case to consider in its own right.

\subsection{Reduced Phase-Space densities and Expectation Values in  Fourier Space}
A system is called homogeneous and isotropic if all expectation values are invariant under translations and rotations, respectively. In other words, they are independent of the absolute position and orientation in space. In particular, this implies for the reduced phase-space densities that
\begin{equation}\label{eq:generalHomogeneous}
\begin{split}
    f_1(x_1, t)=&\,\bar{\rho}\,\varphi(\vec{p}_1, t)\,,\\
    g_2(x_1, x_2, t)=&\,\bar{\rho}^2\,\widetilde g_2(|\vec{q}_1-\vec{q}_2|, \vec{p}_1, \vec{p}_2, t)\,,\\
    g_3(x_1, x_2, x_3, t)=&\,\bar{\rho}^3\,\widetilde g_3(|\vec{q}_1-\vec{q}_2|, |\vec{q}_1-\vec{q}_3|, \vec{p}_1, \vec{p}_2, \vec p_3, t)\,,\\
   \vdots &\,\,\,\,\,\,\,\,
   \end{split}
\end{equation}
where $\varphi(\vec{p})$ is the normalized momentum-distribution function and $\bar{\rho}=N/V$ is the mean particle-number density. For homogeneous systems, most integrals above involve convolutions, which turn into products in Fourier space. We therefore define the Fourier transformed version of the reduced phase-space densities \eqref{eq:generalHomogeneous} as
\begin{equation}\label{eq:generalHomogeneousk_space}
\begin{split}
    f_1(s_1 , t)=&\,(2\pi)^3\dirac{\vec{k}_1}\bar{\rho}\,\varphi(l_1, t)\,,\\
    g_2(s_1, s_2, t)=&\,(2\pi)^3\dirac{\vec{k}_1 + \vec k_2}\bar{\rho}^2\widetilde g_2(k_1, \vec{l}_1, \vec{l}_2, t)\,,\\
    g_3(s_1, s_2, s_3, t)=&\,(2\pi)^3\dirac{\vec{k}_1 + \vec k_2+\vec k_3}\bar{\rho}^3\widetilde g_3(\vec k_1, \vec k_2, \vec{l}_1, \vec{l}_2, \vec l_3, t)\,,\\
   \vdots &\,\,\,\,\,\,\,\,
   \end{split}
\end{equation}
where $s=(\vec k, \vec l)$ is the Fourier conjugate variable to $x=(\vec q, \vec p)$ and we used the same symbol for a function and its Fourier transform, since distinction is made clear by the arguments. The Dirac delta distributions reflect the symmetry of the system by lowering the degrees of freedom in the reduced densities. If the system under consideration initially decomposes as \eqref{eq:generalHomogeneous} or \eqref{eq:generalHomogeneousk_space}, respectively, and the potential respects this symmetry -- which is true if the potential only depends on relative distances \eqref{eq:Hamiltongeneral} -- the system remains homogeneous for all times $t>\ini{t}$. Since it is convenient to re-phrase the field theory in Fourier space, we introduce the Fourier transform of the Klimontovich phase-space density \eqref{eq:Klimontovich}
\begin{equation}
    f(\vec k_1, \vec l_1, t) = \sum_{i=1}^N\e^{-\mi \vec k_1\cdot \vec{\ms{q}}_i(t_1)-\mi \vec l_1\cdot \vec{\ms{p}}_i(t_1)}\,.
\end{equation}
In Fourier space, expectation values as in \eqref{eq:generalmomentummoments} can be computed according to 
\begin{equation}\label{eq:generalmomentummomentsk_space}
    \mean{\hat{\mathcal{T}}\mathcal{O}_1(\vec{k}_1, t_1)\cdots \mathcal{O}_k(\vec{k}_k, t_k)}=\,F_{\mathcal{O}_1}(\mi\nabla_{\vec{l}_1})\cdots F_{\mathcal{O}_k}(\mi\nabla_{\vec{l}_k})\mean{\hat{\mathcal{T}}f(\vec{k}_1, \vec l_1, t_1)\cdots f(\vec{k}_k, \vec l_k, t_k) }\Bigg\rvert_{\vec l_1=\cdots=\vec l_r=0}\,,
\end{equation}
\ie by taking appropriate derivatives \wrt to the variable $\vec l$, conjugate to the momentum. In particular, for pure density cumulants, we can set all $\vec l$'s to zero and find 
\begin{equation}
    G_{\rho\cdots\rho}(\vec k_1, \eta_1, \dots, \vec k_r, \eta_r)=G_{f\cdots f}(\vec k_1,\vec l_1 , \eta_1, \dots, \vec k_r,\vec l_r, \eta_r)\Bigg\rvert_{\vec l_1=\cdots=\vec l_r=0}\,.
\end{equation}

\subsection{Free Cumulants and Tadpole Cancellation}

Turning to the free cumulants, the homogeneous, momentum independent potential together with \eqref{eq:generalHomogeneousk_space} implies 
\begin{equation}\label{eq:momentumconservation}
    G_{f\dots f\mathcal{B}\dots \mathcal{B}}^{(0)}(S_1, \dots, S_r, S_1^\prime, \dots , S_s^\prime)\propto \,(2\pi)^3\dirac{\vec{k_1}+\cdots+\vec{k}_r+\vec{k}_1^\prime+\cdots+\vec k_s^\prime}\prod_{i=1}^s(2\pi)^3\dirac{\vec l_i^\prime}\,,
\end{equation}
where $S=(s, t)$. The above relation enforces a $\vec k$-mode conservation at each vertex, and thus a $\vec k$-mode conservation in the respective Feynman diagrams. This is reminiscent of momentum conservation for a homogeneous system in quantum field theory. Furthermore, it implies that internal $\vec l$-vectors coming from propagators ending on a $\mathcal{B}$-field  are set to zero. We can therefore reduce the degrees of freedom in the free cumulants, by pulling the above Dirac delta distribution out of the corresponding expression. In particular, we can generally write 
\begin{align}\label{eq:propagator_momentumconservation}
    G_f^{(0)}(S_1)=&\,(2\pi)^3\dirac{\vec k_1}\,G_f^{(0)}(\vec l_1, \ini{t})\,,\\[2ex]
    G_{ff}^{(0)}(S_1, S_2)= &\, (2\pi)^3\dirac{\vec{k}_1+\vec{k}_2}G_{ff}^{(0)}(\vec k_1, \vec{l}_1, \vec l_2, t_1, t_2)\,,\label{eq:G_ffFouriernoDelta}\\[2ex]
    G_{f\mathcal{B}}^{(0)}(S_1, S_2)=&\,(2\pi)^3\dirac{\vec{k}_1+\vec{k}_2}(2\pi)^3\dirac{\vec l_2}G_{f\mathcal{B}}^{(0)}(\vec k_1, \vec{l}_1, t_1, t_2)\,,
\end{align}
Where we, once more, use the same symbol on both sides and let the number of arguments distinguish between them. In Appendix \ref{app:freeCumulants} we have listed the general form of all free cumulants for a homogeneous system in terms of the initial reduced phase-space densities \eqref{eq:generalHomogeneousk_space} that are relevant for our discussion. In particular, we find 
\begin{align}\label{eq:GeneralTreeLevelCumulants}
    G_f^{(0)}(\vec l_1, t_1)=&\,\bar{\rho}\,\varphi(\vec l_1, \ini{t})\,,\\[2ex]
    \begin{split}
    G_{ff}^{(0)}(\vec k_1, \vec{l}_1, \vec l_2, t_1, t_2)=&\bar{\rho}\,\varphi \bigg(\vec l_1 + \vec l_2+\vec k_1 \frac{T_{12}}{m}\bigg)+\bar{\rho}^2\,\widetilde g_2\bigg(\vec k_1, \vec l_1 + \vec k_1\frac{T_1}{m}, \vec l_2-\vec k_1\frac{T_2}{m}, \ini{t}\bigg) \,, 
    \end{split}\label{eq:G_ffFourier}\\[2ex]
    G_{f\mathcal{B}}^{(0)}(\vec k_1, \vec{l}_1, t_1, t_2)=&\,\bar{\rho}\,b(1,2)\varphi\left(\vec l_1+\vec{k}_1\frac{T_{12}}{m}, \ini{t}\right)\label{eq:G_fBFourier}\,.
\end{align}
The first term in \eqref{eq:G_ffFourier} corresponds to the Poisson shot-noise and identifies both external particles. We furthermore introduced the short-hand notations 
\begin{equation}\label{eq:abbreviation}
    T_a=t_a-\ini{t}\,,\quad T_{ab}=t_a-t_b\,,\quad b(a,b)=v(k_b, t_b)\vec{k}_b\cdot\bigg(\vec{k}_a\frac{T_{ab}}{m}+\vec{l}_a\bigg)\theta(t_a-t_b)\,,
\end{equation}
where $b(a,b)$ is the Fourier-analogue to the interaction operator \eqref{eq:InteractioOperator} and appears together with every $\mathcal{B}$-field in a mixed cumulant. We therefore find
\begin{equation}\label{eq:cumulantproptoforce}
    G_{f\dots f\mathcal{B}\dots \mathcal{B}}^{(0)}(S_1, \dots, S_r, S_1^\prime, \dots , S_s^\prime)\propto v(k^\prime_{\bar{s}}, t^\prime_{\bar{s}})\,\vec{k}^\prime_{\bar{s}}\quad\quad\forall\bar{s}\in[1,\dots\,,s]\,.
\end{equation}
From the Fourier-mode conservation \eqref{eq:momentumconservation} and \eqref{eq:propagator_momentumconservation}, we can deduce the following property for an amputated tadpole diagram\footnote{Note, that there cannot be a tadpole diagram with an incoming solid arrow, by causality, as explained before.}
\begin{equation}
    \begin{tikzpicture}[baseline={-0.6ex}, scale=0.55]
            \draw[dashed_propagator, stealth-, opacity=0.3] (-2,0)--(-1,0);
            \filldraw[] (-1,0) circle (2pt);
            \draw[propagator] (0,0) circle[radius=1];
            \draw[pattern={Lines[angle=45,distance={3pt}]},pattern color=black]
          (0,0) circle [radius=1cm];
           \draw ( -1.5, 0) node [above] {$\overset{\vec{k}_1}{\leftarrow}$};
        \end{tikzpicture}\propto \dirac{\vec{k}_1}
\end{equation}
If such a diagram is connected to a $\mathcal{B}$-field of any cumulant within a given diagram, forming a tadpole subdiagram, the resulting contribution vanishes. This follows from \eqref{eq:cumulantproptoforce} together with the requirement $v(k, t)\vec{k}\Big\rvert_{\vec{k}=0}=0$\, and the fact, that all internal momenta have to be integrated over\footnote{Thus, the corresponding $\vec{l}$-vectors vanish.}. The corresponding expression in real space reads
\begin{equation}
    v(k, t)\vec{k}\Big\rvert_{\vec{k}=0}=-\mi\int\md^3\vec{q}\,\,\nabla_{\vec{q}}\,v(q,t)=0\,.
\end{equation}
The last equality can be shown by integrating over the angular variables first. Physically, this implies that a homogeneous background cannot exert a force on a particle. We can, therefore, summarize the above discussion in the following rule for homogeneous systems, 
\begin{equation}
    \begin{tikzpicture}[baseline={-0.6ex}, scale=0.55]
            \DeltaRBTree{0,0}{0,2}
            \filldraw[] (0,2) circle (2pt);
            \draw[propagator] (0,2.5) circle[radius=0.5];
            \draw[pattern={Lines[angle=45,distance={3pt}]},pattern color=black]
          (0,2.5) circle [radius=0.5cm];
          \draw[propagator] (0,-1) circle[radius=1];
            \draw[pattern={Lines[angle=45,distance={3pt}]},pattern color=black]
          (0,-1) circle [radius=1cm];
    \draw[propagator, opacity=0.3, reversed_arrow] 
        ({0+cos(-15)*1}, {-1+sin(-15)*1}) -- ++(-15:0.9);
    \draw[propagator, opacity=0.3, reversed_arrow] 
        ({0+cos(-60)*1}, {-1+sin(-60)*1}) -- ++(-60:0.9);
    \draw[dashed_propagator, opacity=0.3, -stealth] 
        ({0+cos(-120)*1}, {-1+sin(-120)*1}) -- ++(-120:1.2);
    \draw[dashed_propagator, opacity=0.3, -stealth] 
        ({0+cos(-165)*1}, {-1+sin(-165)*1}) -- ++(-165:1.2);   
    \draw[loosely dotted, thick] (0,-1) ++(-25:1.7) arc[start angle=-15, end angle=-60, radius=1];
    \draw[loosely dotted, thick] (0,-1) ++(-130:1.7) arc[start angle=-120, end angle=-165, radius=1];

        \end{tikzpicture} \,\,\,= 0\,,
\end{equation}
which justifies the absence of tadpole diagrams in this case.
The only exception to the above rule is when a tadpole diagram is not attached to any other $\mathcal{B}$-field. This can only be the case, if the outgoing leg belongs to an external propagator. Furthermore, the $\widetilde{\Delta}_{f\mathcal{B}}$ contribution vanishes by the above rule, such that only the external $\mathbb{1}$ from \eqref{eq:Delta_rB} survives. Such diagrams will contribute corrections to the one-point momentum distribution function. If we are only interested in information about the density $\rho$, the respective $\vec{l}$ is set to zero and the whole diagram vanishes identically, thus preserving homogeneity. \\

\subsection{Tree-Level Equations for a Homogeneous System}\label{eq:Treelevelhomsys}
For a homogeneous system, the tree-level equation in \eqref{eq:TrueIteration} can be solved following a general procedure. First, we can reduce the degrees of freedom in $\widetilde\Delta_{f\mathcal{B}}$ and $\Delta_{ff}$ by writing, analogously to \eqref{eq:momentumconservation}\footnote{In Fourier space, the identity $\mathbb{1}$ in \eqref{eq:Delta_rB} reads $\mathbb{1}(S_1, S_2)=(2\pi)^3\dirac{\vec{k}_1+\vec{k}_2}(2\pi)^3\dirac{\vec{l}_1+\vec{l}_2}\dirac{t_1-t_2}$},
\begin{align}\label{eq:G_rBwithoutDiracDelta}
    \Delta_{ff}(S_1, S_2)= &\, (2\pi)^3\dirac{\vec{k}_1+\vec{k}_2}\Delta_{ff}(\vec k_1, \vec{l}_1, \vec l_2, t_1, t_2)\,,\\[2ex]
    \widetilde\Delta_{f\mathcal{B}}(S_1, S_2)=&\,(2\pi)^3\dirac{\vec{k}_1+\vec{k}_2}(2\pi)^3\dirac{\vec l_2}\widetilde\Delta_{f\mathcal{B}}(\vec k_1, \vec{l}_1, t_1, t_2)\,,\label{eq:DeltafBFouriernoDelta}\\[2ex]
    \widetilde\Delta_{\rho\mathcal{B}}(\vec{k}_1, t_1, t_2)=&\,\widetilde\Delta_{f\mathcal{B}}(\vec k_1, \vec{l}_1=0, t_1, t_2)\,.
\end{align}
The system of equations that we have to solve for $\Delta_{f\mathcal{B}}$ can now be simplified by performing the integrations over $\vec k$ and $\vec l$ using the Dirac delta distributions. We find
\begin{align}
    \widetilde\Delta_{\rho\mathcal{B}}(\vec{k}_1, t_1, t_2)=&G_{\rho\mathcal{B}}^{(0)}(\vec{k}_1, t_1, t_2)+\int_{t_2}^{t_1}\md t\,G_{\rho\mathcal{B}}^{(0)}(\vec{k}_1, t_1, t)\widetilde\Delta_{\rho\mathcal{B}}(\vec{k}_1, t, t_2)\,,\label{eq:VolterraIntegralEquation}\\[2ex]
   \widetilde \Delta_{f\mathcal{B}}(\vec{k}_1, \vec l_1, t_1, t_2)=&G_{f\mathcal{B}}^{(0)}(\vec{k}_1, \vec l_1, t_1, t_2)+\int_{t_2}^{t_1}\md t\,G_{f\mathcal{B}}^{(0)}(\vec{k}_1, \vec l_1, t_1, t)\widetilde\Delta_{\rho\mathcal{B}}(\vec{k}_1, t, t_2)\,.\label{eq:Delta_rB_homogeneous}
\end{align}
Homogeneity has thus reduced the complexity to a one-dimensional integral equation. This is as far as we can get analytically, in general, for a homogeneous system. Equation \eqref{eq:VolterraIntegralEquation} is a Volterra integral equation of the second kind. The causal structure ensured by the Heaviside function enables a very efficient numerical solution, as -- upon time discretization -- we simply have to invert a lower triangular matrix. In certain rare cases, \eqref{eq:VolterraIntegralEquation} can even be solved analytically. The basic requirement for this to be possible is that the objects appearing in \eqref{eq:VolterraIntegralEquation} are time-translation invariant, \ie only depend on time differences. This requirement is typically fulfilled if the potential is time independent and the Hamilton function in \eqref{eq:Hamiltongeneral}, and thus the total energy, is conserved. For cosmology, we will discuss a special case where an analytical solution is possible even in the case of a time dependent potential. In such cases, the integral in \eqref{eq:VolterraIntegralEquation} becomes a Laplace convolution and may be diagonalized by means of a Laplace transformation. Defining $\tau=t_1-t_2$,  we find 
\begin{equation}\label{eq:LaplaceTrafo}
    \mathcal{L}\Big[\widetilde\Delta_{\rho\mathcal{B}}(\vec{k}_1, \tau)\Big](z)=\mathcal{L}\Big[G^{(0)}_{\rho\mathcal{B}}(\vec{k}_1, \tau)\Big](z)+\mathcal{L}\Big[G^{(0)}_{\rho\mathcal{B}}(\vec{k}_1, \tau)\Big](z)\,\mathcal{L}\Big[\widetilde\Delta_{\rho\mathcal{B}}(\vec{k}_1, \tau)\Big](z)\,,
\end{equation}
where $z$ is the Laplace conjugate variable to $\tau$ and $\mathcal{L}$ is the Laplace transformation operator. The above equation can be solved algebraically and we get the final result
\begin{equation}\label{eq:algebraicDeltaRB}
    \widetilde\Delta_{\rho\mathcal{B}}(\vec{k}_1, \tau)=\mathcal{L}^{-1}\left[\frac{\mathcal{L}\Big[G^{(0)}_{\rho\mathcal{B}}(\vec{k}_1, \tau)\Big](z)}{1-\mathcal{L}\Big[G^{(0)}_{\rho\mathcal{B}}(\vec{k}_1, \tau)\Big](z)}\right](\tau)\,.
\end{equation}

In order to obtain the statistical propagator $\Delta_{ff}$ we use the Fourier representation of \eqref{eq:Delta_ffRealSpace} and insert the respective equations for $\widetilde\Delta_{f\mathcal{B}}$ and $G^{(0)}_{ff}$ to find 
\begin{equation}\label{eq:TreeLevelStatProp}
\begin{split}
    \Delta_{ff}(\vec k_1, \vec{l}_1, \vec l_2, t_1, t_2)=&G^{(0)}_{ff}(\vec k_1, \vec{l}_1, \vec l_2, t_1, t_2)+\int_{\ini{t}}^{t_1}\md t\widetilde\Delta_{f\mathcal{B}}(\vec k_1, \vec{l}_1, t_1, t)\,G^{(0)}_{\rho f}(\vec k_1, \vec l_2, t, t_2)\\&+\int_{\ini{t}}^{t_2}\md t\,G^{(0)}_{f\rho }(\vec k_1, \vec l_1, t_1, t)\,\widetilde\Delta_{\mathcal{B}f}(\vec k_1, \vec{l}_2, t, t_2)\\&+\int_{\ini{t}}^{t_1}\md t\int_{\ini{t}}^{t_2}\md \bar t\,\widetilde\Delta_{f\mathcal{B}}(\vec k_1, \vec{l}_1, t_1, t)\,G^{(0)}_{\rho\rho}(\vec k_1, t, \bar t)\,\widetilde\Delta_{f\mathcal{B}}(\vec k_1, \vec{l}_1, \bar t, t_2)\,,
    \end{split}
\end{equation}
where we made use of the Heaviside functions to restrict the integrations.

\subsection{From Cumulants to Power Spectra}\label{sec:PowerSpectrum}
In the following, we will be interested in the density-fluctuation power spectrum $P_\delta(\vec{k}, t)$ which is generally defined as the Fourier transform of the density-fluctuation pair-correlation function. We, thus, need to establish the connection between the cumulants computed within our formalism and $P_\delta(\vec{k}, t)$.

Using \eqref{eq:G_ff} together with the definition of the density-fluctuation $\delta (\vec{q},t) = \frac{\rho(\vec{q},t)-\bar{\rho}}{\bar{\rho}}$, one can easily verify the relation
\begin{equation}\label{eq:HowToCorrelationFunction}
   \mean{\delta(\vec{q}_1,t)\delta(\vec{q}_2,t)}=\frac{G_{\rho\rho}(\vec q_1, \vec q_2, t)}{\bar{\rho}^2}= \frac{1}{\bar{\rho}^2}\int\md^3\vec{p}_1\,\md^3\vec{p}_2\,G_{ff}(x_1, x_2, t). 
\end{equation}
However, in our particle-based approach, $G_{ff}(x_1, x_2, t)$ does not only contain the true particles correlation, but also the one-particle shot-noise contribution which we discussed in Sec- \ref{sec:ExpectationValues}. In fact, the decomposition \eqref{eq:FreeTwoPoint} into a shot-noise contribution and an irreducible part can be generalized to
\begin{equation}\label{eq:Gff}
    G_{ff}(x_1, t_1, x_2, t_2)=\,\dirac{x_1-x_{\text{cl}}(t_1; x_2, t_2)}\,f_1(\vec{q_1}, \vec p_1, t_1)+g_2(\vec{q_1}, \vec p_1, t_1, \vec{q_2}, \vec p_2, t_2)\,,
\end{equation}
where $f_1$ and $g_2$ denote the full reduced densities. Inserting the above equation into \eqref{eq:HowToCorrelationFunction}, we find for a homogeneous system
\begin{equation}\label{eq:delta_delta}
    \mean{\delta(\vec{q}_1,t_1)\delta(\vec{q}_2,t_2)}=\frac{1}{\bar{\rho}\,T^3_{12}}\varphi\left(\frac{\vec{q}_1-\vec{q}_2}{T_{12}}\right) + \xi(|\vec{q}_1-\vec{q}_2|,t_1,t_2)\,,
\end{equation}
where we have defined the unequal-time pair-correlation function
\begin{equation}
    \xi(|\vec{q}_1-\vec{q}_2|,t_1,t_2) \coloneqq \int \md^3 \vec{p}_1 \md ^3\vec{p}_2\, \widetilde{g}_2(|\vec{q}_1-\vec{q}_2|,\vec{p}_1,\vec{p}_2,t_1,t_2)\,.
\end{equation}
With the usual definition of the  power spectrum $P_\delta(\vec{k}_1, t_1, t_2)$ as the Fourier transform of the pair-correlation function $\xi(|\vec{q}_1-\vec{q}_2|,t_1,t_2)$, we thus find
\begin{equation}
   \frac{G_{ff}(\vec{k}_1, \vec{l}_1, t_1, \vec{k}_2, \vec{l}_2, t_2)}{\bar{\rho}^2}\Bigg\rvert_{\vec l_1=\vec l_2=0} = (2\,\pi)^3\dirac{\vec{k}_1+\vec{k}_2}\left(\frac{1}{\bar{\rho}}\,\varphi(T_{12}\,\vec{k}_1)+P_\delta(\vec{k}_1, t_1, t_2)\right)\,,
\end{equation}
by inserting \eqref{eq:delta_delta} in \eqref{eq:HowToCorrelationFunction}. The above equation implies that we have to subtract the shot-noise contribution from the two-point density cumulant in order to obtain the power spectrum. This prescription is a consequence of the particle-based nature of our approach and would not appear in a fluid-based description. The shot-noise term simply reflects the identification of both external particles with each other. We, thus, find that the power spectrum is given by 
\begin{equation}\label{eq:PS_general}
    (2\,\pi)^3\dirac{\vec{k}_1+\vec{k}_2}\,P_\delta(\vec{k}_1, t_1, t_2) = \frac{G_{ff}(\vec{k}_1, \vec{l}_1, t_1, \vec{k}_2, \vec{l}_2, t_2)}{\bar{\rho}^2}\Bigg\rvert_{\vec l_1=\vec l_2=0} - (2\,\pi)^3\dirac{\vec{k}_1+\vec{k}_2}\,\frac{\varphi(T_{12}\,\vec{k}_1)}{\bar{\rho}}\,.
\end{equation}
With this general prescription we can now compute the power spectrum to arbitrary (loop) order and present the tree-level and one-loop corrected results for cosmic large-scale power spectra in the following sections.
Since the cumulants contain the full phase-space information, the momentum-density power spectrum in analogy to the density-fluctuation power spectrum from $G_{ff}(\vec{k}_1, \vec{l}_1, t_1, \vec{k}_2, \vec{l}_2, t_2)$. With the the momentum-density operator \eqref{eq:densitymomentum} in Fourier space,
\begin{equation}
    \vec{\Pi}(\vec{k},t)=\mi\vec{\nabla}_{\vec{l}}\,f(\vec{k},\vec{l},t)\Bigg\rvert_{\vec l=0}\,,
\end{equation}
and using the general relation in \eqref{eq:generalmomentummomentsk_space}, the two-point correlation function for the momentum-density can be obtained by
\begin{equation}
    \mean{\vec{\Pi}(\vec{k}_1,t_1)\cdot\vec{\Pi}(\vec{k}_2,t_2)} = \mi\vec{\nabla}_{\vec{l}_1}\cdot\mi\vec{\nabla}_{\vec{l}_2}\mean{f(\vec{k}_1,\vec{l}_1,t_1)f(\vec{k}_2,\vec{l}_2,t_2)}\Bigg\rvert_{\vec l_1=\vec l_2=0}\,.
\end{equation}
According to \eqref{eq:generalCumulant}, the two-point momentum-density cumulant is then given by the connected part of the above equation,
\begin{equation}
     G_{\vec{\Pi}\vec{\Pi}}(\vec{k}_1, t_1, \vec{k}_2, t_2) = \mi\vec{\nabla}_{\vec{l}_1}\cdot\mi\vec{\nabla}_{\vec{l}_2}G_{ff}(\vec{k}_1, \vec{l}_1, t_1, \vec{k}_2, \vec{l}_2, t_2)\Bigg\rvert_{\vec l_1=\vec l_2=0}\,.
\end{equation}
Inserting \eqref{eq:Gff} in the equation above and using the definition of the momentum-density power spectrum $P_{\vec\Pi}(\vec{k}_1,t_1,t_1)$ as the Fourier transform of the unequal-time momentum-density correlation function
we find the analogous expression to \eqref{eq:PS_general} for the momentum-density power spectrum
\begin{equation}
\begin{split}\label{eq:momentum_PS}
    (2\,\pi)^3\dirac{\vec{k}_1+\vec{k}_2}\,P_{\vec{\Pi}}(\vec{k}_1, t_1, t_2) =&\, \frac{\mi\vec{\nabla}_{\vec{l}_1}\cdot\mi\vec{\nabla}_{\vec{l}_2}G_{ff}(\vec{k}_1, \vec{l}_1, t_1, \vec{k}_2, \vec{l}_2, t_2)}{\bar{\rho}^2}\Bigg\rvert_{\vec l_1=\vec l_2=0}\\ 
    &- (2\,\pi)^3\dirac{\vec{k}_1+\vec{k}_2}\,\frac{\mi\vec{\nabla}_{\vec{l}_1}\cdot\mi\vec{\nabla}_{\vec{l}_2}\varphi(\vec{l}_1+\vec{l}_2+T_{12}\vec{k}_1)}{\bar{\rho}}\Bigg\rvert_{\vec l_1=\vec l_2=0}\,.
\end{split}
\end{equation}

\section{Newtonian dynamics of correlated particles on an expanding space-time}\label{sec:cosmoSystem}

In the following sections, we specialize to the case of cosmic structure formation, where we aim to model the dark matter distribution in the universe as a self-gravitating ensemble of $N$ classical particles evolving on an expanding background. We define the appropriate Hamilton function and interaction potential which describe the evolution of the individual particles. Once the Hamilton function is defined, the trajectories of the particles are fixed, and the behaviour of the particle ensemble -- whether in equilibrium or out-of-equilibrium -- is only determined by its initial state \cite{Mazenko_2010, Mazenko_2011, Das_2012, Cattaruzza:2010wc}. For self-gravitating systems, the out-of-equilibrium case is generally the more appropriate description \cite{Lynden_bell_1968, HERTEL_1971, Lynden_bell_1999}. Influenced by the gravitational potential, the initial state will evolve over time. We set up a suitable initial phase-space density, which shall contain all relevant correlations needed for the analysis of cosmic structure formation. 

We restrict ourselves to the main steps and key results of the derivations here and refer the reader to \cite{Bartelmann_2014} and Appendix \ref{sec:InitialCondition} for more technical details.

\subsection{Particle Trajectories on an Expanding Background}

We begin by describing the dynamics of a point-like test particle of mass $m$ which moves under the influence of gravitational interaction with a density field. The density field itself consists of $N$ individual particles of the same mass. In physical coordinates, the corresponding Lagrange function reads
\begin{equation}\label{eq:physicalLagrangian}
    \mathcal{L}(\vec{r}, \dot{\vec{r}}, t) = \frac{1}{2}m\dot{\vec{r}}^{\,\,2}-m\Phi(\vec{r}, t)\,,
\end{equation}
where the Newtonian gravitational potential $\Phi(\vec{r}, t)$ is sourced by the mass density $\rho_m(\vec{r}, t)$ through Poisson's equation, 
\begin{equation}
    \Delta_r\Phi(\vec{r}, t)=4\pi G\rho_m(\vec{r}, t) - \Lambda\,,
\end{equation}
with the gravitational constant $G$ and the cosmological constant $\Lambda$. In order to separate the background expansion from the intrinsic particle motion, we switch to comoving coordinates $\vec{r}(t)=a(t)\vec{q}(t)$, where the cosmological scale factor $a(t)$ derives from Friedmann's equation 
\begin{equation}\label{eq:Friedmann}
    \frac{\ddot{a}(t)}{a(t)}=-\frac{4\pi G}{3}\bar{\rho}_m(t)+\Lambda\,.
\end{equation}
Here, $\bar{\rho}_m(t)$ denotes the mean mass density, governing the background expansion. In comoving coordinates the Langrange function \eqref{eq:physicalLagrangian} takes on the form 
\begin{equation}\label{eq:comovingLagrange}
    \mathcal{L}(\vec{q}, \dot{\vec{q}}, t)=\frac{1}{2}ma^2\dot{\vec{q}}^{\,\,2}-m\varphi(\vec{q}, t)\,,
\end{equation}
where the gravitational potential $\varphi(\vec{q}, t)$ in physical coordinates is subject to the Poisson equation 
\begin{equation}\label{eq:PoissonComoving}
    \Delta_q\varphi(\vec{q}, t)=\frac{4\pi G}{a}\left(\rho_m(\vec{q},t)-\bar{\rho}_m\right)=\frac{4\pi G}{a}\bar{\rho}_m\,\delta(\vec{q}, t)\,.
\end{equation}
Thus, $\varphi(\vec{q}, t)$ is sourced by density fluctuations which corresponds exactly the Newtonian limit of general relativity. Since we model the density field by $N$ individual particles, the comoving mass density is given by
\begin{equation}
    \rho_m(\vec{q},t)=m\sum_{i=1}^N\dirac{\vec{q}-\vec{q}_i(t)}\,.
\end{equation}
We can, therefore, find the explicit solution for $\varphi(\vec{q}, t)$, given by 
\begin{align}\label{eq:potential}
    \varphi(\vec{q}, t)=&-\frac{mG}{a(t)}\sum_{i=1}^N\frac{1}{\lvert\vec{q}-\vec{q}_i(t)\rvert}=-\frac{3}{8\pi\bar{\rho}}\,a(t)^2\,H(t)^2\,\Omega_m(t)\,\sum_{i=1}^N\frac{1}{\lvert\vec{q}-\vec{q}_i(t)\rvert}\,.
\end{align}
For later convenience we have expressed the result in terms of the matter-density parameter $\Omega_m$ and the Hubble function $H:=\frac{\dot{a}}{a}$. Note, that the dependence on the particle mass has been replaced by a dependence on the inverse mean particle-number density. For numerical as well as analytical reasons, it is advantageous to use the logarithm of the scale factor as the time coordinate. Therefore, we perform the transformation
\begin{equation}
    \eta=\ln(a(t))\,,\quad \md\eta=H\md t\,,
\end{equation}
which, applied to the Lagrange function \eqref{eq:comovingLagrange}, yields
\begin{equation}
    \mathcal{L}\left(\vec{q}, \frac{\md \vec{q}}{\md\eta}, \eta\right) = \frac{1}{2}m\,\e^{2\eta}\,H(\eta)\,\left(\frac{\md \vec{q}}{\md\eta}\right)^2-m\widetilde\varphi(\vec{q}, \eta)\,, \quad\text{with}\quad \widetilde\varphi(\vec{q}, \eta)=\frac{1}{H(\eta)}\varphi(\vec{q}, \eta)\,.
\end{equation}

We can now construct the corresponding Hamilton function for $N$ mutually interacting particles of equal mass $m$ on this expanding background. We find 
\begin{equation}\label{eq:HamiltonfunctionNParticles}
    \mathcal{H}(\tens{q},\tens{p}, \eta )= \sum_{i=1}^N\frac{\vec{\ms{p}}_i^2}{2\,m\,f(\eta)}+\frac{1}{2}\sum_{i\neq j=1}^N\bar v(|\vec{\ms{q}}_i-\vec{\ms{q}}_j|, \eta)\,,\quad \text{with}\quad \vec{\ms{p}}_i(\eta):=m\,f(\eta)\frac{\md \vec{\ms{q}}}{\md\eta}\,,\quad\text{and}\quad f(\eta)=\e^{2\eta}\,H(\eta)\,,
\end{equation}
where $m\,f(\eta)$ can be interpreted as a time-dependent mass due to the expanding background and the factor of $\frac{1}{2}$ in \eqref{eq:HamiltonfunctionNParticles} ensures that each particle pair contributes only once to the total energy of the system. The pair potential $\bar v(|\vec{\ms{q}}_i-\vec{\ms{q}}_j|, \eta)$ derives from \eqref{eq:potential} and reads 
\begin{equation}
    \bar v(|\vec{\ms{q}}_i-\vec{\ms{q}}_j|, \eta)=-\frac{m}{\bar{\rho}}g(\eta)\,\frac{1}{|\vec{\ms{q}}_i-\vec{\ms{q}}_j|}\,,\quad\quad g(\eta):=\frac{3}{8\pi}f(\eta)\,\Omega_m(\eta)\,.
\end{equation}
As one can see from the definition of the canonical momentum in \eqref{eq:HamiltonfunctionNParticles}, $\vec{\ms{p}}_i$ scales proportional to the particles velocities with the time dependent mass. For our later cosmological application however, the velocities will play a central role, such that it is convenient to work with the velocities instead of the canocical momenta\footnote{Of course, this comes at the cost of not working with the canonical variables in the first place any more. However, this is no problem if one keeps the relation between the canonical conjugate momentum and the velocities in mind, when computing the free cumulants. }. We therefore define the velocity variable 
\begin{equation}
    \vec{\ms{u}}_i(\eta):= \frac{\vec{\ms{p}}_i(\eta)}{mf(\eta)}\,,\quad\text{such that}\quad \vec{\ms{u}}_i(\eta)\equiv\frac{\md \vec{\ms{q}}}{\md\eta}\,.
\end{equation}
The equations of motion for the pair $(\vec{\ms{q}}(\eta), \vec{\ms{u}}(\eta))$ derive from the canonical equations of motion as described by the Hamiltonian function \eqref{eq:HamiltonfunctionNParticles}\,. We therefore find 
\begin{align}
    \frac{\md\vec{\ms{q}}_i}{\md\eta} =\, &\vec{\ms{u}}_i(\eta)\,,\\
    \frac{\md\vec{\ms{u}}_i}{\md\eta} = \, & \frac{\md}{\md\eta}\Bigg(\frac{\vec{\ms{p}}_i(\eta)}{mf(\eta)}\Bigg)=-\sum_{\substack{j=1 \\ j \neq i}}^N {\nabla}_{\vec{\ms{q}}_i} v\left(|\vec{\ms{q}}_i - \vec{\ms{q}}_j|, \eta\right)-  \frac{\vec{\ms{u}}_i(\eta)}{f(\eta)}\frac{\md f(\eta)}{\md\eta}
\end{align}
where we redefined the pair potential to 
\begin{equation}\label{eq:PairPotential}
    v\left(|\vec{\ms{q}}_i - \vec{\ms{q}}_j|, \eta\right):=-\frac{3\,\Omega_m(\eta)}{8\pi\,\bar{\rho}}\cdot\frac{1}{|\vec{\ms{q}}_i-\vec{\ms{q}}_j|}\,.
\end{equation}
We thus see, that working with the particle velocities instead of the canonical momenta, simplifies the time dependence of the pair potential. This is due to the velocities representing the peculiar motion of the particles without the background expansion entering the momenta through the time dependent mass. In this description, the strength of the gravitational interaction is governed by the matter distribution $\Omega_m(\eta)$, which, in case of a purely matter dominated universe (the Einstein-de Sitter universe with $\Omega_m=1$), becomes time independent. The background expansion solely enters through the time variable $\eta$ mapping the system to a self-gravitating gas on a static background. However, if beside matter other components enter the cosmological system $\Omega_m(\eta)<1$ is not constant any more, weakening the interaction strength. This happens, for instance, in a $\Lambda$CDM cosmology where the cosmological constant suppresses gravitational clustering for late times.\\

In the absence of interactions we find an analytic solution for the free trajectories, which we can write as 
\begin{equation}\label{eq:FreeSolution}
\begin{pmatrix}
    \vec{\ms{q}}_i(\eta)\\[1ex]
    \vec{\ms{u}}_i(\eta)
\end{pmatrix}=G\cdot\begin{pmatrix}
    \vini{q}_i \\[1ex]
    \vini{u}_i
\end{pmatrix}\,,\quad\quad\text{with} \quad G=\begin{pmatrix}
    1 & \tau_q(\eta,\ini{\eta})\\[1ex]
    0 & \tau_u(\eta, \ini{\eta})
\end{pmatrix}
\end{equation}
with the initial position $\vini{q}_i$ and velocity $\vini{u}_i$ of the $i$-th particle. Here, we have introduced the functions $\tau_q(\eta_1, \eta_2)$ and $\tau_u(\eta_1, \eta_2)$ defined as
\begin{equation}\label{eq:Tau}
    \tau_q(\eta_1, \eta_2)=f(\eta_2)\,\int_{\eta_2}^{\eta_1}\frac{\md \eta^\prime}{f(\eta^\prime)}\,,\quad\quad  \tau_u(\eta_1, \eta_2)=\frac{f(\eta_2)}{f(\eta_1)}
\end{equation}

\subsection{Initial Conditions for Cosmic Structure Formation}

Having discussed the trajectories of the individual particles, we will now define a suitable initial phase-space density $\varrho_N(\tini{q}, \tini{u}, \ini{\eta})$ describing the probabilistic distribution of the initial particle positions and velocities for our statistical system in the cosmological context. As this derivation has already been presented in \cite{Bartelmann_2014}, we simply state the result here and focus on the relevant discussion for our application. For completeness, a more detailed derivation is provided in Appendix \ref{sec:InitialCondition}. \\

We start from a continuous mass density field in the early matter-dominated epoch. The central objects describing the state of a system are the mass density field $\ini{\rho}_m(\vec{q})$ and the peculiar velocity-density field $\ini{\vec{\Pi}}(\vec{q})$. The cosmological principle, homogeneity and isotropy, implies a constant mean background mass-density and a vanishing mean velocity of the system. We therefore describe the mass-density field by its fluctuation around the mean background,
\begin{equation}
    \ini{\rho}_m(\vec{q})= \ini{\bar{\rho}}_m(1+\ini{\delta}(\vec{q}\,))\,,\quad\text{with}\,\quad \mean{\ini{\delta}(\vec{q}\,)}=0\,.
\end{equation}
As predicted by inflationary models and confirmed to high accuracy by observations of the temperature-fluctuations in the CMB, we assume that the tuple $(\ini{\delta}(\vec{q}),\,\, \ini{\vec{\Pi}}(\vec{q}))$ follows a combined multivariate Gaussian distribution\footnote{Technically, this is due to both fields being related to the Laplacian and the gradient of the same primordial velocity potential which is, in turn, assumed to be a Gaussian random field (see Appendix \ref{sec:InitialCondition}).},
\begin{equation}\label{eq:densityProbability}
    \mathcal{P}(\ini{\delta},\ini{\vec{\Pi}})=\frac{1}{\sqrt{(2\pi)^4\det(C)}}\exp\left(-\frac{1}{2}\begin{pmatrix}
        \ini{\delta}(\vec{q}), &\ini{\vec{\Pi}}(\vec{q})
    \end{pmatrix}\cdot C^{-1}\cdot\begin{pmatrix}
        \ini{\delta}(\vec{q})\\
        \ini{\vec{\Pi}}(\vec{q})
    \end{pmatrix}\right)\,,
\end{equation}
where the covariance matrix is the single parameter characterizing the distribution. It reads 
\begin{equation}\label{eq:CorrelationMatrix}
    C = \begin{pmatrix}
        C_{\delta\delta} & \vec{C}_{\delta\vec{\Pi}}^\top\\
        \vec{C}_{\vec{\Pi}\delta}&C_{\vec{\Pi}\otimes\vec{\Pi}}
    \end{pmatrix}=\begin{pmatrix}
        \mean{\ini{\delta}(\vec{q}_1\,)\ini{\delta}(\vec{q}_2\,)} & \mean{\ini{\delta}(\vec{q}_1\,)\ini{\vec{\Pi}}(\vec{q}_2\,)}^\top\\
        \mean{\ini{\vec{\Pi}}(\vec{q}_1\,)\ini{\delta}(\vec{q}_2\,)}&\mean{\ini{\vec{\Pi}}(\vec{q}_1\,)\otimes\ini{\vec{\Pi}}(\vec{q}_2\,)}
    \end{pmatrix}\,.
\end{equation}
All of the above correlation functions are, due to homogeneity and isotropy, functions of the relative distance $|\vec{q}_1-\vec{q}_2|$ only. Furthermore, they can all be related to the initial power spectrum (see the discussion in Appendix \ref{sec:InitialCondition}), which is defined as the Fourier transform of $C_{\delta\delta}(r)$, 
\begin{equation}
    \ini{P}_{\delta}(k)=\int\md^3\vec{r}\,C_{\delta\delta}(r)\, \e^{-\mi\vec{k}\cdot\vec{r}}\,.
\end{equation}

We now have to map the probability distribution \eqref{eq:densityProbability} of the continuous fields $\ini{\delta}(\vec{q})$ and $\ini{\vec{\Pi}}(\vec{q})$ to the corresponding initial phase-space distribution of the canonical positions and velocities of particles that sample the density field. This is achieved by using conditional probabilities of finding the respective particle at a certain phase-space position, given a value of the density and velocity field at that point. The analysis is performed in detail in Appendix \ref{sec:InitialCondition} and we find
\begin{equation}\label{eq:initphasespacedensity}
    \varrho_N(\tini{q},\tini{u}, \ini{\eta})=V^{-N}\int \frac{\md \tens{t}_u}{(2\pi)^{3N}} \mathcal{C}(\tini{q},\tens{t}_u)\exp\left(-\frac{1}{2}\tens{t}_u\tens{\cdot}\tens{C}_{uu}\tens{\cdot}\tens{t}_u + \mi \tens{t}_u\tens{\cdot}\tini{u}\right)\,,
\end{equation}
where the function $\mathcal{C}(\tini{q},\tens{t}_u)$ contains all density-density and density-velocity correlations and reads
\begin{equation}\label{eq:correlationoperator}
\begin{split}
    \mathcal{C}(\tini{q},\tens{t}_u)=&\prod_{j=1}^N\left(1- \mi\vec{\ms{C}}_{\delta_ju_s}\cdot \vec{\ms{t}}_{u_s}\right)+\sum_{\substack{j,k=1 \\ j \neq k}}^N\ms{C}_{\delta_j\delta_k}\prod_{\substack{l=1 \\ l \neq j,k}}^N\left(1- \mi\vec{\ms{C}}_{\delta_lu_s}\cdot \vec{\ms{t}}_{u_s}\right)\\
    &+\sum_{\substack{j,k=1 \\ j \neq k}}^N\ms{C}_{\delta_j\delta_k}\sum_{\substack{a,b=1 \\ a \neq b\\a,b\neq j,k}}^N\ms{C}_{\delta_a\delta_b}\prod_{\substack{l=1 \\ l \neq j,k, a, b}}^N\left(1- \mi\vec{\ms{C}}_{\delta_lu_s}\cdot \vec{\ms{t}}_{u_s}\right)+\cdots\,.
    \end{split}
\end{equation}
The correspondence between the particle correlations $\ms{C}_{\delta_j\delta_k}\,,\vec{\ms{C}}_{\delta_ju_k}\,,\ms{C}_{u_ju_k}$ and the covariance matrix \eqref{eq:CorrelationMatrix} is established in Appendix \ref{sec:InitialCondition} and reads
\begin{equation}
\begin{split}
    \ms{C}_{\delta_j\delta_k}:=\ms{C}_{\delta\delta}(|\vini{q}_j-\vini{q}_k|)=C_{\delta\delta}(|\vini{q}_j-\vini{q}_k|)\,,\,\,\,&\,\,\,\vec{\ms{C}}_{\delta_j\vec u_k}:=\vec{\ms{C}}_{\delta \vec u}(|\vini{q}_j-\vini{q}_k|)=\vec{C}_{\delta\vec{\Pi}}(|\vini{q}_j-\vini{q}_k|)\\
    \tens{C}_{\vec u\vec u}:=\ms{C}_{\vec u_j\vec u_k}\otimes(e_j\otimes e_k)\,,\,\,\,\,\,\,\text{with}\,\,\,\ms{C}_{\vec u_j\vec u_k}:&=\ms{C}_{\vec u\vec u}(|\vini{q}_j-\vini{q}_k|)= C_{\vec{\Pi}\otimes\vec{\Pi}}(|\vec{\ms{q}}_j-\vec{\ms{q}}_k|)\,.
\end{split}
\end{equation}

As we have seen in Sec. \ref{sec:HomSys}, we require the $s$-particle reduced initial densities in order to evaluate the free cumulants \eqref{eq:timeorderedfB}. These can now be obtained from the initial phase-space density \eqref{eq:initphasespacedensity} as defined in \eqref{eq:reduced_f}. Since the system is homogeneous and isotropic, it is convenient to work directly in Fourier space, with the variables $(\vec k, \vec l)$ conjugate to $(\vec q, \vec u)$. In accordance with \eqref{eq:generalHomogeneousk_space} we thus find 
\begin{equation}\label{eq:initial_reduced_densities}
    \begin{split}
        \varphi(\vec{l}_1,\ini{\eta}) &=  \exp\left(-\frac{\sigma_p^2\,\vec{l}_1^{\,2}}{2}\right)\,,\\[1ex]
        \widetilde{g_2}(\vec{k}_1, \vec{l}_1, \vec{l}_2, \ini{\eta}) &= \mathcal{C}_2(\vec{k}_1,\vec{l}_1, \vec{l}_2 )\,\varphi(\vec{l}_1,\ini{\eta})\,\varphi(\vec{l}_2,\ini{\eta})\,,\\[2ex]
        \widetilde{g_3}(\vec{k}_1, \vec{k}_2, \vec{l}_1, \vec{l}_2, \vec{l}_3, \ini{\eta}) &= \mathcal{C}_3(\vec{k}_1, \vec{k}_2, \vec{l}_1, \vec{l}_2, \vec{l}_3 )\,\varphi(\vec{l}_1,\ini{\eta})\,\varphi(\vec{l}_2,\ini{\eta})\,\varphi(\vec{l}_3,\ini{\eta})\,,
    \end{split}
\end{equation}    
where we have defined the initial velocity dispersion
\begin{equation}
    \sigma_u^2\equiv \ms{C}_{uu}(0)=\frac{1}{3}\int\frac{\md^3\vec{k}}{(2\pi)^3}\frac{P_{\delta}^{(i)}(k)}{k^2}\,.
\end{equation}
and introduced the functions $\mathcal{C}_n$ describing all connected phase-space correlations between $n$ particles at initial time which are obtained from the general expression \eqref{eq:correlationoperator}. For $n=1$, this function reduces to $\mathcal{C}_1(\vec{l_1})=1$ and we simply end up with a Gaussian-shaped momentum-distribution function for the one-particle density.
For $n=2$ we get
\begin{equation}
\begin{split}
    \mathcal{C}_2(\vec{k}_1,\vec{l}_1, \vec{l}_2 )= \int \md^3 \vec{r}_{12}\, \e^{-\mi \vec{k}_1\cdot \vec{r}_{12}}\Big[\Big(1 + \ms{C}_{\delta\delta}({r}_{12})&-\mi\vec{\ms{C}}_{\delta u}({r}_{12})\cdot\left(\vec{l}_1-\vec{l}_2\right)\\
    &+\left(\vec{\ms{C}}_{\delta u}({r}_{12})\cdot\vec{l}_1\right)\left(\vec{\ms{C}}_{\delta u}({r}_{12})\cdot\vec{l}_2\right)\Big)\e^{-\vec{l}_1\cdot \ms{C}_{uu}({r}_{12})\cdot\vec{l}_2}-1\Big]\,,
\end{split}
\end{equation}
and the expression $\mathcal{C}_3(\vec{k}_1, \vec{k}_2,\vec{l}_1, \vec{l}_2, \vec{l}_3 )$ can be obtained analogously. 
The expansion of \eqref{eq:correlationoperator} -- or $\mathcal{C}_2$ and $\mathcal{C}_3$ respectively -- strongly resembles a Mayer-cluster expansion. In fact, it has been shown in \cite{Fabis_2018} that the initial distribution can be constructed diagrammatically by considering all possible density-density, density-momentum and momentum-momentum correlations between $N$-particles. In the absence of those correlations, the initial distribution reduces to an uncorrelated Boltzmann gas, putting the system in thermal equilibrium initially. Initial correlations, thus, lead to an out-of-equilibrium state.

\section{Two-Point Statistics for Cosmic Large-Scale Structures}\label{sec:Results}
As discussed in \ref{sec:HomSys} two-point statistics of the cosmic density and velocity field is described by the two-point phase-space density cumulant $G_{ff}(\vec{k}_1, \vec{l}_1, \vec{l}_2, t_1, t_2)$. We have shown for instance in \eqref{eq:PS_general} that the density-fluctuation power spectrum can be obtained by marginalizing over the momentum (or velocity) information in $G_{ff}(\vec{k}_1, \vec{l}_1, \vec{l}_2, t_1, t_2)$ , \ie setting $\vec{l}_1=\vec{l}_2=0$.

According to \eqref{eq:FullCumulants}, at tree-level, the two-point phase-space density cumulant $G_{ff}(\vec{k}_1, \vec{l}_1,\vec{l}_2, t_1, t_2)$ corresponds to the statistical propagator $\Delta_{ff}(\vec{k}_1, \vec{l}_1,\vec{l}_2, t_1, t_2 )$ given by \eqref{eq:TreeLevelStatProp}. 
We, thus, require the free cumulants $G^{(0)}_{ff}$ and $G^{(0)}_{f\mathcal{B}}$ in \eqref{eq:G_ffFourier} and \eqref{eq:G_fBFourier}. These are given by 
\begin{align}
\begin{split}
     G^{(0)}_{ff}(\vec{k}_1, \vec{l}_1, \vec{l}_2, t_1, t_2) =&\, \bar{\rho}\,\e^{-\frac{1}{2}\sigma^2_p\left(\widetilde T_{1}\vec{l}_1+\widetilde T_{2}\vec{l}_2+(T_1-T_2)\,\vec{k}_1\right)^2}\\
     &+ \bar{\rho}^2\,\mathcal{C}_2(\vec{k}_1, \widetilde T_{1}\vec{l}_1 + T_1\vec{k}_1, \widetilde T_{2}\vec{l}_2-T_2\vec{k}_1\,)\e^{-\frac{1}{2}\sigma^2_p\left(\left(\widetilde T_{1}\vec{l}_1+T_1\vec{k}_1\right)^2 + \left(\widetilde T_{2}\vec{l}_2-T_2\vec{k}_1\right)^2\right)}\,,\label{eq:G_ff_cosmo}\end{split}\\[2ex]
      G^{(0)}_{f\mathcal{B}}(\vec{k}_1, \vec{l}_1, t_1, t_2) =&\, -\bar{\rho}\, v(\vec{k}_1,\eta_2)\,\left(T_{12}\vec{k}_1^2\,+\widetilde T_{12}\vec{k}_1\cdot\vec{l}_1\right)\e^{-\frac{1}{2}\sigma^2_p\left(\widetilde T_{1}\vec{l}_1+(T_1-T_2)\,\vec{k}_1\right)^2 }\theta\left(\eta_1-\eta_2\right)\,,\label{eq:G_fB_cosmo}
\end{align}
once the the initial reduced phase-space densities provided in \eqref{eq:initial_reduced_densities} are inserted.
Above, we have introduced the short-hand notation 
\begin{align}
    T_{a}\coloneqq\tau_q(\eta_a, \ini{\eta})\,,\quad\quad\quad T_{ab}\coloneqq\tau_q(\eta_a, \eta_b)\,,\\[2ex]
    \widetilde T_{a}\coloneqq\tau_u(\eta_a, \ini{\eta})\,,\quad\quad\quad \widetilde T_{ab}\coloneqq\tau_u(\eta_a, \eta_b)\,,
\end{align}
in analogy to \eqref{eq:abbreviation} and used the Fourier transform of the interaction potential \eqref{eq:PairPotential}, 
\begin{equation}\label{eq:PairPotential_k}
    v(\vec{k},\eta) = -\frac{3\,\Omega_m(\eta)}{2\,\bar{\rho}\,\vec{k}^2}\,.
\end{equation}
The functions $\tau_q(\eta_1, \eta_2)$ and $\tau_p(\eta_1, \eta_2)$ are given by \eqref{eq:Tau} as the integral over the function $f(\eta)=\e^{2\eta}H(\eta)$ which contains the Hubble function. Therefore, cosmology enters through the Hubble function -- together with an appropriate set of cosmological parameters -- and an initial power spectrum which fixes the initial conditions. \\ 

Setting the interaction potential $v(\vec{k},\eta)\equiv0$, we recover the density-fluctuation power spectrum in the free theory from \eqref{eq:G_ff_cosmo} by setting $\vec{l}_1=\vec{l}_2=0$ according to \eqref{eq:PS_general}, 
\begin{equation}\label{eq:free_PS}
    P^{(0)}_\delta(\vec{k}_1,t_1,t_2) = \mathcal{C}_2(\vec{k}_1, T_1\vec{k}_1, -T_2\vec{k}_1\,)\e^{-\frac{1}{2}\sigma^2_p\vec{k}_1^{2}\left(\,T_1^2+ T_2^2\right)}\,.
\end{equation}
As expected, the shot-noise contribution cancels exactly and only the true two-particle correlation given by the second term in \eqref{eq:G_ff_cosmo} remains.\\

To obtain the tree-level expression \eqref{eq:TreeLevelStatProp} for the statistical propagator $\Delta_{ff}$, we first need to solve the integral equation for the causal propagator \eqref{eq:Delta_rB_homogeneous}. As discussed, it is sufficient to solve the integral equation for $\widetilde\Delta_{\rho\mathcal{B}}(\vec{k}_1, t_1, t_2)$,
\begin{equation}\label{eq:cosmo_volterra}
    \widetilde\Delta_{\rho\mathcal{B}}(\vec{k}_1, \eta_1, \eta_2) = G_{\rho\mathcal{B}}^{(0)}(\vec{k}_1, \eta_1, \eta_2)+\int_{\eta_2}^{\eta_1}\md \eta\,G_{\rho\mathcal{B}}^{(0)}(\vec{k}_1, \eta_1, \eta)\widetilde\Delta_{\rho\mathcal{B}}(\vec{k}_1, \eta, \eta_2)
\end{equation}
since the interaction potential does not depend on momenta. The cumulants $G_{\rho\mathcal{B}}^{(0)}$ and $G_{\rho f}^{(0)}$ required in the computation can be obtained from \eqref{eq:general_free_f_cumulants} by setting the appropriate conjugate vectors $\vec{l}$ to zero.

\subsection{Tree-level Results for an Einstein-de Sitter Cosmology in the Large-Scale Limit}\label{sec:tree-level_EdS}

In an Einstein-de Sitter (EdS) cosmology ($\Omega_m = 1$) our system simplifies considerably and we can perform most of the integrations analytically as has been shown in \cite{Lilow_2019}. We present the calculations for this special case in order to illustrate the structure of the theory in more detail. For instance, we find 
\begin{equation}
    H(\eta)=H_0\,\e^{-\frac{3}{2}\eta}\,,\quad\quad f(\eta)=H_0\,\e^{\frac{1}{2}\eta}\,,
\end{equation}
which implies,
\begin{equation}
    \tau_q(\eta_1, \eta_2)=2\left(1-\e^{-\frac{1}{2}\left(\eta_1-\eta_2\right)}\right)\,,\quad\tau_u(\eta_1, \eta_2)=\e^{-\frac{1}{2}\left(\eta_1-\eta_2\right)}\,,\quad\text{and}\quad v(k, \eta)=-\frac{3}{2\,\bar{\rho}\,k^2}\,,
\end{equation}
for the functions $\tau_q(\eta_1, \eta_2)$ and $\tau_p(\eta_1, \eta_2)$ defined in \eqref{eq:Tau} and the Fourier transform of the pair potential \eqref{eq:PairPotential_k}. Importantly, the pair potential is now independent of the time coordinate. Furthermore, $\tau_q(\eta_1, \eta_2)$ and $\tau_p(\eta_1, \eta_2)$ are time-translation invariant. This is clear, since the evolution of the universe is now perfectly described by a powerlaw in the scale factor $a$, which becomes translation invariant when described with $\eta=\log(a)$. The only remaining term spoiling the time translation invariance in $G_{\rho\mathcal{B}}^{(0)}$ is the exponential damping factor originating from the velocity dispersion. Assuming that we are only interested in scales for which $k^2\ll \frac{1}{\sigma_p^2}$, we can approximate this by $1$, thus neglecting the small scale damping. Furthermore, we linearize $\mathcal{C}_2$ in $G_{ff}^{(0)}$ in terms of the initial power spectrum as higher order effects will also only be relevant on smaller scales. In this \textit{large-scale limit} 
free cumulants in \eqref{eq:G_ff_cosmo} take on the simple form\footnote{In principle, we could also have neglected the exponential containing the $\vec{l}$ damping. However, this term does not enter the analytical integration, such that we do not need to approximate it.}
\begin{align}\label{eq:cumulants_EdS_LS}
\begin{split}
    G^{(0,\text{LS})}_{ff}(\vec{k}_1, \vec{l}_1, \vec{l}_2, t_1, t_2) = &\, \bar{\rho}\,\e^{-\frac{1}{2}\sigma^2_p\left(\widetilde T_{1}\vec{l}_1+\widetilde T_{2}\vec{l}_2\right)^2}\\
     &+\bar{\rho}^2\left(1+T_1+\widetilde T_1\frac{\vec k_1\cdot\vec l_1}{k_1^2}\right)\left(1+T_2-\widetilde T_2\frac{\vec k_1\cdot\vec l_2}{k_1^2}\right)P_{\delta}(k_1, \ini{\eta})\e^{-\frac{1}{2}\sigma^2_p\left(\widetilde T_{1}^2\vec{l}_1^{\,2} + \widetilde T_{2}^2\vec{l}_2^{\,2}\right)}\end{split}\\[2ex]
    G^{(0,\text{LS})}_{f\mathcal{B}}(\vec{k}_1, \vec{l}_1, t_1, t_2) &=\frac{3}{2}\left(T_{12}+\widetilde T_{12}\frac{\vec k_1\cdot\vec{l}_1}{k_1^2}\right)\e^{-\frac{1}{2}\sigma^2_p\widetilde T_{1}^2\vec{l}_1^{\,2}}\,,
\end{align}
where again, the first term in $G^{(0,\text{LS})}_{ff}$ represents the shot noise contribution and scales with one power less of $\bar{\rho}$ than the second term. Furthermore, $G^{(0,\text{LS})}_{f\mathcal{B}}$ is independent of $\bar{\rho}$ as the potential scales itself with $\bar{\rho}^{-1}$. Following the discussion in \ref{eq:Treelevelhomsys} we find the completely time-translation invariant expression for $G_{\rho\mathcal{B}}^{(0)}$,
\begin{equation}\label{eq:G_rB_EdS}
\begin{split}
    G^{(0,\text{LS})}_{\rho\mathcal{B}}(\vec{k}_1, \eta_1,\eta_2) &= 3\,\left(1 - \e^{-\frac{1}{2}\left(\eta_1-\eta_2\right)}\right)\,\theta(\eta_1-\eta_2)\,,
\end{split}
\end{equation}
and compute its Laplace transform 
\begin{equation}
    \mathcal{L}\left[G^{(0,\text{LS})}_{\rho\mathcal{B}}(\vec{k}_1, \bar{\eta})\right](z) = 3\left(\frac{1}{z}-\frac{1}{\frac{1}{2}+z}\right)\,,
\end{equation}
where $z$ is the Laplace conjugate variable to $\bar\eta = \eta_1-\eta_2$. The causal propagator $\Delta_{\rho\mathcal{B}}$ can now be solved algebraically as described in equation \eqref{eq:algebraicDeltaRB}. It is given by the inverse Laplace transform
\begin{equation}\label{eq:LaplaceVolterra}
    \begin{split}
    \widetilde{\Delta}^{\text{LS}}_{\rho\mathcal{B}}(\vec{k}, \bar{\eta}) &= \mathcal{L}^{-1}\left[\frac{\mathcal{L}\left[G_{\rho\mathcal{B}}^{(0,\text{LS})}(\vec{k}, \bar{\eta})\right](s)}{1-\mathcal{L}\left[G_{\rho\mathcal{B}}^{(0,\text{LS})}(\vec{k}, \bar{\eta})\right](s)}\right](\bar{\eta}) = \mathcal{L}^{-1}\left[\frac{3}{2z^2+z-3}\right](\bar{\eta})\\ 
    &= \frac{3}{5}\left(\e^{\bar{\eta}}-\e^{-\frac{3}{2}\bar{\eta}}\right)\theta(\bar{\eta})\,,
    \end{split}
\end{equation}
It is worth noting that $G^{(0,\text{LS})}_{\rho\mathcal{B}}$ and consequently $\widetilde{\Delta}^{\text{LS}}_{\rho\mathcal{B}}$ are independent of $\vec{k}$. The scale-dependence of $G^{(0,\text{LS})}_{\rho\mathcal{B}}$ is cancelled by the gravitational potential which is proportional to $\frac{1}{k^2}$. Thus, the causal propagator $\widetilde{\Delta}^{\text{LS}}_{\rho\mathcal{B}}$ is merely a time-dependent scaling factor which only increases initial structures monotonically over time.\\

The tree-level two-point cumulant is then given by 
\begin{align}\label{eq:G_ff_tree_EdS_LS}
    G^{(\text{tree, LS})}_{ff}(\vec{k}_1,\vec{l}_1,\vec{l}_2, \eta_1,\eta_2) \equiv &\Delta^{(\text{LS})}_{ff}(\vec{k}_1,\vec{l}_1,\vec{l}_2, \eta_1,\eta_2)\\[2ex]
    \begin{split}
    =&\, \bar{\rho}\,F(\vec{k}_1, \vec{l}_1, \eta_1)F(\vec{k}_1, \vec{l}_2, \eta_2)\e^{-\frac{1}{2}\sigma^2_p\left(\widetilde T_{1}^2\vec{l}_1^{\,2} + \widetilde T_{2}^2\vec{l}_2^{\,2}\right)} \\[2ex]
    &+\bar{\rho}^2\e^{\eta_1+\eta_2-2\ini{\eta}}\left(1+\frac{\vec{k}_1\cdot\vec{l}_1}{k_1^2}\right)\left(1-\frac{\vec{k}_1\cdot\vec{l}_2}{k_1^2}\right)P_\delta({k}_1, \ini{\eta})\e^{-\frac{1}{2}\sigma^2_p\left(\widetilde T_{1}^2\vec{l}_1^{\,2} + \widetilde T_{2}^2\vec{l}_2^{\,2}\right)}
    \end{split}
\end{align}
where $F(\vec{k}, \vec{l}, \eta)$ is defined as 
\begin{equation}
    F(\vec{k}_1, \vec{l}_1, \eta_1)=\frac{3}{5}\left(\left(1+\frac{\vec{k}_1\cdot\vec{l}_1}{k_1^2}\right)\e^{\eta_1-\ini{\eta}}+\left(\frac{2}{3}-\frac{\vec{k}_1\cdot\vec{l}_1}{k_1^2}\right)\e^{-\frac{3}{2}(\eta_1-\ini{\eta})}\right)
\end{equation}
which yields the density-fluctuation power spectrum at tree-level upon setting $\vec{l}_1=\vec{l}_2=0$. Note, that the first term which scales with $\bar{\rho}$ in the above expression originates from the shot noise contribution from the free cumulant $G^{(0,\text{LS})}_{ff}$. However, by \eqref{eq:PS_general} we only subtract the free shot-noise contribution as the one-point phase-space function does not get tree-level corrections from the homogeneous potential. This correction will therefore only cancel the shot noise contribution coming from the free $G_{ff}^{(0)}$ cumulant in the expansion \ref{eq:Treelevelhomsys}. The remaining terms do in fact contribute to the true two-particle correlation and thus to the density fluctuation power spectrum. They represent correlations between the two external particles which are only connected by an interaction and not by an initial correlation. Thus, inside a given diagram, loop or tree-level, some of the shot noise terms will indeed contribute to the correlation function by identifying different particles in the diagram. In our scenario, however, we sample a density field with an arbitrarily high density of particles\footnote{Keeping the mass density constant, of course, see the discussion earlier.}. We therefore send $\bar{\rho}\rightarrow\infty$. Since we have to divide by $\bar{\rho}^2$ in \eqref{eq:PS_general} we only keep terms that come with the highest $\bar{\rho}$ contributions. In that limit, which we call the \textit{continuum limit}, we find 
\begin{equation}\label{eq:Delta_rr_EdS_LS}
    P_\delta^{(\text{tree, LS})}(\vec{k}_1, \eta_1,\eta_2) = \e^{\eta_1+\eta_2- 2\ini{\eta}}\ini{P}_\delta(k)
\end{equation}
For equal times, \ie $\eta = \eta_1 = \eta_2$, we can bring \eqref{eq:Delta_rr_EdS_LS} into the more familiar form,
\begin{equation}
    P_\delta^{(\mathrm{tree},\mathrm{LS})}(\vec{k}, \eta) = \frac{D^2_+(\eta)}{D^2_+(\ini{\eta})}\ini{P}_\delta(k)
\end{equation}
by using that for an EdS cosmology the linear growth factor known from Eulerian SPT is given by $D_+(\eta)=\e^{\eta}$.\\

Having the full phase-space information at our disposal in \eqref{eq:G_ff_tree_EdS_LS}, we can now just as easily obtain velocity power spectra by taking appropriate derivatives with respect to the conjugate velocity variables $\vec{l}$, as described in \eqref{eq:generalmomentummomentsk_space}. For instance, using \eqref{eq:momentum_PS} we directly find the expression for the velocity-density power spectrum,
\begin{equation}
    P^{(\mathrm{tree},\mathrm{LS})}_{\vec{\Pi}}(\vec{k}_1,\eta_1,\eta_2) =  \frac{\mi \nabla_{\vec{l}_1}\cdot \mi \nabla_{\vec{l}_2}\, G^{(\text{tree, LS})}_{ff}(\vec{k}_1,\vec{l}_1,\vec{l}_2, \eta_1,\eta_2)}{\bar{\rho}}\Big\vert_{\vec{l}_1=\vec{l}_2=0} = \e^{\eta_1+\eta_2- 2\ini{\eta}}\frac{\ini{P}_\delta(k_1)}{\vec{k}^2_1}\,.
\end{equation}
Expressing the above relation in terms of the growth factor for an EdS cosmology, we find the familiar result which corresponds to the solution found in Eulerian SPT for equal times,
\begin{equation}
    P^{(\mathrm{tree},\mathrm{LS})}_{\vec{\Pi}}(\vec{k}_1,\eta) =  \frac{D^2_+(\eta)}{D^2_+(\ini{\eta})}\frac{\ini{P}_\delta(k)}{\vec{k}^2_1}\,.
\end{equation}

However, it should be noted that these results are only obtained in the large-scale limit upon linearizing the free cumulants. Without this simplifying assumption, $G^{(0,\text{LS})}_{\rho\mathcal{B}}$ and therefore $\widetilde{\Delta}^{\text{LS}}_{\rho\mathcal{B}}$ are scale-dependent functions and the evolution of structure is a highly non-linear process, already at tree-level. In this case, a fully analytical solution is no longer possible and the integral equations for the causal and statistical propagators have to be solved numerically as described in the following section.

\subsection{Numerical Results for the Tree-Level Density Power Spectrum}
In general, the Volterra integral equation \eqref{eq:cosmo_volterra} has to be solved numerically upon time-discretisation. The causal structure of $G^{(0)}_{\rho B}$ reduces the matrix equation to a lower triangular matrix, which can be inverted by forward substitution. After numerically solving the time-integrals in \eqref{eq:TreeLevelStatProp}, we find the full tree-level result for the statistical propagator $\Delta_{ff}(\vec{k}_1,\vec{l}_1,\vec{l}_2, \eta_1,\eta_2)$ and thus $G^{(\text{tree})}_{ff}(\vec{k}_1,\vec{l}_1,\vec{l}_2, \eta_1,\eta_2)$ from which we obtain the density-fluctuation power spectrum $P^{(\text{tree})}_\delta(\vec{k}_1,\eta_1,\eta_2)$ according to \eqref{eq:PS_general}. 

\begin{figure}
    \centering
    \begin{subfigure}{0.45\linewidth}
        \centering
        \includegraphics[width=\linewidth]{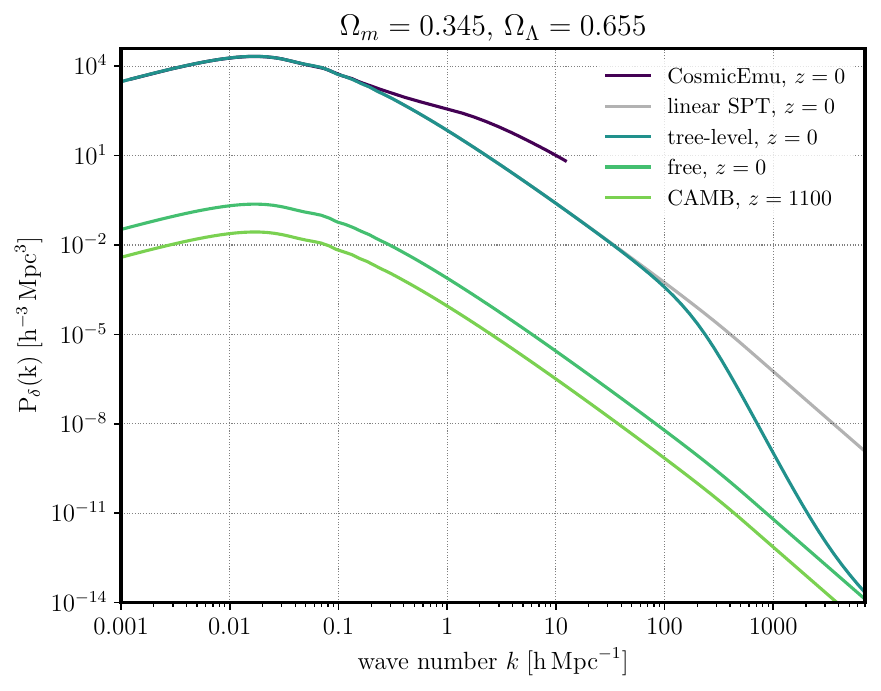}
    \end{subfigure}
    \begin{subfigure}{0.45\linewidth}
        \centering
        \includegraphics[width=\linewidth]{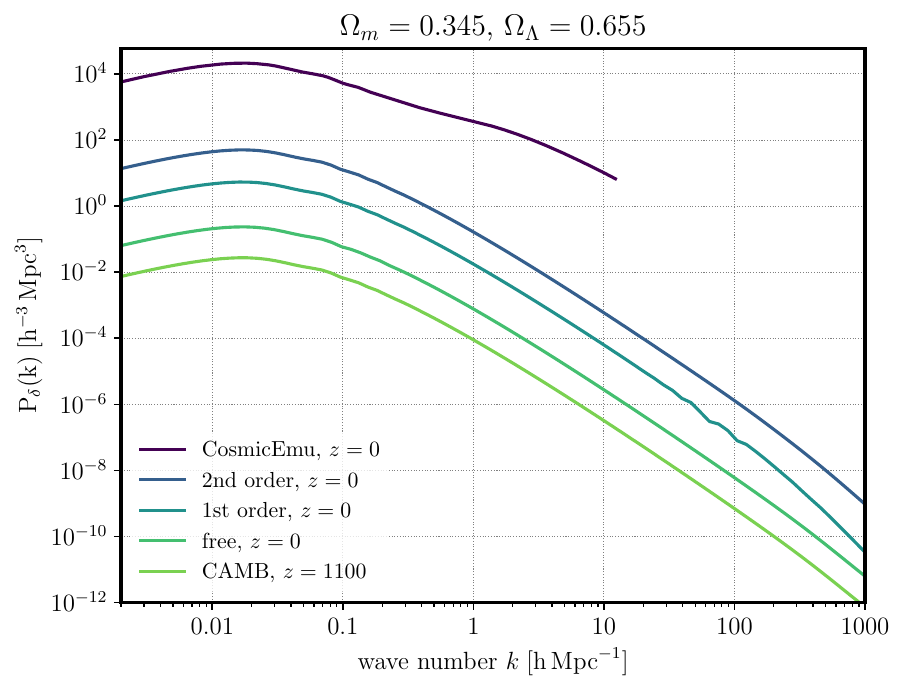}
    \end{subfigure}
    \caption{(\textit{Left panel:}) We show the tree-level result for the density-fluctuation power spectra $P_\delta(k)$ for a standard $\Lambda$CDM cosmology. The initial power spectrum is taken from CAMB \cite{camb} at $z=1100$ and is evolved to $z=0$ according to the tree-level theory. We also show the free evolution given by \eqref{eq:free_PS}. 
    On very small scales $k\,>\,100\,\mathrm{h}\,\mathrm{Mpc^{-1}}$ structures are washed out due to the initial velocity dispersion of particles so that the tree-level power spectrum approaches the free-streaming result. (\textit{Right panel:}) We show $P_\delta(k)$ for the canonical PT for comparison.
    We show the non-linear power spectrum as predicted by numerical simulation (Cosmic Emulator \cite{CosmicEmu}) for reference.
    }
    \label{fig:tree-level}
\end{figure}

In Fig. \ref{fig:tree-level} we show the numerical tree-level result for an equal-time density-fluctuation power spectrum $P^{(\text{tree})}_\delta(\vec{k},\eta)$ for a standard $\Lambda$CDM cosmology, evolved from the time of CMB decoupling at redshift $z=1100$ to the present at $z=0$ which was also found in \cite{Lilow_2019}. We compare our result to the free evolution given by \eqref{eq:free_PS} and the power spectrum obtained from numerical simulations. 
For reference, we also include the linear power spectrum given by 
\begin{equation}
    P^{(\text{lin})}_\delta(\vec{k},\eta) = \frac{D^2_+(\eta)}{D^2_+(\ini{\eta})}\ini{P}_\delta(k)\,,
\end{equation}
where $D_+(\eta)$ is the linear growth factor for a $\Lambda$CDM cosmology.
For wave numbers $k \leq 100\,\mathrm{h}\,\mathrm{Mpc^{-1}}$, the tree-level result exactly follows the linear power spectrum $P^{(\text{lin})}_\delta$. On these scales $\sigma_p \sim 10^{-5}$ and we are well within the regime which is well described by the large-scale limit $\vec{k}^2 \ll \frac{1}{\sigma^2_p}$. In this regime, the effects of the exponential damping in \eqref{eq:G_ff_cosmo} are negligible and $G^{(0)}_{\rho\mathcal{B}}$ as well as the causal propagator $\Delta^{(\text{tree})}_{\rho\mathcal{B}}$ are independent of $k$ and only a time dependent scaling function. On these scales, the amplitudes of initial density-fluctuation simply grow as a function of time only. In general, however, the integral equation which is solved in order to find the tree-level evolution, incorporates non-linear effects in $k$, leading to deviations from the linear growth on smaller scales. There, for $k \geq 100\,\mathrm{h}\,\mathrm{Mpc^{-1}}$, the large-scale limit can no longer be applied, and the tree-level power spectrum is dominated by the exponential damping in \eqref{eq:G_ff_cosmo}, which prevents structures from growing. As $k\rightarrow \infty$, the tree-level power spectrum approaches the free evolution described by \eqref{eq:free_PS}, since
\begin{equation}
    \Delta_{\rho\rho}(\vec{k}_1, \eta_1, \eta_2) \rightarrow G^{(0)}_{\rho\rho}(\vec{k}_1, \eta_1, \eta_2) \quad \text{for} \quad k\rightarrow \infty\,, 
\end{equation}
as can be seen from \eqref{eq:TreeLevelStatProp}.
\\

In Figure \ref{fig:tree-level} we also present results from the canonical microscopic PT\footnote{See the end of Sec.\ref{sec:ExpectationValues} and \cite{Daus2024, Heisenberg:2022uhb, Pixius:2022hqs} for technical details and cosmological application.}. As discussed in Sec.\ref{sec:propagators} the tree-level theory resums all contributions that lead to deviations of the free one-point density $G_f^{(0)}$ due to a single interaction. Iterating the integral equation \eqref{eq:Delta_rB} once, yields the leading order contribution to the first order microscopic perturbation theory. Thus, solving the integral equation exactly resums an infinite subseries of the microscopic perturbation theory. \\

While the tree-level result correctly reproduces the shape and amplitude of the power spectrum from numerical $N$-body simulation on large scales, its amplitude falls below the $N$-body results on scales $k \geq 0.1\,\mathrm{h}\,\mathrm{Mpc^{-1}}$. The tree-level iteration \eqref{eq:cosmo_volterra}, which accounts for all contributions deflecting the one-particle phase-space distribution via a single interaction, fails to capture sufficient gravitational interactions to describe the growth of structures on smaller scales. While, as previously discussed, the tree-level equation for gravity scales the initial structure in a $k$-independent way, this behaviour changes for higher-order, more complicated diagrams that appear in the loop corrections, which we will discuss in the following section.

\subsection{Numerical Results for the One-Loop Density Power Spectrum}

In order to compute the full one-loop correction to the two-point phase space density cumulant, one has to evaluate all diagrams that appear in the self-energy \eqref{eq:selfenergy} and enter \eqref{eq:1LoopPropagator}. This is tedious work, as on top of the thirteen diagrams that have to be evaluated, each $\Delta_{f\mathcal{B}}$ propagator appearing inside the loops or in the outer legs, consists of two terms according to \eqref{eq:Delta_rB} that have to be integrated separately. Each self-energy consists of one or two propagators, that are attached to higher-order cumulants, in a similar manner as in the tree-level computation of $\Delta_{ff}$. We list all cumulants necessary for the integration in Appendix \ref{app:freeCumulants}. Consider for instance the full loop contribution with external propagators associated to the self-energy from \eqref{eq:diagram1},
\begin{equation}
    \begin{tikzpicture}[baseline={(0,-0.6ex)}, scale=0.55]
        \DeltaRBTree{-3,0}{-1,0}
        \ThreePointVertices{-1,0}{1,0}
        \DeltaRRTreeDown
        \DeltaRRTreeUp
        \DeltaRBTree{3,0}{1,0}
    \end{tikzpicture}
    = (2\,\pi)^3\dirac{\vec{k}_1+\vec{k}_2}\int_{\ini{\eta}}^\infty \md \bar{\eta}_1 \md \eta'_1 
    \Delta_{\rho\mathcal{B}}(\vec{k}_1, \eta_1, \bar{\eta}_1)\, 
    \Sigma_{\rho\rho}^{(\text{1-loop,1})}(\vec{k}_1,\bar{\eta}_1, \eta'_1)\, 
    \Delta_{\mathcal{B}\rho}(-\vec{k}_1, \eta'_1, \eta_2)
\end{equation}
where we have defined the one-loop self-energy contribution to the first diagram as
\begin{equation}
    \begin{split}
        \Sigma_{\rho\rho}^{(\text{1-loop,1})}(\vec{k},\bar{\eta}_1, \eta'_1) = \frac{1}{2}\int \md \bar{\eta}_2 \md \bar{\eta}_3 \md \eta'_2 \md \eta'_3 \int \frac{\md^3 \vec{k}'}{(2\,\pi)^3}\, &G^{(0)}_{\rho \mathcal{B}\mathcal{B}}(\vec{k}, -\vec{k}', \bar{\eta}_1, \bar{\eta}_2, \bar{\eta}_3)\,\Delta_{\rho\rho}(\vec{k}',\bar{\eta}_2,\eta'_2)\\&\times \Delta_{\rho\rho}(\vec{k}-\vec{k}',\eta_3,\eta'_3)\,G^{(0)}_{\rho \mathcal{B}\mathcal{B}}(-\vec{k}, \vec{k}', \eta'_1, \eta'_2, \eta'_3)\;.
    \end{split}
\end{equation}
In analogy to quantum field theory, all loop diagrams represent convolutions of the propagators, which, in the above case, are of $\Delta_{\rho\rho}$-type and carry initial two-point correlation. Thus, we see, that the above diagram convolves structure on small scales with structure on larger scales, and hence, introduce information beyond the linear scaling of tree-level\footnote{As we have seen, apart from the exponential damping due to the velocity dispersion, the tree-level theory only evolves the initial correlation linearly.}. Together with the external propagators, the remaining integrals contain 6 time integrations, one angular and one radial $k$-integration that can numerically be solved without further complications.\\

\begin{figure}
    \centering
    \begin{subfigure}{0.45\linewidth}
        \centering
        \includegraphics[width=\linewidth]{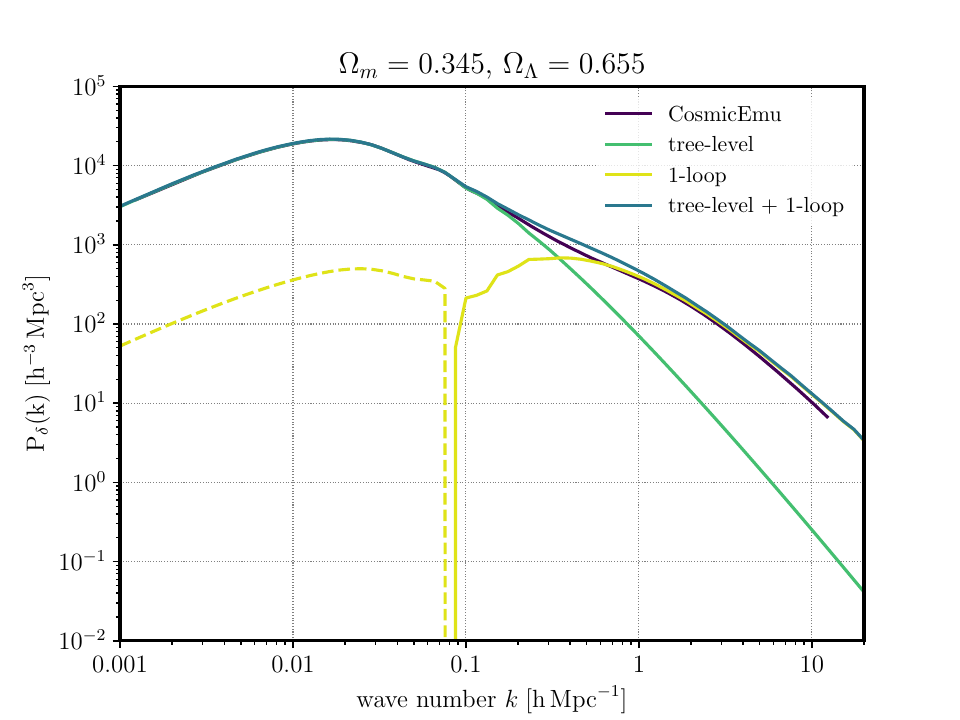}
    \end{subfigure}
    \begin{subfigure}{0.45\linewidth}
        \centering
        \includegraphics[width=\linewidth]{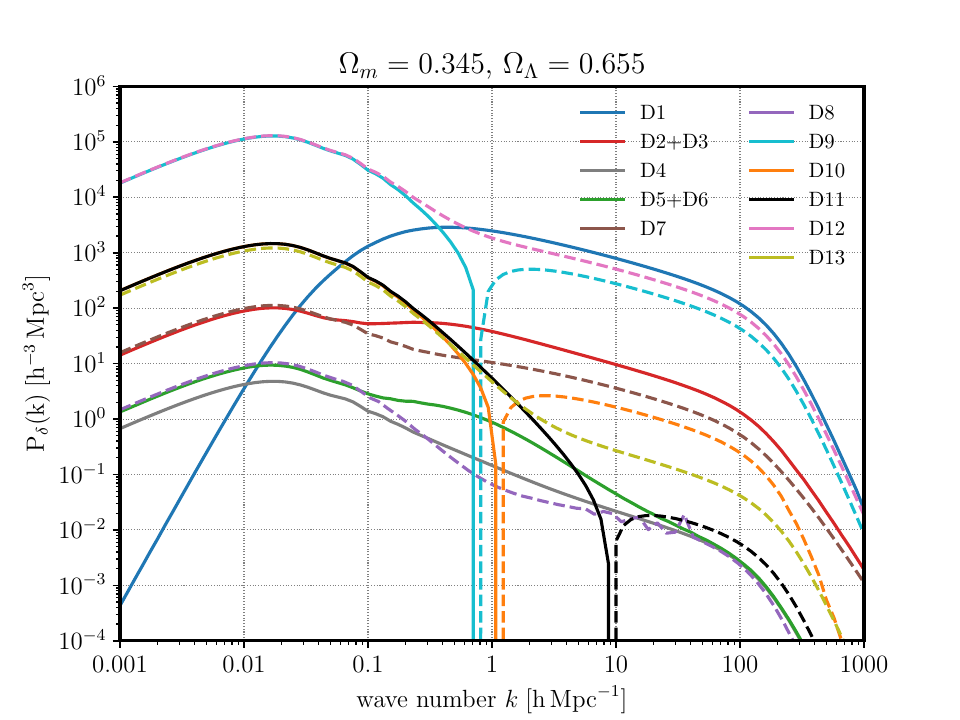}
    \end{subfigure}
     \caption{(\textit{Left panel:}) We show the tree-level and one-loop contributions for the density-fluctuation power spectrum $P_\delta(k)$ for a standard $\Lambda$CDM cosmology at $z=0$. The initial power spectrum is taken from CAMB \cite{camb} at $z=1100$. We compare our full result (tree-level $+$ one-loop) to the non-linear power spectrum obtained in simulations generated with Cosmic Emulator\cite{CosmicEmu}. (\textit{Right panel:}) We show all thirteen individual one-loop contributions, labelled by $\mathrm{D}1$ to $\mathrm{D}13$ in the same order as they appear in \eqref{eq:1LoopSelfenergyff} and \eqref{eq:1LoopSelfenergyfB}. Dashed lines indicate negative values.}
     \label{fig:allLoops}
\end{figure}

In Figure \ref{fig:allLoops} we present the result of all individual loop diagrams and their total contribution to the density power spectrum. 
We can identify the three leading loop contributions that are given by the three diagrams,
\begin{align}\label{diag:Drr1}
   \begin{tikzpicture}[baseline={-0.6ex}, scale=0.55]
            \draw (-2.5,0.5) node {$\mathrm{D}1$};
            \DeltaRBTree{-3,0}{-1,0}
            \ThreePointVertices{-1,0}{1,0}
            \DeltaRRTreeDown
            \DeltaRRTreeUp
            \DeltaRBTree{3,0}{1,0}
        \end{tikzpicture}\,,\; 
        \begin{tikzpicture}[baseline={-0.6ex}, scale=0.55]
            \draw (-2.5,0.5) node {$\mathrm{D}9$};
            \DeltaRBTree{-3,0}{-1,0}
            \ThreePointVertices{-1,0}{1,0}
            \DeltaRBTreeDownL
            \DeltaRRTreeUp
            \DeltaRRTree{3,0}{1,0}
        \end{tikzpicture}\,,\; 
        \begin{tikzpicture}[baseline={-0.6ex}, scale=0.55]
            \draw (-1.6,0.5) node {$\mathrm{D}12$};
            \DeltaRBTree{-2,0}{0,0}
            \FourPointVertex{0,0}
            \DeltaRRTreeRound
            \DeltaRRTree{2,0}{0,0}
        \end{tikzpicture}\,.
\end{align}
Those diagrams appear to only include vertices containing one $\rho$- and several $\mathcal{B}$-fields, \ie $G_{\rho\mathcal{B}\mathcal{B}}^{(0)}$ and $G_{\rho\mathcal{B}\mathcal{B}\mathcal{B}}^{(0)}$. Thus, the internal structure of the diagram only affects the free one-point phase space density by several interactions. As can be seen, this process dominates over processes in which higher phase-space densities are deformed by interaction. For instance the eighth diagram in \eqref{eq:selfenergy}, corresponding to the deflection of the free three-point phase space density cumulant by a single interaction, represents one of the weakest contributions. On the other hand, diagram D1 contains the deflection of two one-point phase space densities by two interactions and is by far the most dominant diagram that leads to structure growth on smaller scales. \\

We observe that some diagrams lead to significant corrections on larger scales. This seems counter intuitive as we already have seen, that structure growth on larger scales is fully described by the tree-level theory. However, those corrections involve finely tuned cancellations with other diagrams such that the overall contribution to the large scale amplitude is several magnitudes below the tree-level contribution. Although, the loop contributions in Fig. \ref{fig:allLoops} appear in groups which are separated by orders of magnitude, it is in general not possible to simply neglect those contributions with lower amplitude due to the above cancellations. \\

The sum of all 1-loop corrections leads to the shape of the non-linear power spectrum which we compare against numerical simulations in Fig. \ref{fig:allLoops}. While the one-loop corrections qualitatively reproduce the CosmicEmu power spectrum, the deviations are of the order of $10\%-20\%$ on scales $k>0.2\,\mathrm{h}\,\mathrm{Mpc^{-1}}$. The usual approach would be, of course, to include higher-order loop corrections in the hope of achieving improvement on these scales. However, a loop-expansion may not be suitable in our case for two reasons: (a) Typically, higher-order loop corrections will yield significant contributions on increasingly smaller scales. The deviation from numerical results on intermediate scales around $k \sim 0.5\,\mathrm{h}\,\mathrm{Mpc^{-1}}$ will likely not improve significantly due to higher loop corrections. (b) The strong cancellations between different one-loop contributions indicate that our perturbative loop-expansion is not suitable to capture relevant physical processes. We should also bear in mind that perturbative expansions will typically break down and start to diverge after a few orders in PT. This problem is well-known in the context of QFT and has also been demonstrated for Eulerian SPT. This indicates that certain non-perturbative effects have to be considered, that cannot be captured by perturbative approximations. For these reasons non-perturbative schemes such as 2PI, DSE and FRG as discussed in Sec.\ref{sec:selfenergy} have to be explored. The goal hereby is to restructure perturbation theory in such a way that relevant effects may be resummed similarly to how our macroscopic PT fully resums microscopic effects which lead to structure growth on large scales.

\section{Conclusion and Outlook}
In this paper, we have presented a field theory approach which allows us to describe cosmic structure formation based on Newtonian particle dynamics. 
The main advantage of this formulation is that it constitutes a resummation scheme for the well-known BBGKY-hierarchy. Therefore, the complete microscopic information of the system is preserved and can be included order by order in the perturbative expansion. The perturbative loop-expansion implicitly introduces a natural truncation criterion for the underlying BBGKY-hierarchy with each loop order. This sets our approach apart from Eulerian SPT which introduces a (fixed) truncation of the BBGKY-hierarchy leading to the Vlasov equation in order to derive the set of hydrodynamical equations on which it is based. This truncation leads to a permanent loss of microscopic degrees of freedom which is absent in our approach.\\

Our main result in this paper is the cosmic density-fluctuation power spectrum at tree and one-loop level of our field theory. We compare our result to the numerical result obtained with Cosmic Emulator \cite{CosmicEmu} for the same set of cosmological parameters and initial conditions. On large scales, $k<0.1 \,\mathrm{h}\,\mathrm{Mpc^{-1}}$, where the power spectrum is dominated by the tree-level contribution, our analytical results exactly match those of numerical simulations. However, the intermediate regime around $k \sim 0.5 \,\mathrm{h}\,\mathrm{Mpc^{-1}}$ is not well described at one-loop level. The deviation from numerical simulations reaches a maximum of $\sim 20\%$ in this scale regime. On smaller scales, $k > 1 \,\mathrm{h}\,\mathrm{Mpc^{-1}}$, the deviation from numerical results decreases. 
We have reasoned at the end of Sec.\ref{sec:Results} that higher-order loop corrections are unlikely to provide better agreement with numerical simulations on intermediate scales. This is due to the fact that -- at least for a well-behaved perturbative expansion -- increasing orders of loop corrections become important on increasingly smaller scales. In addition, we have observed finely-tuned cancellation between contributions from different one-loop digrams -- a telltale sign that a loop-expansion is not the ideal expansion scheme in our case.  We have, therefore, argued that non-perturbative methods such as 2PI, Dyson-Schwinger equations or the functional renormalisation group approach are more suitable to capture the relevant effects on these scales. Such non-perturbative methods will lead to a restructuring of the macroscopic perturbation theory and -- given an appropriate truncation -- allow us to resum certain classes of diagrams. Such a restructuring and resummation of perturbation theory has allowed us to go from the canonical microscopic PT to the field theory formulation in Sec.\ref{sec:macroscopicFieldTheory}. In this case, the tree-level field theory allowed us to resum all microscopic effects that lead to structure growth on large scales. The practical application of non-perturbative methods requires a suitable truncation scheme. Studying one-loop diagrams and understanding the microscopic processes behind them on a particle-level can help us to devise a sensible truncation procedure. The goal is to identify those processes which determine the shape of the power spectrum in the intermediate scale regime where we observe the largest deviations from numerical simulations. Since our approach contains the full microscopic information of our system by construction, these processes must be accessible within our description.

\begin{acknowledgments}
We are immensely grateful to Robert Lilow for many fruitful discussions and for sharing his thoughts, insights and experience which helped to improve the work presented here.
Furthermore, we would like to thank Matthias Bartelmann, Thomas Gasenzer and Bjoern Malte Schaefer for providing their expertise and valuable input on many aspects of the research presented in this paper. During this work, TD was funded by the Studienstiftung des deutschen Volkes. EK was funded by the Deutsche Forschungsgemeinschaft (DFG, German Research Foundation) -- $452923686$. 
This work was funded in part by the Deutsche Forschungsgemeinschaft (DFG, German Research Foundation) under Germany’s Excellence Strategy EXC-2181/1 - 390900948 (the Heidelberg STRUCTURES Cluster of Excellence).
\end{acknowledgments}

\appendix

\section{Initial Conditions}\label{sec:InitialCondition}
We start from a continuous mass density field in the early matter-dominated epoch. The central objects describing the state of a system are the mass density field $\ini{\rho}_m(\vec{q})$ and the peculiar velocity density field $\ini{\vec{\Pi}}(\vec{q})$. The cosmological principle, homogeneity and isotropy, is responsible for a constant mean background mass density and a vanishing mean peculiar velocity of the system, \ie 
\begin{align}
    \mean{\begin{pmatrix}
        \ini{\rho}_m(\vec{q})\\
        \ini{\vec{\Pi}}(\vec{q})
    \end{pmatrix}}=\begin{pmatrix}
        \ini{\bar{\rho}}_m\\
        \vec{0}
    \end{pmatrix}\,.
\end{align}
 We can thus describe the mass density field by its fluctuations around the mean background density and introduce the initial density contrast $\ini{\delta}(\vec{q}\,)$\, defined as,
\begin{equation}
    \ini{\rho}_m(\vec{q})= \ini{\bar{\rho}}_m(1+\ini{\delta}(\vec{q}\,))\,,\quad\text{with}\,,\quad \mean{\ini{\delta}(\vec{q}\,)}=0\,.
\end{equation}
During that early time period, the formation of structures is very well described by linear Lagrangian perturbation theory. 
Using linear evolution and the continuity equation, the velocity and mass density-contrast fields can be related to the same initial velocity potential $\ini{\psi(\vec{q}})$,
\begin{align}
    \ini{\delta}(\vec{q})&= -\ini{D_+}\Delta\ini{\psi(\vec{q}})\,,\label{eq:initdensity}\\
    \ini{\vec{\Pi}}(\vec{q})&= \ini{\dot{D}}_+\nabla\ini{\psi(\vec{q}}) = \ini{H}\ini{D}_+\nabla\ini{\psi(\vec{q}})\label{eq:initmomentum}\,,
\end{align}
where we used that $\dot{D}_+$ can be solved by the ansatz $\dot{D}_+\propto a$ for the growing mode in the radiation dominated Universe at early enough times.
As predicted by inflationary models and confirmed to high accuracy by observations of the temperature-fluctuations in the CMB, we assume the initial velocity potential $\ini{\psi}(\vec{q})$ to be a Gaussian random field. This property is inherited by all its derivatives, and consequently, due to \eqref{eq:initdensity} and \eqref{eq:initmomentum} also by the initial mass and velocity density, which thus follow a joint multivariate Gaussian distribution. The statistical properties of $(\ini{\delta}(\vec{q}),\,\, \ini{\vec{\Pi}}(\vec{q}))$ are thus fully described by the mean value and the covariance matrix. 
We are finally left with the multivariate Gaussian distribution for the density contrast and the velocity-density field, 
\begin{equation}\label{eq:joint_probability_delta_pi}
    \mathcal{P}(\ini{\delta},\ini{\vec{\Pi}})=\frac{1}{\sqrt{(2\pi)^4\det(C)}}\exp\left(-\frac{1}{2}\begin{pmatrix}
        \ini{\delta}(\vec{q}), &\ini{\vec{\Pi}}(\vec{q})
    \end{pmatrix}\cdot C^{-1}\cdot\begin{pmatrix}
        \ini{\delta}(\vec{q})\\
        \ini{\vec{\Pi}}(\vec{q})
    \end{pmatrix}\right)\,,
\end{equation}
where the covariance matrix is the single remaining parameter characterizing the distribution. It reads 
\begin{equation}
    C = \begin{pmatrix}
        C_{\delta\delta} & \vec{C}_{\delta\vec{\Pi}}^\top\\
        \vec{C}_{\vec{\Pi}\delta}&C_{\vec{\Pi}\otimes\vec{\Pi}}
    \end{pmatrix}=\begin{pmatrix}
        \mean{\ini{\delta}(\vec{q}_1\,)\ini{\delta}(\vec{q}_2\,)} & \mean{\ini{\delta}(\vec{q}_1\,)\ini{\vec{\Pi}}(\vec{q}_2\,)}^\top\\
        \mean{\ini{\vec{\Pi}}(\vec{q}_1\,)\ini{\delta}(\vec{q}_2\,)}&\mean{\ini{\vec{\Pi}}(\vec{q}_1\,)\otimes\ini{\vec{\Pi}}(\vec{q}_2\,)}
    \end{pmatrix}\,.
\end{equation}
Homogeneity and isotropy further imply that all of the correlation functions above are functions of relative distances $|\vec{q}_1-\vec{q}_2|$ only and that $\vec{C}_{\delta\vec{\Pi}}(0)=\vec{0}$ \cite{Fabis_2018}. The above correlation functions can be related to the two-point density-fluctuation correlation function,
\begin{equation}
    C_{\delta\delta}(|\vec{q}_1-\vec{q}_2|)=\mean{\ini{\delta}(\vec{q}_1\,)\ini{\delta}(\vec{q}_2\,)}\,,
\end{equation}
or its Fourier transform, the initial power spectrum $\ini P_{\delta}(k)$, as usual defined as 
\begin{equation}
    \mean{\ini{\delta}(\vec{k}_1)\ini{\delta}(\vec{k}_2)}=(2\pi)^3\dirac{\vec{k}_1+\vec{k}_2}\ini P_{\delta}(k_1)\,.
\end{equation}
The density-momentum and the momentum-momentum correlation matrices are then given by
\begin{align}
    \vec{C}_{\vec{\Pi}\delta}(|\vec{q}_1-\vec{q}_2|) &= \int\md^3\vec{k}_1\,\frac{\vec{k}_1}{k_1^2}\,\ini P_{\delta}(k_1)\,\e^{\mi\vec{k}_1\cdot(\vec{q}_1-\vec{q}_2)}\\
    C_{\vec{\Pi}\otimes\vec{\Pi}}(|\vec{q}_1-\vec{q}_2|) &= \int\md^3\vec{k}_1\,\frac{\vec{k}_1\otimes\vec{k}_1}{k_1^4}\,\ini P_{\delta}(k_1)\,\e^{\mi\vec{k}_1\cdot(\vec{q}_1-\vec{q}_2)}\,.
\end{align}
\\

In order to obtain an initial distribution for individual particles in phase space from the above continuous density field, we fix the total mass $M$ of the mass density field and discretize it into $N$ particles on a volume $V$, such that $M=Nm$, with the particle mass $m$. In the end, we will be interested in taking the thermodynamic limit $N\rightarrow\infty, \,\,V\rightarrow\infty$, while keeping the mean particle number density $\bar{\rho}=N/V=\mathrm{const.}$. The relation between the mean particle number density and the mean mass density is again given by $\bar{\rho}_m=m\bar{\rho}$. 
The process of drawing individual particles from the probability density distribution \eqref{eq:densityProbability} is performed by marginalising the conditional probability $\mathcal{P}(\ini{\ms{x}}_j|\ini{\delta}(\vec{\ms{q}}_j),\,\ini{\vec{\Pi}}(\vec{\ms{q}}_j))$ of finding particle $j$ at phase space position $\ini{\ms{x}}_j=(\vini{q}_j, \vini{u}_j)$ given a specific density contrast $\ini{\delta}_j=\ini{\delta}(\vini{q}_j)$ and momentum field value  $\ini{\vec{\Pi}}_j=\ini{\vec{\Pi}}(\vini{q}_j)$ at that position, over the joint probability $\mathcal{P}(\ini{\delta}_j,\ini{\vec{\Pi}}_j)$,
\begin{equation}\label{eq:initial_probability_q_p}
    \mathcal{P}(\ini{\ms{x}}_j)=\int\md\ini{\delta}_j\,\md^3\ini{\Pi}_j\,\mathcal{P}(\ini{\ms{x}}_j|\ini{\delta}_j,\,\ini{\vec{\Pi}}_j)\,\mathcal{P}(\ini{\delta}_j,\ini{\vec{\Pi}}_j)\,.
\end{equation}
Since the density field is decomposed into individual particles, the probability of finding particle $j$ at position $\ini{\vec{\ms{q}}}_j$ given the density $\ini{\rho}(\vini{q}_j) = \bar{\rho}\left(1-\ini{\delta}(\vini{q}_j)\right)$ at that point is simply 
\begin{equation}\label{eq:conditional_probability_delta_p}
    \mathcal{P}(\ini{\ms{q}}_j|\ini{\delta}_j) = \frac{\ini{\rho}(\vini{q}_j)}{N} = \frac{\bar{\rho}}{N}\left(1-\ini{\delta}(\vini{q}_j)\right)\,.
\end{equation}
Together with the probability $\mathcal{P}(\ini{\ms{u}}_j|\ini{\vec{\Pi}}_j)=\dirac{\ini{\vec{\ms{u}}}_j-\ini{\vec{\Pi}}(\vini{q}_j)}$ for a particle at $\ini{\vec{\ms{q}}}_j$ to have velocity $\ini{\vec{\ms{u}}}_j$, given the velocity field $\ini{\vec{\Pi}}(\vini{q}_j)$ at this point, the joint conditional probability has the form
\begin{equation}\label{eq:conditional_probability_delta_p}
    \mathcal{P}(\ini{\ms{x}}_j|\ini{\delta}_j,\,\ini{\vec{\Pi}}_j) = \frac{\bar{\rho}}{N}\left(1-\ini{\delta}(\vini{q}_j)\right)\,\dirac{\ini{\vec{\ms{u}}}_j-\ini{\vec{\Pi}}(\vini{q}_j)}\,.
\end{equation}
Generalising to $N$ particles and inserting \eqref{eq:conditional_probability_delta_p} and \eqref{eq:joint_probability_delta_pi} in \eqref{eq:initial_probability_q_p} we can perform the integration over $\ini{\vec{\Pi}}(\vini{q}_j)$ right away, 
\begin{equation}
\begin{split}
    \varrho_N(\tini{x},\ini{\eta}) = V^{-N}\!\!\int \md \tini{t}_u\,&\exp\left(-\frac{1}{2}\tini{t}_u\tens{\cdot}\tens{C}_{uu}\tens{\cdot}\tini{t}_u + \mi \tini{t}_u\tens{\cdot}\tini{u}\right)\int \md \tini{\ms t}_z 
     \exp\left(-\frac{1}{2}\tini{\ms t}_{z}\cdot \tens{\ms C}_{\delta\delta}\cdot\tini{\ms t}_z - \tini{\ms t}_z\tens{\cdot} \tens{\ms C}_{\delta\Pi}\tens{\cdot}\tini{\ms t}_u\right)\\
     &\times\exp\left(\mi\sum_j\ini{\ms t}_{z_j}\right)\prod_{j=1}^{N} \int \md z_j\,z_j\,\exp\left(-\mi\sum_j\ini{\ms t}_{z_j}\, z_j\right)\,,
\end{split}
\end{equation}
where we have substituted $z_j = 1 - \ini{\delta}_j$ and take a Fourier transform  introducing the Fourier conjugates $\ini{\vec{\ms t} }_z$ and $\ini{\vec{\ms t} }_u$ to $z$ and $\ini{\vec{\ms u}}$, respectively.  
After performing the integration over $z_j$ and integrating the resulting expression by parts, we obtain
\begin{equation}\label{eq:initial_correlation_before_int_tz}
\begin{split}
    \varrho_N(\tini{x},\ini{\eta}) =& V^{-N}\!\! \int \md \tini{t}_u\,\exp\left(-\frac{1}{2}\tini{t}_u\tens{\cdot}\tens{C}_{uu}\tens{\cdot}\tini{t}_u + \mi \tini{t}_u\tens{\cdot}\tini{u}\right)(-2\,\pi\mi)^N\times\\
    &\times\int \md \tini{\ms t}_z \prod_{j=1}^{N}\frac{\partial}{\partial\, \ini{\ms t}_{z_j}}
     \exp\left(-\frac{1}{2}\tini{\ms t}_{z}\cdot \tens{\ms C}_{\delta\delta}\cdot\tini{\ms t}_z - \tini{\ms t}_z\tens{\cdot} \tens{\ms C}_{\delta\Pi}\tens{\cdot}\tini{\ms t}_u+\mi\sum_j\ini{\ms t}_{z_j}\right)\dirac{\ini{\ms t}_{z_j}}\,.
\end{split}
\end{equation}
Since \eqref{eq:initial_correlation_before_int_tz} is at most quadratic in $\tini{\ms t}_z$, only first- and second-order derivatives can appear in the above expression. 
Using this property and performing the integration over $\tini{\ms t}_z$ finally allows us to
write the initial phase-space density in the compact form
\begin{equation}
    \varrho_N(\tini{x},\ini{\eta})=V^{-N}\int \md \tini{t}_u \mathcal{C}(\tini{q},\tini{t}_u)\exp\left(-\frac{1}{2}\tini{t}_p\tens{\cdot}\tens{C}_{uu}\tens{\cdot}\tini{t}_u + \mi \tini{t}_u\tens{\cdot}\tini{u}\right)\,,
\end{equation}
with the correlation function 
\begin{align}
    \mathcal{C}(\tini{q},\tini{t}_p)=&\prod_{j=1}^N\left(1- \vec{\ms{C}}_{\delta_ju_k}\cdot t_{u_k}\right)+\sum_{\substack{j,k=1 \\ j \neq k}}^N\ms{C}_{\delta_j\delta_k}\prod_{\substack{l=1 \\ l \neq j,k}}^N\left(1- \vec{\ms{C}}_{\delta_lu_s}\cdot t_{u_s}\right)\\
    &+\sum_{\substack{j,k=1 \\ j \neq k}}^N\ms{C}_{\delta_j\delta_k}\sum_{\substack{a,b=1 \\ a \neq b\\a,b\neq j,k}}^N\ms{C}_{\delta_j\delta_k}\prod_{\substack{l=1 \\ l \neq j,k, a, b}}^N\left(1- \vec{\ms{C}}_{\delta_lu_s}\cdot t_{u_s}\right)+\cdots\,.\nonumber
\end{align}
 The correlation matrices $\ms{C}_{\delta_j\delta_k}\,,\vec{\ms{C}}_{\delta_jp_k}\,,\ms{C}_{p_jp_k}$ generalize the entries of the covariance matrix \eqref{eq:CorrelationMatrix} to the $6N$-dimensional phase-space by evaluating \eqref{eq:CorrelationMatrix} at the respective particle positions,
\begin{equation}
    \begin{split}
        \ms{C}_{\delta_j\delta_k}=C_{\delta\delta}(|\vec{\ms{q}}_j-\vec{\ms{q}}_k|)\,,\,\,\,&\,\,\,\vec{\ms{C}}_{\delta_ju_k}=\vec{C}_{\delta\vec{\Pi}}(|\vec{\ms{q}}_j-\vec{\ms{q}}_k|)\,,\\
    \tens{C}_{uu}=\ms{C}_{u_ju_k}\otimes(e_j\otimes e_k)\,,\,\,\,&\,\,\,\text{with}\,\,\,\ms{C}_{u_ju_k}=C_{\vec{\Pi}\otimes\vec{\Pi}}(|\vec{\ms{q}}_j-\vec{\ms{q}}_k|)\,.
    \end{split}
\end{equation}
By going to spherical coordinates, we can write 
\begin{equation}
    \begin{split}
    \vec{\ms{C}}_{\delta_ju_k}&= C_{\delta\vec{\Pi}}(\ms{r})\,e_\ms{r}\,,\\ 
    \tens{C}_{uu} &= C_{\vec{\Pi}\otimes\vec{\Pi}}(\ms{r})\,e_\ms{r}\otimes e_\ms{r} = C^{xx}_{\vec{\Pi}\otimes\vec{\Pi}}(\ms{r})\,(\mathbb{1} - e_\ms{r}\otimes e_\ms{r}) + C^{zz}_{\vec{\Pi}\otimes\vec{\Pi}}(\ms{r})\,e_\ms{r}\otimes e_\ms{r}\,, 
\end{split}
\end{equation}
with the radial unit vector $e_\ms{r}$ and radial distance $\ms{r}=|\vec{\ms{q}}_j-\vec{\ms{q}}_k|$. The matrix $\tens{C}_{uu}$ thus takes on a diagonal form
\begin{align}
    \tens{C}_{uu} 
    =\begin{pmatrix}
        C^{xx}_{\vec{\Pi}\otimes\vec{\Pi}}(\ms{r}) & 0 & 0\\
        0 & C^{xx}_{\vec{\Pi}\otimes\vec{\Pi}}(\ms{r}) & 0\\
        0 & 0 & C^{zz}_{\vec{\Pi}\otimes\vec{\Pi}}(\ms{r})
    \end{pmatrix}\,.
\end{align}
The auto-correlation entries of $\tens{C}_{uu}$ are simply given by $C^{xx}_{\vec{\Pi}\otimes\vec{\Pi}}(0) = C^{zz}_{\vec{\Pi}\otimes\vec{\Pi}}(0) = \sigma^2_u$ with the the velocity dispersion
\begin{equation}
    \sigma^2_u \coloneq  \frac{1}{3}\int\frac{\md^3 k}{(2\,\pi)^3}\frac{\ini{P}_\delta(k)}{k^2}\;.
\end{equation}

\section{Free Cumulants}\label{app:freeCumulants}
Following \cite{Daus2024}, we provide general expressions for the free cumulants for a homogeneous system which are necessary in order to compute the density-fluctuation power spectrum at tree- and one-loop level. By inserting the appropriate trajectories and initial reduced phase-space densities derived in Sec.\ref{sec:cosmoSystem}, the general expressions are be specified to the cosmological system. In \eqref{eq:G_ff_cosmo} and \eqref{eq:G_fB_cosmo} the tree-level cumulants for cosmology are explicitly given. Higher-order free cumulants for cosmology are then obtained analogously from the general expressions below.
\begin{align}
    G_f^{(0)}(S_1)=&\,(2\pi)^3\dirac{\vec k_1}\bar{\rho}\,\varphi(\vec l_1, \ini{t})\,,\\[2ex]
    \begin{split}
    G_{ff}^{(0)}(S_1, S_2)= &\, (2\pi)^3\dirac{\vec{k}_1+\vec{k}_2}\left[\bar{\rho}\,\varphi \bigg(\vec l_1 + \vec l_2+\vec k_1 \frac{T_{1}}{m}+\vec k_2 \frac{T_{2}}{m}\bigg)\right.\\
    &\left.+\bar{\rho}^2\,\widetilde g_2\bigg(\vec k_1,  \vec l_1 + \vec k_1\frac{T_1}{m}, \vec k_2, \vec l_2+\vec k_2\frac{T_2}{m}, \ini{t}\bigg)\right]\,,
    \end{split}
\end{align}
\begin{align}
    \begin{split}
    G^{(0)}_{fff}(S_1,S_2,S_3) &=(2\pi)^3\dirac{\vec{k}_1+\vec{k}_2+\vec{k}_3}\left[\bar{\rho}\,\varphi \bigg(\vec l_1 + \vec l_2 +\vec l_3 + \vec k_1 \frac{T_{1}}{m}+\vec k_2 \frac{T_{2}}{m}+\vec k_3 \frac{T_{3}}{m}\bigg)\right.\\
    &\left.+\bar{\rho}^2\,\varphi \bigg(\vec l_1 + \vec k_1 \frac{T_{1}}{m}\bigg)\widetilde g_2\bigg(\vec k_2,  \vec l_2 + \vec k_2\frac{T_2}{m}, \vec k_3, \vec l_3+\vec k_3\frac{T_3}{m}, \ini{t}\bigg) \right.\\
    &\left.+\bar{\rho}^2\,\varphi \bigg(\vec l_2 + \vec k_2 \frac{T_{2}}{m}\bigg)\widetilde g_2\bigg(\vec k_1,  \vec l_1 + \vec k_1\frac{T_1}{m}, \vec k_3, \vec l_3+\vec k_3\frac{T_3}{m}, \ini{t}\bigg)\right.\\
    &\left.+\bar{\rho}^2\,\varphi \bigg(\vec l_3 + \vec k_3 \frac{T_{3}}{m}\bigg)\widetilde g_2\bigg(\vec k_1,  \vec l_1 + \vec k_1\frac{T_1}{m}, \vec k_2, \vec l_2+\vec k_2\frac{T_2}{m}, \ini{t}\bigg)
    \right.\\
    &\left.+\bar{\rho}^2\,\widetilde g_3\bigg(\vec k_1,  \vec l_1 + \vec k_1\frac{T_1}{m}, \vec k_2, \vec l_2+\vec k_2\frac{T_2}{m},\vec k_3, \vec l_3+\vec k_3\frac{T_3}{m} \ini{t}\bigg)
    \right]\,,
    \end{split}\\[2ex]
    G_{f\mathcal{B}}^{(0)}(S_1, S_2)=&\,(2\pi)^3\dirac{\vec{k}_1+\vec{k}_2}(2\pi)^3\dirac{\vec l_2}b(1,2)\,G_f^{(0)}(\vec{k}_1+\vec{k}_2, \vec{l}_1+\vec{k}_1\frac{T_1}{m} + \vec{l}_2+\vec{k}_2\frac{T_2}{m})\,,\\[2ex]
    \begin{split}
    G^{(0)}_{f\mathcal{B}\mathcal{B}}(S_1, S_2, S_3) = &\, (2\pi)^3\dirac{\vec{k}_1+\vec{k}_2+\vec{k}_3}(2\pi)^3\dirac{\vec l_2}\dirac{\vec l_3}\bigg(b(1,2) + b(3,2)\bigg)\times\\
    &\times\bigg(b(1,3) + b(2,3)\bigg) G_f^{(0)}(\vec{k}_1+\vec{k}_2+\vec{k}_3, \vec{l}_1+\vec{k}_1\frac{T_1}{m} + \vec{l}_2+\vec{k}_2\frac{T_2}{m} +\vec{l}_3+\vec{k}_3\frac{T_3}{m})\,,
    \end{split}\\[2ex]
    \begin{split}
    G^{(0)}_{f\mathcal{B}\mathcal{B}\mathcal{B}}(S_1,S_2,S_3,S_4) &=(2\pi)^3\dirac{\vec{k}_1+\vec{k}_2+\vec{k}_3+\vec{k}_4}(2\pi)^3\dirac{\vec l_2}\dirac{\vec l_3}\dirac{\vec l_4}\times\\
    &\times\bigg(b(1,2)+b(3,2)+ b(4,2)\bigg)\bigg(b(1,3)+b(2,3)+ b(4,3)\bigg)\bigg(b(1,4)+b(2,4)+ b(3,4) \bigg)\times\\
    &\times \,G_f^{(0)}(\vec{k}_1+\vec{k}_2+\vec{k}_3+\vec{k}_4, \vec{l}_1+\vec{k}_1\frac{T_1}{m}+\vec{k}_2\frac{T_2}{m}+\vec{k}_3\frac{T_3}{m} +\vec{k}_4\frac{T_4}{m})\,,
    \end{split}\\[2ex]
    \begin{split}
    G^{(0)}_{ff\mathcal{B}}(S_1,S_2,S_3) &= \, (2\pi)^3\dirac{\vec{k}_1+\vec{k}_2+\vec{k}_3}(2\pi)^3\dirac{\vec l_3}\times\\&\times\left[b(1,3)\,G^{(0)}_{ff}(\vec{k}_1 +\vec{k}_3 , \vec{l}_1+\vec{k}_1\frac{T_1}{m} + \vec{k}_3\frac{T_3}{m}, \vec{k}_2,\vec{l}_2+\vec{k}_2\frac{T_2}{m})\right.\\ 
    &\left.+ b(2,3)\,G^{(0)}_{ff}(\vec{k}_1, \vec{l}_1+\vec{k}_1\frac{T_1}{m}, \vec{k}_2 +\vec{k}_3 , \vec{l}_2+\vec{k}_2\frac{T_2}{m}+\vec{k}_3\frac{T_3}{m})\right]\,,
    \end{split}\\[2ex]
    \begin{split}
    G^{(0)}_{ff\mathcal{B}\mathcal{B}}(S_1,S_2,S_3,S_4) &=(2\pi)^3\dirac{\vec{k}_1+\vec{k}_2+\vec{k}_3+\vec{k}_4}(2\pi)^3\dirac{\vec l_3}\dirac{\vec l_4}\left[\bigg(b(1,3)b(1,4)+b(1,3)b(3,4)\right.\\
    &\left.+b(4,3)b(1,4)\bigg) G^{(0)}_{ff}(\vec{k}_1+\vec{k}_3+\vec{k}_4, \vec{l}_1+\vec{k}_1\frac{T_1}{m}+\vec{k}_3\frac{T_3}{m}+\vec{k}_4\frac{T_4}{m}, \vec{k}_2, \vec{l}_2+\vec{k}_2\frac{T_2}{m})\right.\\
    &\left.+\bigg(b(2,3)b(2,4)+b(2,3)b(3,4)+b(4,3)b(2,4)\bigg)\times\right.\\
    &\left.\times G^{(0)}_{ff}(\vec{k}_1, \vec{l}_1+\vec{k}_1\frac{T_1}{m}, \vec{k}_2+\vec{k}_3+\vec{k}_4, \vec{l}_2+\vec{k}_2\frac{T_2}{m}+\vec{k}_3\frac{T_3}{m}+\vec{k}_4\frac{T_4}{m})\right.\\
    &\left.+b(1,3)b(2,4)\,G^{(0)}_{ff}(\vec{k}_1+\vec{k}_3, \vec{l}_1+\vec{k}_1\frac{T_1}{m}+\vec{k}_3\frac{T_3}{m}, \vec{k}_2+\vec{k}_4, \vec{l}_2+\vec{k}_2\frac{T_2}{m}+\vec{k}_4\frac{T_4}{m})\right.\\
    &\left.+b(2,3)b(1,4)\,G^{(0)}_{ff}(\vec{k}_1+\vec{k}_4, \vec{l}_1+\vec{k}_1\frac{T_1}{m}+\vec{k}_4\frac{T_4}{m}, \vec{k}_2+\vec{k}_3, \vec{l}_2+\vec{k}_2\frac{T_2}{m}+\vec{k}_3\frac{T_3}{m})
    \right]\,,
    \end{split}
\end{align}
\begin{align}
    \begin{split}
    G^{(0)}_{fff\mathcal{B}}(S_1,S_2,S_3,S_4) &=(2\pi)^3\dirac{\vec{k}_1+\vec{k}_2+\vec{k}_3+\vec{k}_4}(2\pi)^3\dirac{\vec l_4}\times\\
    &\times\left[b(1,4)\,G^{(0)}_{fff}(\vec{k}_1+\vec{k}_4,\vec{l}_1+\vec{k}_1\frac{T_1}{m}+\vec{k}_4\frac{T_4}{m}, \vec{k}_2, \vec{l}_2+\vec{k}_2\frac{T_2}{m}, \vec{k}_3, \vec{l}_3+\vec{k}_3\frac{T_3}{m})\right.\\
    &\left.+b(2,4)\,G^{(0)}_{fff}(\vec{k}_1,\vec{l}_1+\vec{k}_1\frac{T_1}{m}, \vec{k}_2+\vec{k}_4, \vec{l}_2+\vec{k}_2\frac{T_2}{m}+\vec{k}_4\frac{T_4}{m}, \vec{k}_3, \vec{l}_3+\vec{k}_3\frac{T_3}{m})\right.\\
    &\left.+b(3,4)\,G^{(0)}_{fff}(\vec{k}_1,\vec{l}_1+\vec{k}_1\frac{T_1}{m}, \vec{k}_2, \vec{l}_2+\vec{k}_2\frac{T_2}{m}, \vec{k}_3+\vec{k}_4, \vec{l}_3+\vec{k}_3\frac{T_3}{m}+\vec{k}_4\frac{T_4}{m})\right]\,.
    \end{split}
\end{align}

\bibliography{main.bib}

\end{document}

%% file: f_tree_final.pdf_tex
\begingroup%
  \makeatletter%
  \providecommand\color[2][]{%
    \errmessage{(Inkscape) Color is used for the text in Inkscape, but the package 'color.sty' is not loaded}%
    \renewcommand\color[2][]{}%
  }%
  \providecommand\transparent[1]{%
    \errmessage{(Inkscape) Transparency is used (non-zero) for the text in Inkscape, but the package 'transparent.sty' is not loaded}%
    \renewcommand\transparent[1]{}%
  }%
  \providecommand\rotatebox[2]{#2}%
  \newcommand*\fsize{\dimexpr\f@size pt\relax}%
  \newcommand*\lineheight[1]{\fontsize{\fsize}{#1\fsize}\selectfont}%
  \ifx\svgwidth\undefined%
    \setlength{\unitlength}{156.89310269bp}%
    \ifx\svgscale\undefined%
      \relax%
    \else%
      \setlength{\unitlength}{\unitlength * \real{\svgscale}}%
    \fi%
  \else%
    \setlength{\unitlength}{\svgwidth}%
  \fi%
  \global\let\svgwidth\undefined%
  \global\let\svgscale\undefined%
  \makeatother%
  \begin{picture}(1,1.02058962)%
    \lineheight{1}%
    \setlength\tabcolsep{0pt}%
    \put(0,0){\includegraphics[width=\unitlength,page=1]{f_tree_final.pdf}}%
    \put(0.83137847,0.97699595){\color[rgb]{0,0,0}\makebox(0,0)[lt]{\lineheight{1.25}\smash{\begin{tabular}[t]{l}$X_1$\end{tabular}}}}%
  \end{picture}%
\endgroup%

%% file: ff_tree_final.pdf_tex
\begingroup%
  \makeatletter%
  \providecommand\color[2][]{%
    \errmessage{(Inkscape) Color is used for the text in Inkscape, but the package 'color.sty' is not loaded}%
    \renewcommand\color[2][]{}%
  }%
  \providecommand\transparent[1]{%
    \errmessage{(Inkscape) Transparency is used (non-zero) for the text in Inkscape, but the package 'transparent.sty' is not loaded}%
    \renewcommand\transparent[1]{}%
  }%
  \providecommand\rotatebox[2]{#2}%
  \newcommand*\fsize{\dimexpr\f@size pt\relax}%
  \newcommand*\lineheight[1]{\fontsize{\fsize}{#1\fsize}\selectfont}%
  \ifx\svgwidth\undefined%
    \setlength{\unitlength}{265.51266888bp}%
    \ifx\svgscale\undefined%
      \relax%
    \else%
      \setlength{\unitlength}{\unitlength * \real{\svgscale}}%
    \fi%
  \else%
    \setlength{\unitlength}{\svgwidth}%
  \fi%
  \global\let\svgwidth\undefined%
  \global\let\svgscale\undefined%
  \makeatother%
  \begin{picture}(1,0.50634351)%
    \lineheight{1}%
    \setlength\tabcolsep{0pt}%
    \put(0,0){\includegraphics[width=\unitlength,page=1]{ff_tree_final.pdf}}%
    \put(-0.00055827,0.48488654){\color[rgb]{0,0,0}\makebox(0,0)[lt]{\lineheight{1.25}\smash{\begin{tabular}[t]{l}$X_1$\end{tabular}}}}%
    \put(0.91700377,0.48488654){\color[rgb]{0,0,0}\makebox(0,0)[lt]{\lineheight{1.25}\smash{\begin{tabular}[t]{l}$X_2$\end{tabular}}}}%
  \end{picture}%
\endgroup%